\documentclass{aa}

\usepackage{txfonts}
\usepackage{graphicx}
\usepackage{natbib}
\bibpunct{(}{)}{;}{a}{}{,}

\begin{document}

\title{MASSIV: Mass Assembly Survey with SINFONI in VVDS\thanks{This work is based mainly on observations collected at the European Southern Observatory (ESO) Very Large Telescope (VLT), Paranal,
Chile, as part of the Programs 179.A-0823, 177.A-0837, 78.A-0177,
75.A-0318, and 70.A-9007. This work also benefits from data products
produced at TERAPIX and the Canadian Astronomy Data Centre as
part of the Canada-France-Hawaii Telescope Legacy Survey, a collaborative project of NRC and CNRS.}}
\subtitle{V. The major merger rate of star-forming galaxies at $0.9 < z < 1.8$ from IFS-based close pairs}
\titlerunning{MASSIV: the major merger rate at $0.9 < z < 1.8$ from IFS-based close pairs}

\authorrunning{C. L\'opez-Sanjuan, et al.}

\author{C.~L\'opez-Sanjuan\inst{1,2} 
\and O.~Le F\`evre\inst{1} 
\and L.~A. M. Tasca\inst{1}
\and B.~Epinat\inst{1,3,4}
\and P.~Amram\inst{1}
\and T.~Contini\inst{3,4}
\and B.~Garilli\inst{5}
\and M.~Kissler-Patig\inst{6}
\and J.~Moultaka\inst{3,4}
\and L.~Paioro\inst{5}
\and V.~Perret\inst{1}
\and J.~Queyrel\inst{3,4}
\and L.~Tresse\inst{1}
\and D.~Vergani\inst{7}
\and C.~Divoy\inst{3,4}}

\institute{Aix Marseille Universit\'e, CNRS, LAM (Laboratoire d'Astrophysique de Marseille) UMR 7326, 13388, Marseille, France 
\and
Centro de Estudios de F\'{\i}sica del Cosmos de Arag\'on, Plaza San Juan 1, planta 2, 44001, Teruel, Spain\\ 
\email{clsj@cefca.es}
\and
Institut de Recherche en Astrophysique et Plan\'etologie (IRAP), CNRS, 14 avenue \'Edouard Belin, 31400, Toulouse, France 
\and
IRAP, Universit\'e de Toulouse, UPS-OMP, Toulouse, France 
\and
IASF-INAF, via Bassini 15, 20133, Milano, Italy 
\and
ESO, Karl-Schwarzschild-Str.2, 85748, Garching b. M\"unchen, Germany 
\and
INAF-IASFBO, Via P. Gobetti 101, 40129 Bologna, Italy 
}

\date{Received 24 August 2012 / Accepted 10 March 2013}

\abstract
{The contribution of the merging process to the early phase of galaxy assembly at $z > 1$ and, in particular, to the build-up of the red sequence, still needs to be accurately assessed.}
{We aim to measure the major merger rate of star-forming galaxies at $0.9 < z <1.8$, using close pairs identified from integral field spectroscopy (IFS).}
{We use the velocity field maps obtained with SINFONI/VLT on the MASSIV sample, selected from the star-forming population in the VVDS. We identify physical pairs of galaxies from the measurement of the relative velocity and the projected separation ($r_{\rm p}$) of the galaxies in the pair. Using the well constrained selection function of the MASSIV sample, we derive at a mean redshift up to $z = 1.54$ the gas-rich major merger fraction (luminosity ratio $\mu = L_2/L_1 \geq 1/4$), and the gas-rich major merger rate using merger time scales from cosmological simulations.}
{We find a high gas-rich major merger fraction of $20.8^{+15.2}_{-6.8}$\%, $20.1^{+8.0}_{-5.1}$\%, and $22.0^{+13.7}_{-7.3}$\% for close pairs with $r_{\rm p} \leq 20h^{-1}$ kpc in redshift ranges $z = [0.94, 1.06], [1.2, 1.5)$, and $[1.5, 1.8)
$, respectively. This translates into a gas-rich major merger rate of $0.116^{+0.084}_{-0.038}$~Gyr$^{-1}$, $0.147^{+0.058}_{-0.037}$~Gyr$^{-1}$, and $0.127^{+0.079}_{-0.042}$~Gyr$^{-1}$ at $z = 1.03, 1.32$, and $1.54$, respectively. Combining our results with previous studies at $z < 1$, the gas-rich major merger rate evolves as $(1+z)^{n}$, with $n = 3.95 \pm 0.12$, up to $z = 1.5$. From these results we infer that $\sim35$\% of the star-forming galaxies with stellar masses $\overline{M}_{\star} = 10^{10}-10^{10.5}\ M_{\odot}$ have undergone a major merger since $z \sim 1.5$. We develop a simple model that shows that, assuming that all gas-rich major mergers lead to early-type galaxies, the combined effect of gas-rich and dry mergers is able to explain most of the evolution in the number density of massive early-type galaxies since $z \sim 1.5$, with our measured gas-rich merger rate accounting for about two-thirds of this evolution.}
{Merging of star-forming galaxies is frequent at around the peak in star formation activity. Our results show that gas-rich mergers make an important contribution to the growth of massive galaxies since $z \sim 1.5$, particularly on the build-up of the red sequence.}


\maketitle

\section{Introduction}

Understanding the mechanisms involved in the mass assembly of
galaxies and their relative role over cosmic time is an important
open topic in modern astrophysics. In particular, the evolution of
the red sequence, which includes passive galaxies dominated by
old stellar populations and an early-type (E/S0) morphology, imposes 
fundamental constraints on the formation and evolution models.

The stellar mass density in the red sequence has increased
by a factor of $\sim 10$ in the 2.5 Gyr between $z = 2$ and $z = 1$, but
only by a factor of $\sim 2$ in the last $7 - 8$ Gyr of cosmic history 
\citep[e.g.,][]{arnouts07,vergani08,ilbert10}. Major mergers, 
the merger of two galaxies with similar stellar masses, is an efficient 
mechanism for creating new passive, early-type galaxies 
\citep[e.g.,][]{naab06ss,rothberg06a,rothberg06b,hopkins08ss,rothberg10,bournaud11}. 
Thus, the knowledge of the merger rate at $z > 1$ is important input when
estimating the relative contribution of merging and cold-gas accretion \citep[e.g.,][]{dekel06} 
in the early assembly of galaxies and, in particular, the role of merging 
in the build-up of the red sequence.

The evolution of the merger rate since $z \sim 1$ is now well
constrained by direct observations. The early measurements using photometric pairs 
\citep{patton97,lefevre00}
or post-merger morphological signatures \citep{conselice03,jogee09} have
been superseded by spectroscopic measurements confirming
physical pairs from the redshift measurement of both components of a major 
merger with a luminosity/mass ratio $\mu \geq 1/4$
\citep[e.g.,][]{lin08,deravel09,deravel11}, as well as for
minor mergers down to $\mu = 1/10$ \citep{clsj11mmvvds}.
With a parametrization of the merger rate's evolution following
$\propto (1 + z)^{n}$, it is observed that the major merger rate's evolution
depends on the luminosity and on the mass of the galaxy sample \citep[e.g.,][]{deravel09}, where massive galaxies 
with $M_{\star} > 10^{11}\ M_{\odot}$ have a higher merger rate,
but with little redshift evolution ($n \sim 0 - 2$), while lower mass
galaxies with $M_{\star} = 10^{9}-10^{11}\ M_{\odot}$ have a lower merging rate but with
stronger redshift evolution ($n \sim 3 - 4$). This mass dependency
seems to explain some of the apparent discrepancy of merger
rate measurements made from observations targeting different
mass samples.

Beyond $z \sim 1$, direct measurements of the merger rate are
still limited. Previous attempts to measure the major merger rate
at $z > 1$ have focused on the identification of merger remnants
from morphological signatures \citep{conselice08,conselice11,bluck12}, 
on the study of projected close pairs 
\citep{ryan08,bluck09,williams11,man12,marmol12,law12}, or on indirect
estimations \citep{cameron12,puech12}. These studies find a high merger 
rate to $z\sim2-3$ but with a large scatter between different measurements.
However, these results are up to now solely based on photometric measurements 
which are increasingly hard to correct for contamination
along the line of sight as redshift increases. Another complication stems from the morphological 
evolution of galaxies, with show more irregular morphologies at high redshifts, and a wavelength
dependency with more multi-component objects present when observed
in the rest-frame UV \citep{law07}, with some of these components possibly 
related to strong star-forming regions rather
than to different dynamical components.

To improve on this situation, it is necessary to obtain spectroscopic confirmation of the physical 
nature of the photometric pairs at $z \gtrsim 1$. In the last years, NIR Integral Field Spectrographs (IFSs), 
like SINFONI on the VLT or OSIRIS on the Keck, have opened the possibility for a systematic 
study of the dynamical field around high redshift galaxies in the optical rest-frame.
Some examples are the MASSIV\footnote{http://www.ast.obs-mip.fr/users/contini/MASSIV/} 
(Mass Assembly Survey with SINFONI in VVDS, \citealt{massiv1}) survey at $0.9 < z < 1.8$,
the SINS\footnote{http://www.mpe.mpg.de/~forster/SINS/sins\_nmfs.html} (Spectroscopic Imaging survey 
in the Near-infrared with SINFONI, \citealt{sins}) survey at $z \sim 2$, or
the Keck-OSIRIS \citep{keckosiris}, the AMAZE (Assessing
the Mass-Abundance redshift -Z- Evolution, \citealt{amaze}) and the LSD (Lyman-break galaxies 
Stellar populations and Dynamics, \citealt{lsd}) surveys at $z \sim 3$.

The MASSIV survey has been designed to target the peak
of the star-formation rate at $0.9 < z < 1.8$, filling the gap
between higher redshift ($z \sim 2$) IFS surveys with those at
$z < 1$, e.g., IMAGES (Intermediate MAss Galaxy Evolution Sequence, \citealt{images}). 
The MASSIV survey has targeted
84 star-forming galaxies at $0.9 < z < 1.8$ with SINFONI , drawn
from the VVDS\footnote{http://cesam.oamp.fr/vvdsproject/} 
(VIMOS VLT Deep Survey, \citealt{lefevre05}) survey. 
MASSIV has been used as a unique opportunity
to study in detail the dynamical state of $0.9 < z < 1.8$ galaxies 
\citep{massiv2}, their metallicity gradients \citep{massiv3}, 
or the evolution of the fundamental mass-size-velocity
relations since $z \sim 1.2$ \citep{massiv4}.

In this paper, using the MASSIV survey, we present for the fist time 
a measurement of the gas-rich major merger rate of star-forming galaxies 
from kinematical close pairs at $0.9 < z < 1.8$. 
Thanks to the large field-of-view of IFS we have access
to the complete surrounding volume of the galaxies when searching 
for close kinematical companions. In addition, the well-defined selection 
of sources from the VVDS and the well controlled selection
function of MASSIV observations ensures the study of a representative 
population of star-forming galaxies at these redshifts
\citep[see][for details]{massiv1}. This all together enables the measurement 
of average volume quantities like the merger fraction and rate.

The paper is organised as follows: in Sect.~\ref{data} we summarise the 
MASSIV data set used to identify merging pairs, and in
Sect.~\ref{method} we develop the methodology to measure the merger fraction 
from IFS data. We report the gas-rich major merger fraction in MASSIV in Sect.~\ref{ffmassiv}, 
and derive the gas-rich major merger rate in Sect.~\ref{mrmassiv}. We discuss the implication 
of our results in Sect.~\ref{discussion}. Finally, we present our conclusions 
in Sect.~\ref{conclusion}. We use $H_0 = 100h$ km s$^{-1}$ Mpc$^{-1}$, $h = 0.7$,
$\Omega_{\rm m} = 0.3$, and $\Omega_{\Lambda} = 0.7$ throughout this paper. All magnitudes
refer to the AB system. The stellar masses assume a \citet{salpeter55} 
initial mass function (IMF).

\begin{figure*}[t]
\centering
\includegraphics[width = 9cm]{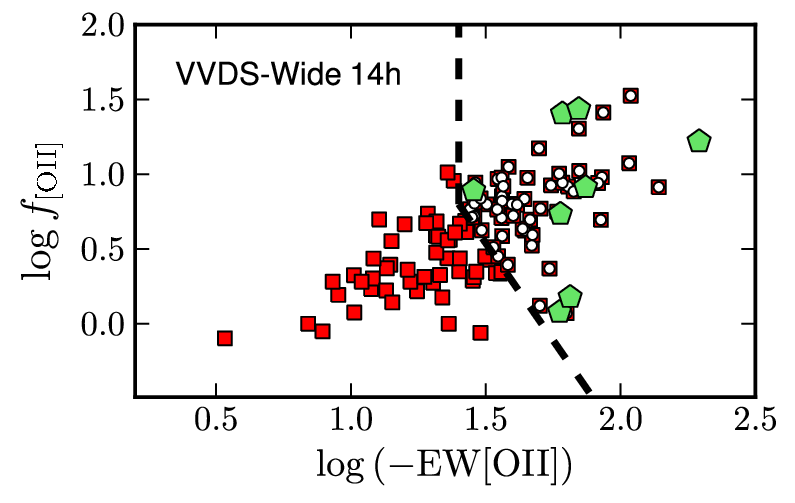}
\includegraphics[width = 9cm]{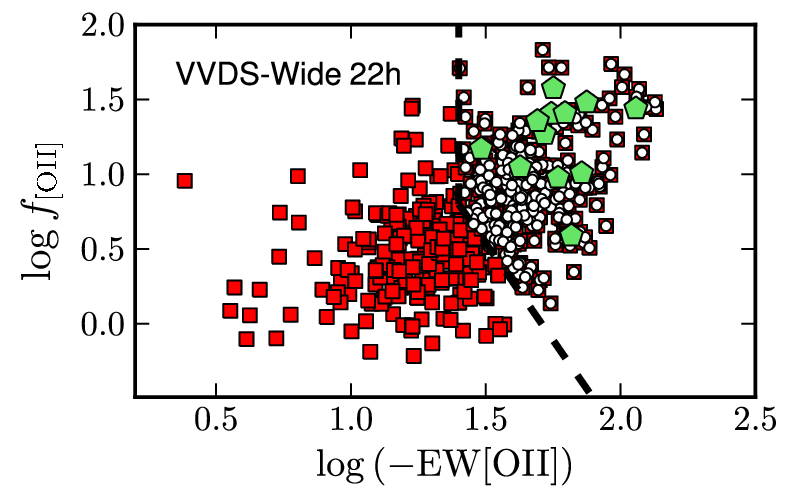}
\includegraphics[width = 9cm]{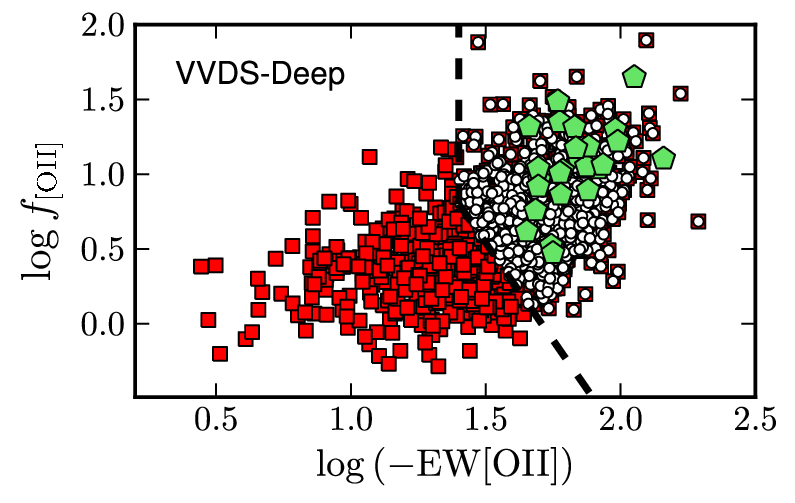}
\includegraphics[width = 9cm]{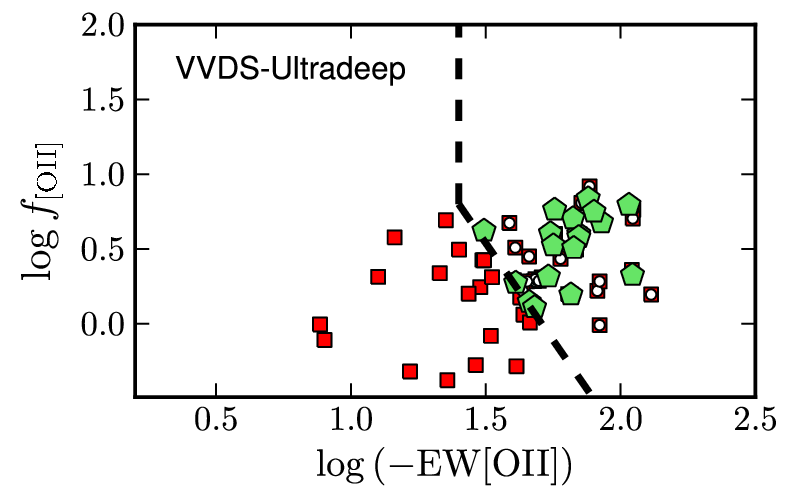}
\caption{MASSIV selection in the [$\ion{O}{ii}$]$\lambda3727$ flux [10$^{-17}$ erg s$^{-1}$ cm$^{-2}$] vs equivalent width [$\AA$] plane (see \citealt{massiv1},
for details). Red squares are the VVDS sources at $0.94 < z < 1.5$ with individual $[\ion{O}{ii}$]$\lambda3727$ line measurement in the VVDS-Wide 14h (top-left), VVDS-Wide 22h (top-right), VVDS-Deep (bottom-left), 
and VVDS-Ultradeep surveys (bottom-right). The dashed lines mark the selection of MASSIV star-forming galaxies. 
White dots are those VVDS sources that fulfil the MASSIV selection. 
Green pentagons are the MASSIV galaxies observed with SINFONI/VLT. 
[{\it A colour version of this plot is available at the electronic edition}].}
\label{wmassiv}
\end{figure*}

\section{The MASSIV data set}\label{data}
The galaxy sample studied in this paper is the final release of the
MASSIV project (ESO Large Programme 179.A-0823; PI.: T.
Contini). A full description of the sample can be found in \citet{massiv1}. 
We briefly summarise some properties of this sample of 84 galaxies below. 
The galaxies were selected from the
VVDS in the RA = 2h area of the deep ($I_{\rm AB} \leq 24$; \citealt{lefevre05}) 
and ultradeep ($I_{\rm AB} \leq 24.75$; Le F\`evre et al., in prep.)
surveys, and in the RA = 14h and RA = 22h areas of the wide survey
($I_{\rm AB} \leq 22.5$; \citealt{vvdswide}).

The MASSIV sources are a subsample of the VVDS star-forming 
population at $z > 0.9$. The star-forming selection 
was performed on the measured intensity
of [$\ion{O}{ii}$]$\lambda3727$ emission line in the VIMOS spectrum (see
Lamareille et al. 2009; Vergani et al. 2008) or, for the cases
where the [$\ion{O}{ii}$]$\lambda3727$ emission line was out of the VIMOS spectral
range (i.e., for $z \gtrsim 1.5$ galaxies), on the $UV$ flux based on their observed
photometric $UBVRIK$ spectral energy distribution and/or
$UV$ rest-frame spectrum. The star formation criteria ensure that rest-frame
optical emission lines H$\alpha$ and [$\ion{N}{ii}$]$\lambda6584$, or in 
a few cases [$\ion{O}{iii}$]$\lambda5007$, will be bright enough to be
observed with SINFONI in the NIR $J$ (sources at $z < 1.1$) and $H$ 
(sources at $z > 1.1$) bands. In addition, to ensure that the H$\alpha$ 
emission line will be detected
with SINFONI with a sufficient signal-to-noise ($S/N$) ratio in a reasonable
exposure time, an additional selection in the [$\ion{O}{ii}$]$\lambda3727$ equivalent width
of $0.9 < z < 1.5$ sources is imposed. The final selection of parent VVDS star-forming galaxies 
in the [$\ion{O}{ii}$]$\lambda3727$ flux vs equivalent width plane at $0.9 < z < 1.5$ 
is shown in Fig.~\ref{wmassiv}.

The 84 MASSIV sources were randomly selected from the star-forming population in the VVDS and
also fulfil two important observational constraints, (i) the observed wavelength of H$\alpha$ line 
falls 9$\AA$ away from strong OH night-sky lines, in order to avoid heavy contamination 
of the galaxy spectrum by sky-subtraction residuals. And (ii) a bright star ($R < 18$ mag) is 
close enough to the target to observe it at higher spatial resolution with the adaptive optics (AO) 
system of SINFONI. 

The most stringent selection criterion used to build the
MASSIV sample is certainly the requirement of a minimum
[$\ion{O}{ii}$]$\lambda3727$ equivalent width. This sensitivity
limit is likely to translate into an overall bias towards younger
and more actively star-forming systems. To check for this possible
effect, \citet{massiv1} compare the properties of the
MASSIV sample with those of the global VVDS sample, which is 
purely apparent magnitude-selected without any colour selection and 
thus representative of the overall star-forming population of galaxies 
at high redshifts. They conclude that the final
MASSIV sample provides a good representation of ``normal"
star-forming galaxies at $0.9 < z < 1.8$ in the stellar mass 
regime $M_{\star} = 10^{9} - 10^{11}\ M_{\odot}$, with a median 
star formation rate $SFR \sim 30\ M_{\odot}$~yr$^{-1}$ and 
a detection limit of $\sim 5\ M_{\odot}$~yr$^{-1}$.

The observations have been performed between April
2007 and January 2011. Most (85\%) of the galaxies in the sample 
have been observed in a seeing-limited mode (with a spatial
sampling of 0.125 arcsec/pixel). However, eleven galaxies have
been acquired with AO assisted with a laser guide
star (AO/LGS, seven with 0.05 and four with 
0.125 arcsec/pixel spatial sampling). The data reduction was performed 
with the ESO SINFONI pipeline (version 2.0.0), using the standard master 
calibration files provided
by ESO. The absolute astrometry for the SINFONI data cubes
was derived from nearby bright stars also used for point spread
function (PSF) measurements. Custom \texttt{IDL} and \texttt{Python} scripts
have been used to flux calibrate, align, and combine all the individual exposures. 
For each galaxy a non sky-subtracted cube
was also created, mainly to estimate the effective spectral resolution. 
For more details on data reduction, we refer to \citet{massiv2}.

We use the H$\alpha$ emission line (or [$\ion{O}{iii}$]$\lambda5007$
in a few cases) in the SINFONI data cubes to derive the kinematical maps 
(flux, velocity field and velocity dispersion map) of
the MASSIV galaxies. To estimate their dynamical properties, we assume 
that the ionised gas rotates in a thin disc with two regimes for the rotation velocity,
a solid body shape in the innermost regions and a plateau in the outskirts.
Using a $\chi^2$ minimization we produce
seeing-corrected velocity and dispersion maps of galaxies with
geometrical inputs weighted for the $S/N$ ratio of each
pixel. We estimate the geometrical parameters used in the fitting
model on the $i-$band best-seeing CFHTLS Megacam images for
all galaxies \citep{cfhtlsT06}, except for VVDS-Wide 14h galaxies that were covered 
with the CFHT-12K/CFHT camera \citep{lefevre04}.
We use the GALFIT software \citep{peng02} that convolves
a PSF with a model galaxy image based on the initial parameter
estimates fitting a \citet{sersic68} profile. Residual maps from the fitting
were used to optimise the results.
At the end of the fitting procedure, GALFIT converges 
into a final set of parameters such as the centre, the position angle, 
and the axial ratio. The $i-$band images were also
used to correct for SINFONI astrometry, using the relative position 
of the PSF star. The morphology and kinematics maps of
the 50 first-epoch galaxies are presented in \citet{massiv2}
together with a more extensive discussion on the model fitting
procedure. The 34 second-epoch galaxies, already included in
the present work, will be presented in a future paper.

We obtain the stellar mass of MASSIV galaxies from a
spectral energy distribution (SED) fit to the photometric and
spectroscopic data with \citet{BC03} stellar population synthesis 
models using the GOSSIP2 Spectral Energy
Distribution tool \citep{franzetti08}. We assume a \citet{salpeter55} IMF, 
and a set of delayed exponential star formation histories with 
galaxy ages in the range from 0.1 to 15 Gyr. 
As input for the SED fitting,  in addition to
the VVDS spectra, we use the multi-band photometric 
observations available in the VVDS fields \citep[see][for further details]{massiv1}.
Following \citet{walcher08} we adopt the probability distribution function 
to obtain the stellar mass.

In all, the final 84 MASSIV galaxies are representative of the normal
star-forming ($SFR \gtrsim 5 M_{\odot}$~yr$^{-1}$) population
of $M_{\star} = 10^{9} - 10^{11}\ M_{\odot}$ galaxies at $0.9 < z < 1.8$.

\section{Measuring the merger fraction from IFS data}\label{method}
We define as a close pair two galaxies with a projected separation in 
the sky plane $r_{\rm p}^{\rm min} \leq r_{\rm p} \leq r_{\rm p}^{\rm max}$ and a 
rest-frame relative velocity along the line of sight $\Delta v \leq \Delta v^{\rm max}$.
We used $\Delta v^{\rm max} = 500\ {\rm km\, s^{-1}}$ and $r_{\rm p}^{\rm max} = 20-30h^{-1}$ kpc
(see Sect.~\ref{areacorr}, for details), while setting $r_{\rm p}^{\rm min} = 0$.
We therefore searched for close
companions in the kinematical maps of the MASSIV sources, analysing and classifying 
the sample using the velocity field and
the velocity dispersion map \citep[see][for details
about the classification]{massiv2}. Note that spectroscopic redshifts from the 
VVDS sources outside MASSIV are not used in this close pair search and only those sources 
detected in the SINFONI data cubes are taken into account. 
We find 20 close pair candidates in the
MASSIV data cubes, and we study these systems in detail to
select major (luminosity difference between both components 
$\mu = L_2/L_1 \geq 1/4$) close pairs (Sect.~\ref{ffmassiv}). 

If there was $N_{\rm p}$ major close pairs in our sample, the major
merger fraction is
\begin{equation}
f_{\rm MM} = \frac{N_{\rm p}}{N},\label{ff}
\end{equation}
where $N$ is the number of principal galaxies targeted for the survey. 
We named principal galaxy the source in the pair closest to the kinematical 
centre of the targeted system, even if it is not
the brightest/more massive galaxy in the pair. In addition, 
we assume that all our close pairs are gas-rich: as shown by \citet{massiv4}, 
the gas fraction of the MASSIV sources is above 10\%, with a median value 
of $\sim30$\%.
The simple definition in Eq.~(\ref{ff}) 
is only valid for volume-limited samples. While our sample is not only
luminosity-limited but spectroscopically defined, we
must take into account the different selection effects, both in the
VVDS parent samples and in MASSIV, in our computation of
the merger fraction.

\subsection{Accounting for selection effects in the VVDS}
Since a fraction of the total number of potential targets in the
VVDS fields have been spectroscopically observed and since the redshifts 
are not measured with 100\% accuracy, we must correct
for the VVDS Target Sampling Rate ($TSR$) and the Redshift
Success Rate ($SSR$), computed as a function of redshift and
source magnitude. The $SSR$ has been assumed independent of
galaxy type, as demonstrated up to $z \sim 1$ in \citet{zucca06}.
Every VVDS source has a redshift confidence flag \citep[see][for details]{lefevre05}, 
that can be flag = 4 (redshift 99\% secure), flag
= 3 (97\% secure), flag = 2 (87\% secure), flag = 9 (redshift from
a single emission line, 90\% secure), flag = 1 (50\% secure), or flag = 0 (no
redshift information). As several VVDS-Deep galaxies 
with flag = 2 have been re-observed in the VVDS-Ultradeep
survey, providing a robust measurement of their redshift, this offers 
the opportunity to estimate the reliability of VVDS-Deep
flag = 2 sources. We thus define a weight $w_{29}$ to take this into
account. We also define the weight $w_{29}$ for flag = 9 sources by
comparison with the latest photometric redshifts in the VVDS-Deep field 
(see \citealt{cucciati10gr}, for details about the latest
photometric data set in the 2h field). By definition, the $w_{29}$ weight
is equal to 1 for flag = 3 and 4 sources in VVDS-Deep, and for
all sources in VVDS-Wide and VVDS-Ultradeep. We derived
the spectroscopic completeness weight for each galaxy $i$ in the
VVDS catalogue as
\begin{equation}
w^i_{\rm VVDS}(z,I_{\rm AB},{\rm flag}) = \frac{w_{29}^{i}(z,I_{\rm AB},{\rm flag})}{TSR^{i}(z,I_{\rm AB}) \times SSR^{i}(z,I_{\rm AB})}.
\end{equation}

The $TSR$, $SSR$ and $w_{29}$ on VVDS-Deep and VVDS-Ultradeep 
were measured in previous works \citep{ilbert06,cucciati12}. 
We assume $TSR = 0.22$ in VVDS-Wide
fields \citep{vvdswide}, and we detail the computation of the
$SSR$ in VVDS-Wide fields in Appendix~\ref{ssrvvds}.

\subsection{Accounting for selection effects in the MASSIV survey}
The MASSIV sources were randomly drawn from the star-forming 
population in the VVDS (Sect.~\ref{data}). We correct for three basic selection
effects in MASSIV.
\begin{itemize}
\item The selection weight, $w_{\rm sel}$. We define this weight as the fraction 
of star-forming galaxies that fulfil the MASSIV selection in the 
[$\ion{O}{ii}$]$\lambda3727$ flux [10$^{-17}$ erg s$^{-1}$ cm$^{-2}$] vs equivalent width [$\AA$] plane at $z < 1.5$ (Fig.~\ref{wmassiv}). The weight $w_{\rm sel}$ tells us how 
representative the MASSIV selection of the global star-forming
population in each of the VVDS surveys (Wide, Deep and
Ultradeep) is, and gives more importance to the more representative samples. 
The VVDS-Deep and Ultradeep have $w_{\rm sel} \sim 0.67$ at $z < 1.5$, 
while the VVDS-Wide fields have
$w_{\rm sel} \sim 0.53$ in the same redshift range. We assume $w_{\rm sel} = 1$
at $z \geq 1.5$, where the selection is based on colour/spectral
properties and all the star-forming galaxies in the VVDS are
thus pre-selected;

\item the MASSIV IFS Rate ($MIR$) is defined as the fraction of
galaxies that fulfil the MASSIV selection and which were
finally observed with SINFONI (Fig.~\ref{wmassiv}). The $MIR$ ranges
from 0.43 for VVDS-Ultradeep to 0.05 for VVDS-Deep;

\item the MASSIV Success Rate ($MSR$) is the fraction of observed
sources with a reliable kinematical classification. This fraction 
is always high, $MSR \gtrsim 0.8$.
\end{itemize}

Finally, the MASSIV weight is
\begin{equation}
w^j_{\rm MASSIV}(x,z) = \frac{w_{\rm sel}\,(x,z)}{MIR^j(x,z) \times MSR^j(x,z)},
\end{equation}
where the index $j$ spans for the MASSIV sources and $x$ refers to the VVDS survey 
(Wide at 14h, Wide at 22h, Deep or Ultradeep) to which the source belongs.

\begin{figure}[t]
\centering
\resizebox{\hsize}{!}{\includegraphics{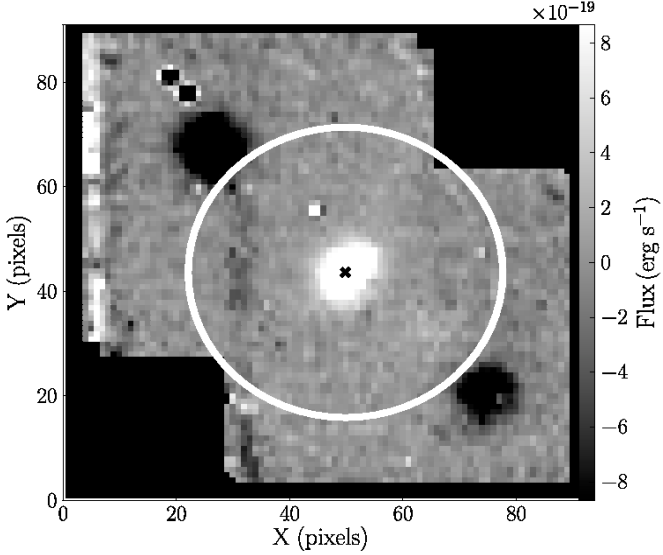}}
\caption{Typical field-of-view of a MASSIV source. We show the flux of
the source 220397579 in the five channels of its reduced, flux calibrated,
data cube centred on the position of the H$\alpha$ emission line (scale in the 
right). The pixel spatial scale is 0.125$\arcsec$, that at 
the redshift of the source ($z = 1.0379$) corresponds to $\sim$1 kpc. 
The source is at the centre of the image (black cross), while the white circle delimits
the $20h^{-1}$ kpc area around the source. In this particular case, 99\% of
this area is covered with MASSIV data and the two negatives of the
source (black regions) are also excluded.}
\label{areacorrfig}
\end{figure}

\subsection{Area correction}\label{areacorr}
Thanks to the large field-of-view of SINFONI, we have access to 
the complete surrounding volume of the principal galaxy
when searching for close companions. This is a major advantage 
with respect to long-slit spectroscopic surveys, in which the
observed number of close pairs is proportional to $TSR^{2}$, 
diminishing the statistics. However, the final reduced MASSIV data
cubes cover a finite area in the sky plane. To deal with this, we
define the area weight as
\begin{equation}
w_{\rm area}^{j}(z,r_{\rm p}^{\rm max}) = \frac{A_{\rm r}}{A_{\rm MASSIV}},
\end{equation}
where $A_{\rm r}$ is the area subtended in the sky plane by a circle of
radius $r_{\rm p}^{\rm max}$ at the redshift of the source $j$, and $A_{\rm MASSIV}$ 
is the area of the same circle covered by the reduced SINFONI mosaic
(Fig.~\ref{areacorrfig}). We state that the optimum search radius is 
$r_{\rm p}^{\rm max} = 20h^{-1}$ kpc. This choice of $r_{\rm p}^{\rm max}$ 
minimises the area correction, with
$w_{\rm area} \sim 1$ in all cases. In addition, the two negatives of the main
source produced by the offset observing procedure and that appear in the extremes 
of the reduced SINFONI mosaic are then excluded from the search area.

Finally, the corrected gas-rich major merger fraction is
\begin{equation}
f_{\rm MM} = \frac{\sum_k^{N_{\rm p}} w^k_{\rm VVDS} w^k_{\rm MASSIV} w^k_{\rm area}}{\sum_j^{N} w^j_{\rm VVDS} w^j_{\rm MASSIV}},
\end{equation}
where the index $j$ and $k$ spans respectively for all MASSIV galaxies and
for the MASSIV galaxies with a major close companion. The error budget in the major 
merger fraction is dominated by the low statistics. We use the Bayesian approach
from \citet{cameron11} to measure the statistical error in the raw
major merger fraction, i.e., in $N_{p}/N$, then scale it with the weighting scheme above.

\begin{table*}
\caption{Close pair candidates at $0.9 < z < 1.8$ in the MASSIV sample.}
\label{mergertab}
\begin{center}
\begin{tabular}{lcccccccc}
\hline\hline\noalign{\smallskip}
ID  &  $z$  &  $r_{\rm p}$    &  $\Delta v$   &  $\Delta m_i$  &  $\Delta m_{K_{\rm s}}$  &  Classification\\
\noalign{\smallskip}
      &       & ($h^{-1}$ kpc)  & (km s$^{-1}$) &              &                    &                 \\
\hline
\noalign{\smallskip}
020294045     &    1.0028   &    2.9     &   180    &    0.4           &    0.3     &   {\bf Major merger}  \\
020386743     &    1.0487   &    3.7     &    80    &    1.8           &  $\cdots$  &   No major merger     \\
020461235     &    1.0349   &    2.8     &    15    &    1.7           &  $\cdots$  &   No major merger     \\
140096645     &    0.9655   &  $\cdots$  & $\cdots$ &    $> 1.5$       &  $\cdots$  &   No major merger     \\
220397579     &    1.0379   &   14.4     &   340    &    0.4           &   -1.4     &   {\bf Major merger}  \\
220544394     &    1.0101   &    7.1     &    50    &    1.3           &  $\cdots$  &   {\bf Major merger}  \\
\noalign{\smallskip}
\hline
\noalign{\smallskip}
020167131     &    1.2246   &   15.2     &   130    &    0.2           &    0.1     &   {\bf Major merger}  \\
020218856     &    1.3103   &  $\cdots$  & $\cdots$ &    $> 1.5$       &  $\cdots$  &   No major merger     \\
020240675     &    1.3270   &  $\cdots$  & $\cdots$ &    $> 1.5$       &  $\cdots$  &   No major merger     \\
020283083     &    1.2818   &    3.8     &     5    &    0.7           &  $\cdots$  &   {\bf Major merger}  \\
020283830     &    1.3949   &    8.5     &   500    &    1.9           &  $\cdots$  &   No major merger     \\
020465775     &    1.3583   &    3.6     &    40    &    0.7           &  $\cdots$  &   {\bf Major merger}  \\
220376206     &    1.2445   &   13.4     &   400    &    2.4           &  $\cdots$  &   No major merger     \\
220544103     &    1.3973   &    5.0     &    75    &    -1.1          &  $\cdots$  &   {\bf Major merger}  \\
910154631     &    1.3347   &    4.2     &   130    &    0.8           &  $\cdots$  &   {\bf Major merger}  \\
910296626     &    1.3558   &   12.1     &   165    &    -0.1          &   -0.2     &   {\bf Major merger}  \\
910337228     &    1.3955   &    9.5     &   220    &    1.4           &  $\cdots$  &   {\bf Major merger}  \\
\noalign{\smallskip}
\hline
\noalign{\smallskip}
020116027     &    1.5302   &   26.8     &   100    &    0.7           &    0.5     &   {\bf Major merger}  \\
910186191     &    1.5399   &   12.7     &   450    &    -0.2          &   -2.4     &   {\bf Major merger}  \\
910274060     &    1.5694   &    3.4     &    10    &    0.2           &  $\cdots$  &   {\bf Major merger}  \\
\hline
\end{tabular}
\end{center}
\end{table*}

\section{Mergers classification and the gas-rich major merger fraction in MASSIV}\label{ffmassiv}

In this section we measure, for the first time, the major merger
fraction at $0.9 < z < 1.8$ from spectroscopically-confirmed
close pairs. The MASSIV observational strategy defines three
natural redshift bins (Sect.~\ref{data}). The low redshift MASSIV sources were
observed in the $J$ band, while the higher redshift ones in the
$H$ band. This translates to a gap in the redshift distribution at
$z \sim 1.1$. In addition, the selection function of MASSIV targets
changes at $z = 1.5$, providing another redshift boundary. 
We take advantage of these natural separations in the data
to estimate the gas-rich major merger fraction in three redshift ranges,
$z_{\rm r,1} = [0.94, 1.06]$, $z_{\rm r,2} = [1.2, 1.5)$, and $z_{\rm r,3} = [1.5, 1.8)$. 
We restrict our study to those galaxies with $I_{\rm AB} \leq 23.9$ to ensure completeness 
in the detection of close pairs (see Appendix~\ref{maglim}, for details).

We follow the steps described bellow to split the close pairs
candidates in the MASSIV data cubes into major and no major
mergers:

\begin{enumerate}
\item As described in \citet{massiv2}, we had performed a
classification of the MASSIV sources based on the shape of
the velocity field (regular or irregular) and the close environment 
(isolated or not-isolated). From this classification, we
had pre-selected as close pair candidates the non-isolated
sources and those identified as mergers from the velocity
field \citep[see][for more details]{massiv2}. We identified
20 close pair candidates (Table~\ref{mergertab}).

\item To study in detail these close pair candidates we used the
deepest $i-$band images in the VVDS fields: CFTH12K (14h-field, 
exposure time of $t_{\rm exp} = 3.6$ ks, \citealt{lefevre04}),
CFHTLS-Wide (22h field, $t_{\rm exp} = 4 - 10$ ks, \citealt{cfhtlsT06}), 
and CFTHLS-Deep (02h field, $t_{\rm exp} \sim 300$ ks,
\citealt{cfhtlsT06}). We run SExtractor on the systems
with well separated sources, and GALFIT (with two S\'ersic
components) on the blended ones, to estimate the luminosity
difference in the i band between both sources, $\Delta m_{i} = m_{i,2} - m_{i,1}$. 
We took $\Delta m_{i} \leq 1.5$ (factor four or less in luminosity)
to identify major mergers.
The observed $i$ band corresponds to $\sim 300 - 350$ nm rest-frame 
in the redshift range of our sample. We stress that for
the blended sources, we run GALFIT v3.0 \citep{galfit3}
without imposing any constraint to the parameters of the fit
and we only used the information from the kinematical maps
to set the initial positions of the sources. We present the residual
maps of these blended sources in Appendix~\ref{galres}.

\item  We confronted the images and the two component fits from
GALFIT with the velocity field and the velocity dispersion
map of the sources. We compared the distribution of H$\alpha$
emission and the geometry of the velocity field to the rest-frame 
UV continuum (or rest-frame visible when NIR images are available). 
The presence of two components with position and geometry concordant 
in the velocity field and in the continuum images, is a strong indication 
of the reality of the pair.

\item  We run SExtractor in the residual image from GALFIT with the principal
source subtracted to obtain a second estimation of $m_{i,2}$, while
with the companion source subtracted to estimate $m_{i,1}$. Then
we compared the $\Delta m_{i}$ derived from the GALFIT modeling with that from these
SExtractor estimations. We found good agreement between both measurements 
(difference of $\sim$0.2 mag or less). The major merger classification 
did not change from the initial estimate.

\item Finally, we also explored $\Delta m_{K_{\rm s}}$
for well separated galaxies and for one blended source (020294045). 
We used the $K_{\rm s}-$band images from UKIDSS-DXS survey (22h field, \citealt{ukidss}) 
and WIRDS (2h field, \citealt{wirds}). The
observed $K_{\rm s}$ band corresponds to $\sim 0.8 - 1$ $\mu$m rest-frame in
the redshift range of our sample, and is a better tracer of the
stellar mass content of the galaxies. After this second check
there is not change in the major merger classification,
except for one close pair (source 910186191), supporting the
previous $i-$band results.
\end{enumerate}

We converged to the steps above after exploring different possibilities.
The stellar mass ratio between the two galaxies in a close pair is the best parameter to classify 
such system as a major merger. However, the estimation of the stellar mass in the blended sources and in some 
well separated pairs is not feasible due to spatial resolution and photometric depth limitations. 
We concluded that the luminosity difference in the $i$ band is an homogeneous criterion applicable 
to the whole MASSIV sample and have additional benefits (i) the observed $i$ band corresponds 
to the UV continuum of the source,
which is related with the star formation of the galaxy and thus with its H$\alpha$ emission. Hence,
both pieces of information should provide a consistent picture about the system under study. (ii) 
The MASSIV PSF is $0.5-0.8\arcsec$, similar to the typical seeing in the $i$ band, $\sim 0.74\arcsec$, 
minimising spatial resolution differences. (iii) The 3$\sigma$ detection magnitude of the CFTLHS $i-$band
images in the 2h field is $\sim$26 (AB). Thanks to this depth we are able to detect the outskirts of the 
fainter MASSIV galaxies, making feasible  the decomposition of the blended sources. We tried the two 
components fit in redder bands, but the lower $S/N$ lead in general to poor constraints. And (iv) the
VVDS parent samples are $i$-band magnitude selected, so the completeness of the MASSIV close pair 
sample is well defined in the $i$ band (see Appendix~\ref{maglim}, for further details). In conclusion, 
even if $\Delta m_{i}$ is not the optimal parameter to select major merger systems, it is the best
 practical one with the current data sets.

In the next sections we present the decomposition and the
classification of the 20 close pair candidates in the
MASSIV data cubes (Table~\ref{mergertab}).



	\begin{figure*}[t!]
	\resizebox{0.32\hsize}{!}{\includegraphics{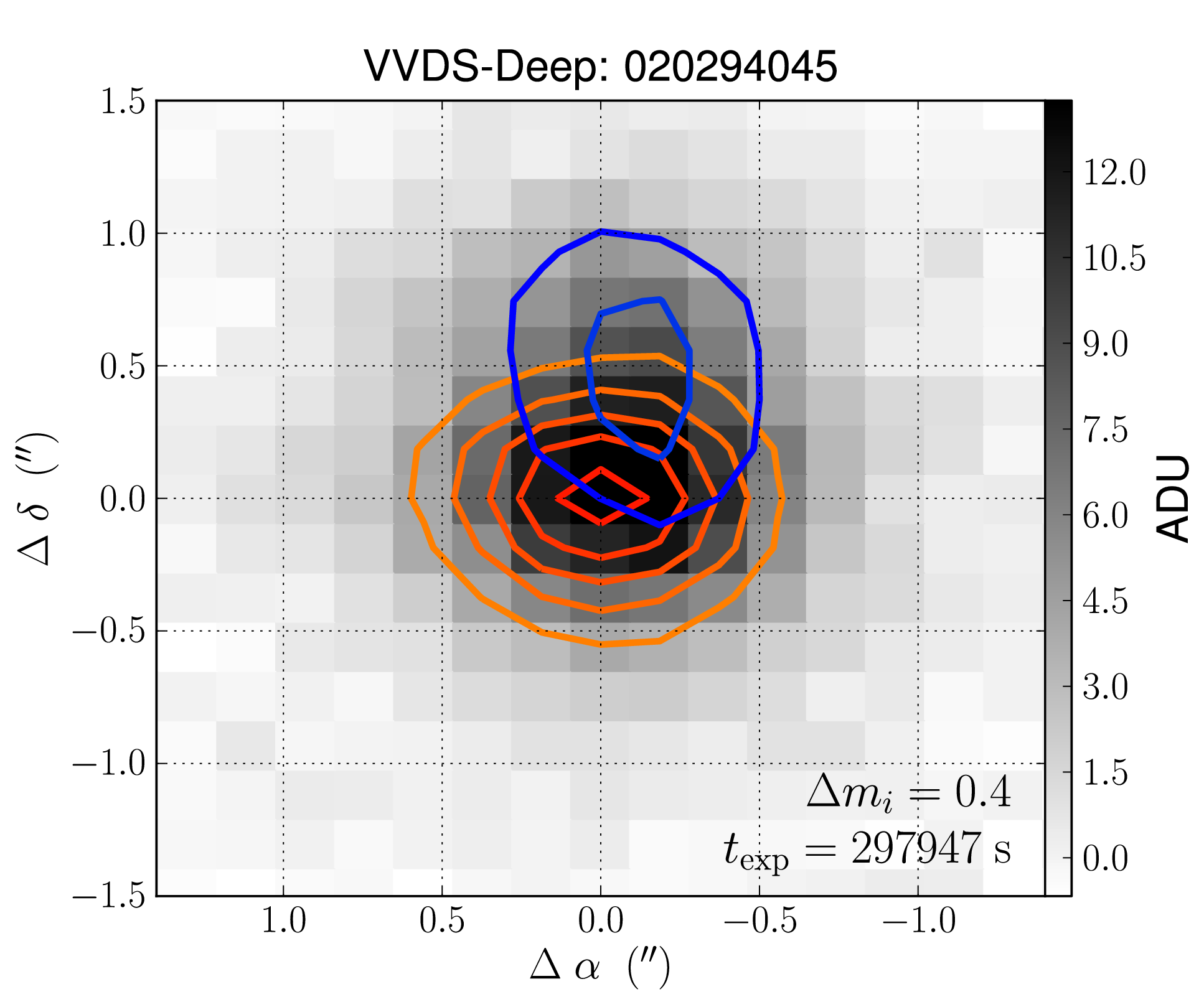}}
	\resizebox{0.32\hsize}{!}{\includegraphics{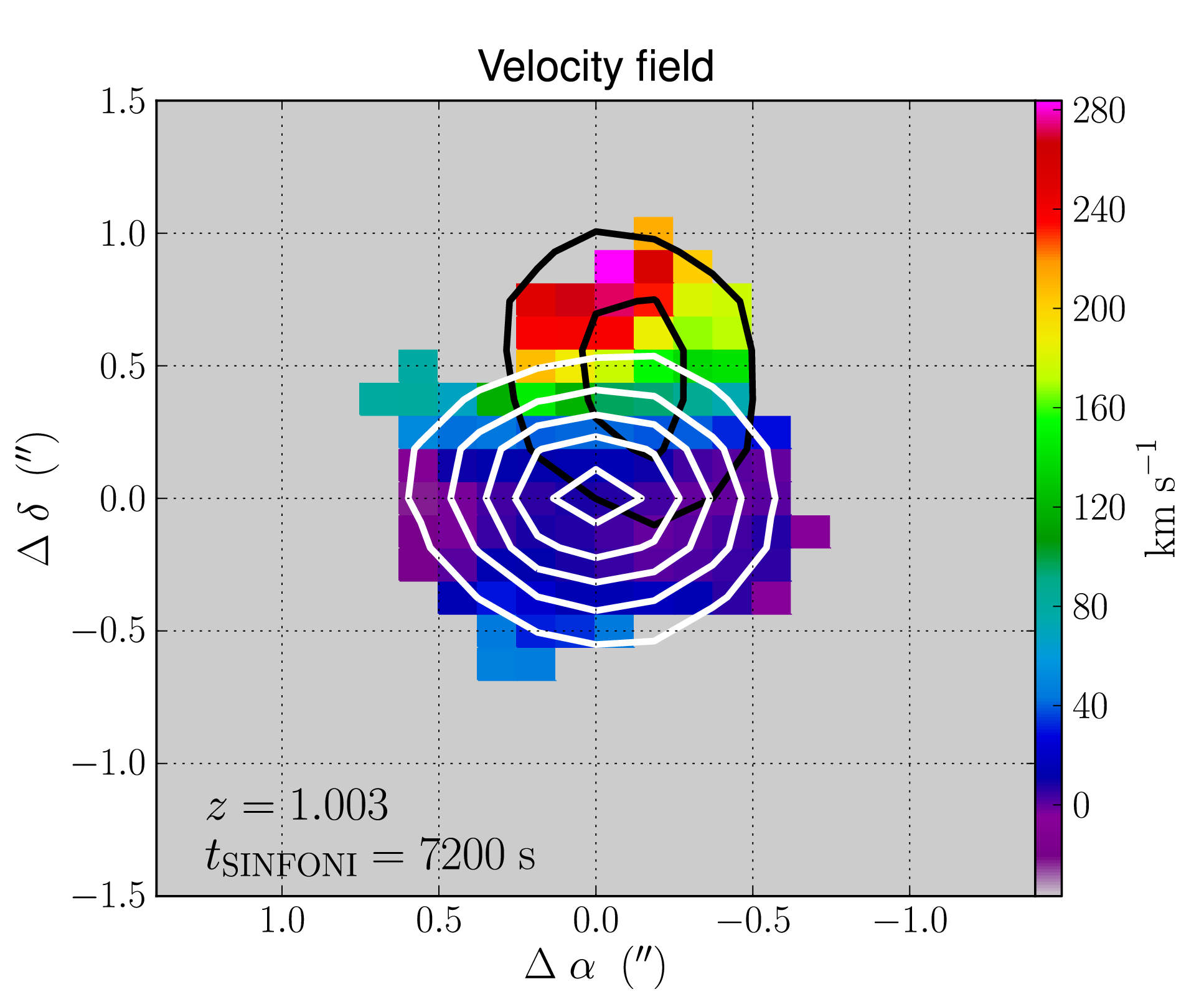}}
	\resizebox{0.32\hsize}{!}{\includegraphics{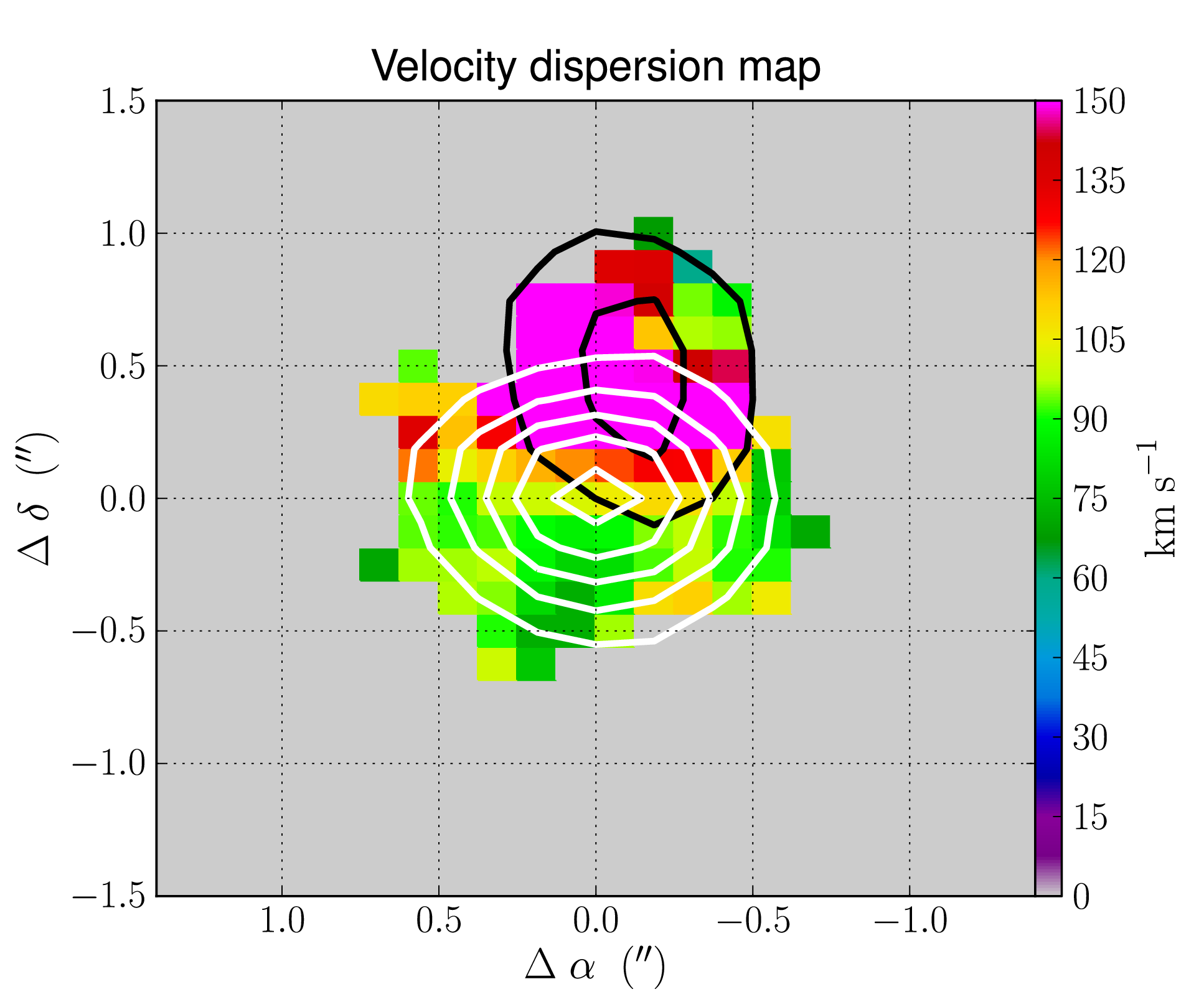}}
	\caption{The $i-$band image ({\it left panel}, scale in ADU in the right), 
		the velocity field ({\it central panel}, scale in km s$^{-1}$ in the right), 
		and the velocity dispersion map ({\it right panel}, scale in km s$^{-1}$ in the right) 
		of the MASSIV source 020294045 (major merger). On each map, North is up and East is left. 
		The level contours mark the isophotes of the two components obtained with GALFIT in 
		the $i-$band image. The principal galaxy (red/white) is the one closer to the 
		kinematical centre of the system and sets the origin in right 			
		ascension ($\alpha$) and declination ($\delta$), while the companion (blue/black) 
		is the secondary component.
		The outer contour marks the 3.43 ADU (7$\sigma_{\rm sky}$) isophote.
		The next contours mark brighter isophotes in 1.96 ADU (4$\sigma_{\rm sky}$ ) steps. 
		The luminosity difference between both components in the $i$ band, $\Delta m_{i}$, 
		as well as the CFHTLS exposure time, $t_{\rm exp}$, are shown in the left panel. 
		The redshift of the source and the on-source SINFONI exposure time, $t_{\rm SINFONI}$, 
		are shown in the central panel. 
		[{\it A colour version of this plot is available at the electronic edition}].}
	\label{src_020294045_i}
	\end{figure*}

	\begin{figure*}[t!]
	\resizebox{0.32\hsize}{!}{\includegraphics{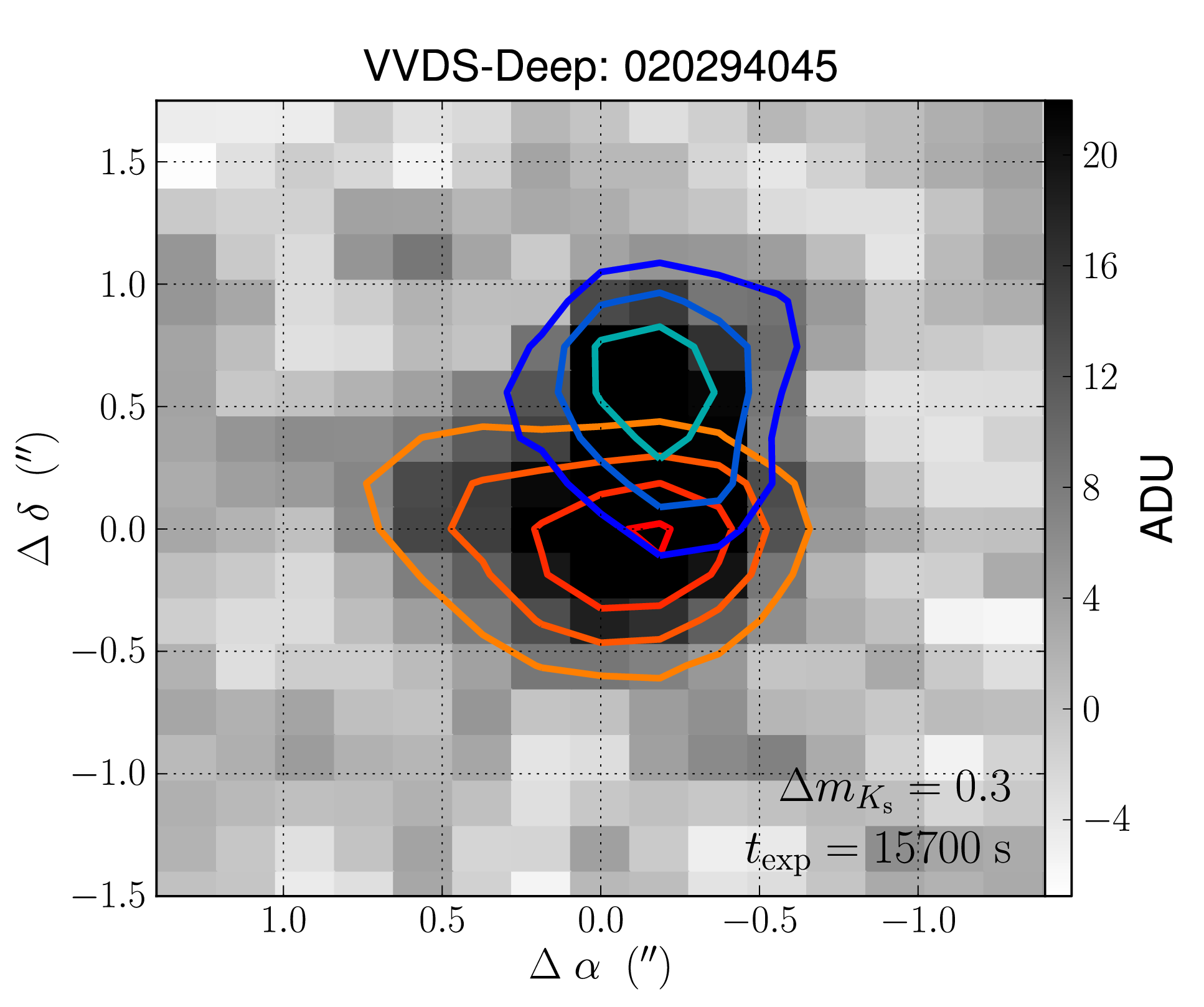}}
	\resizebox{0.32\hsize}{!}{\includegraphics{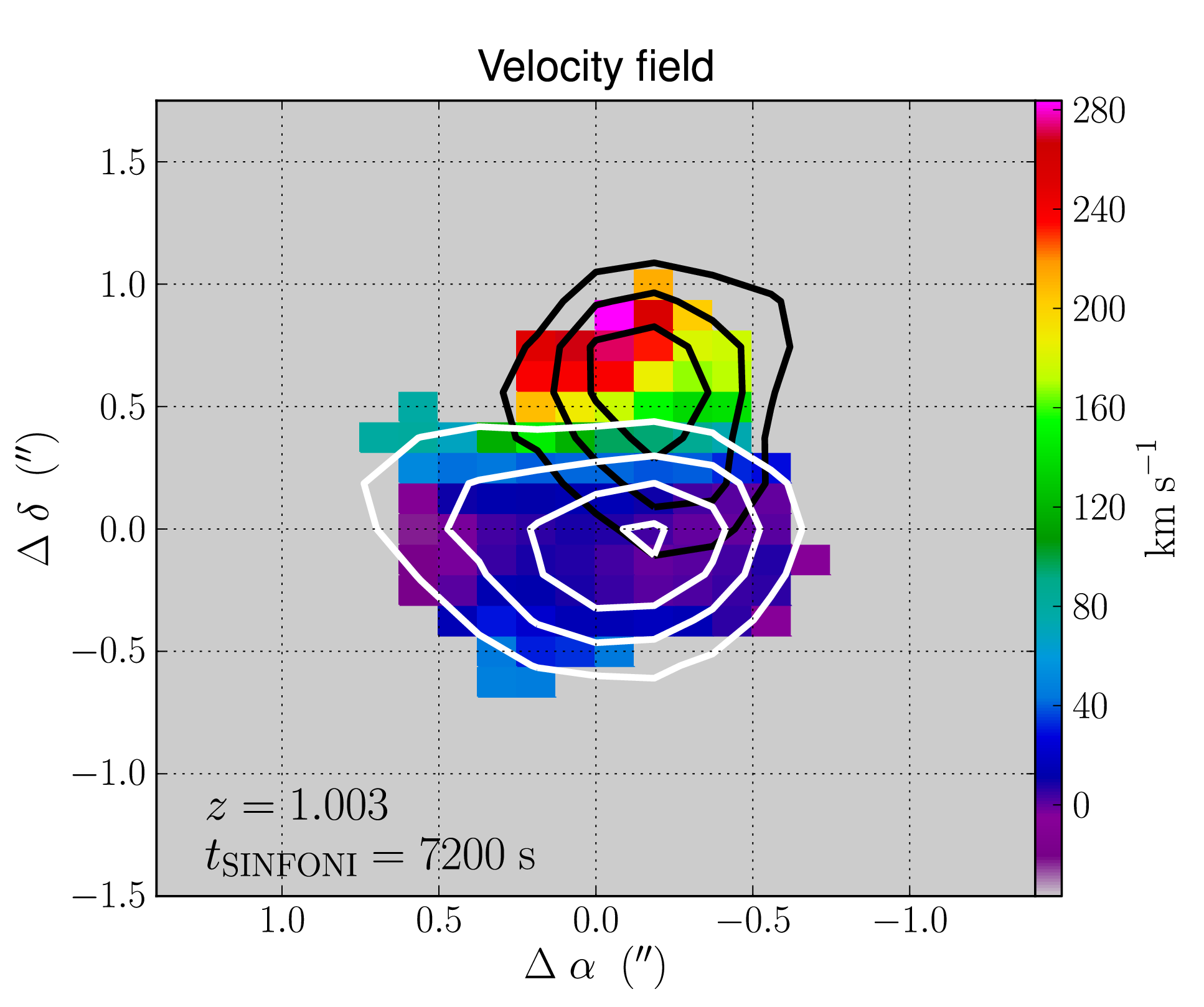}}
	\resizebox{0.32\hsize}{!}{\includegraphics{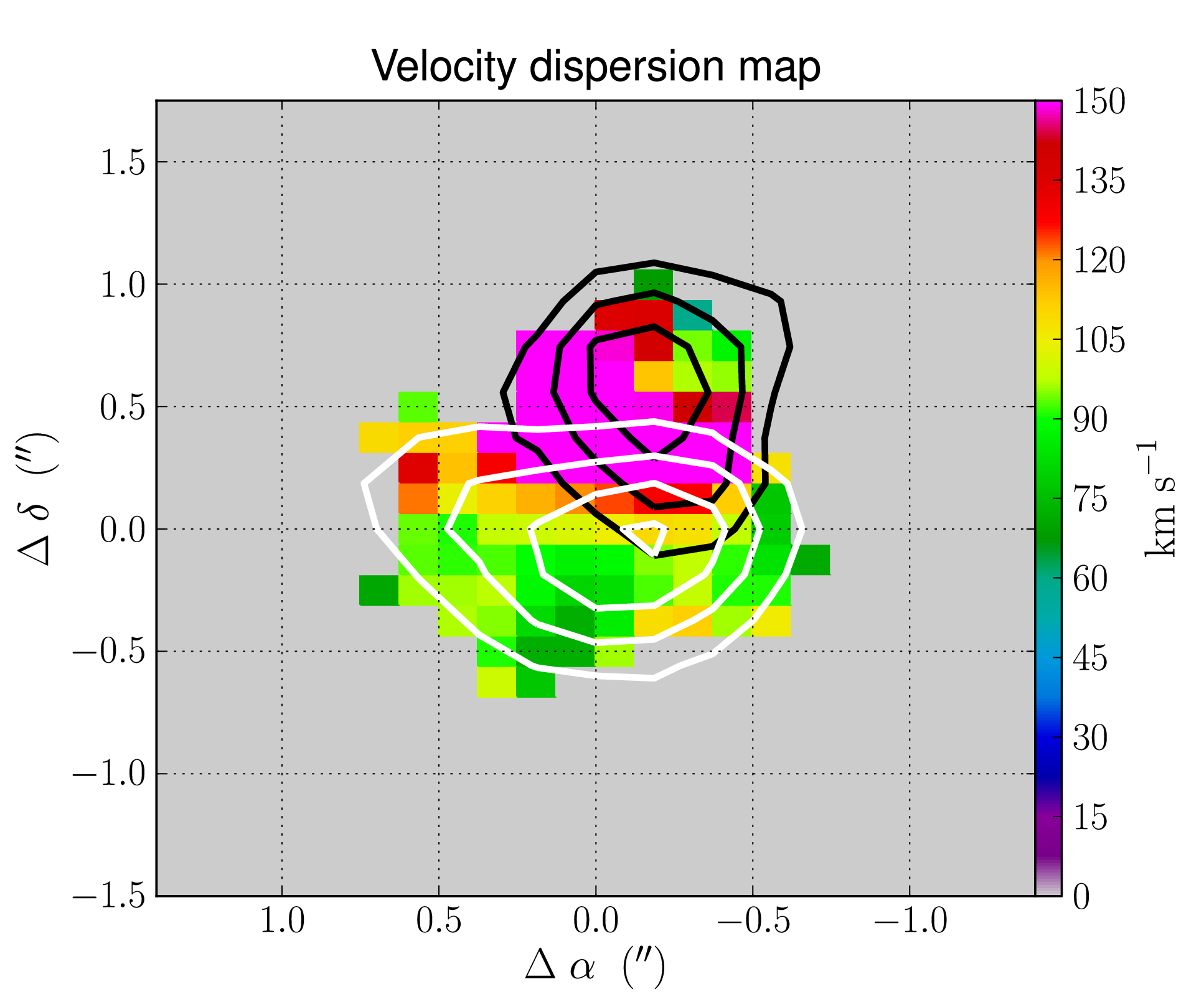}}
	\caption{The same as Fig.~\ref{src_020294045_i}, but for the $K_{\rm s}-$band image ({\it left panel}) of the MASSIV 			source 020294045 (major merger). The outer contour marks the 6.76 ADU (2$\sigma_{\rm sky}$) isophote, 
		while brighter isophotes 
		increase in 5.07 ADU (1.5$\sigma_{\rm sky}$) steps. The flux difference between both components in the 
		$K_{\rm s}$ band, $\Delta m_{K_{\rm s}}$, as well as the WIRDS exposure time, $t_{\rm exp}$, are show in the 
		left panel. [{\it A colour version of this plot is available at the electronic edition}].}
	\label{src_020294045_ks}
	\end{figure*}


	\begin{figure*}[t!]
	\resizebox{0.32\hsize}{!}{\includegraphics{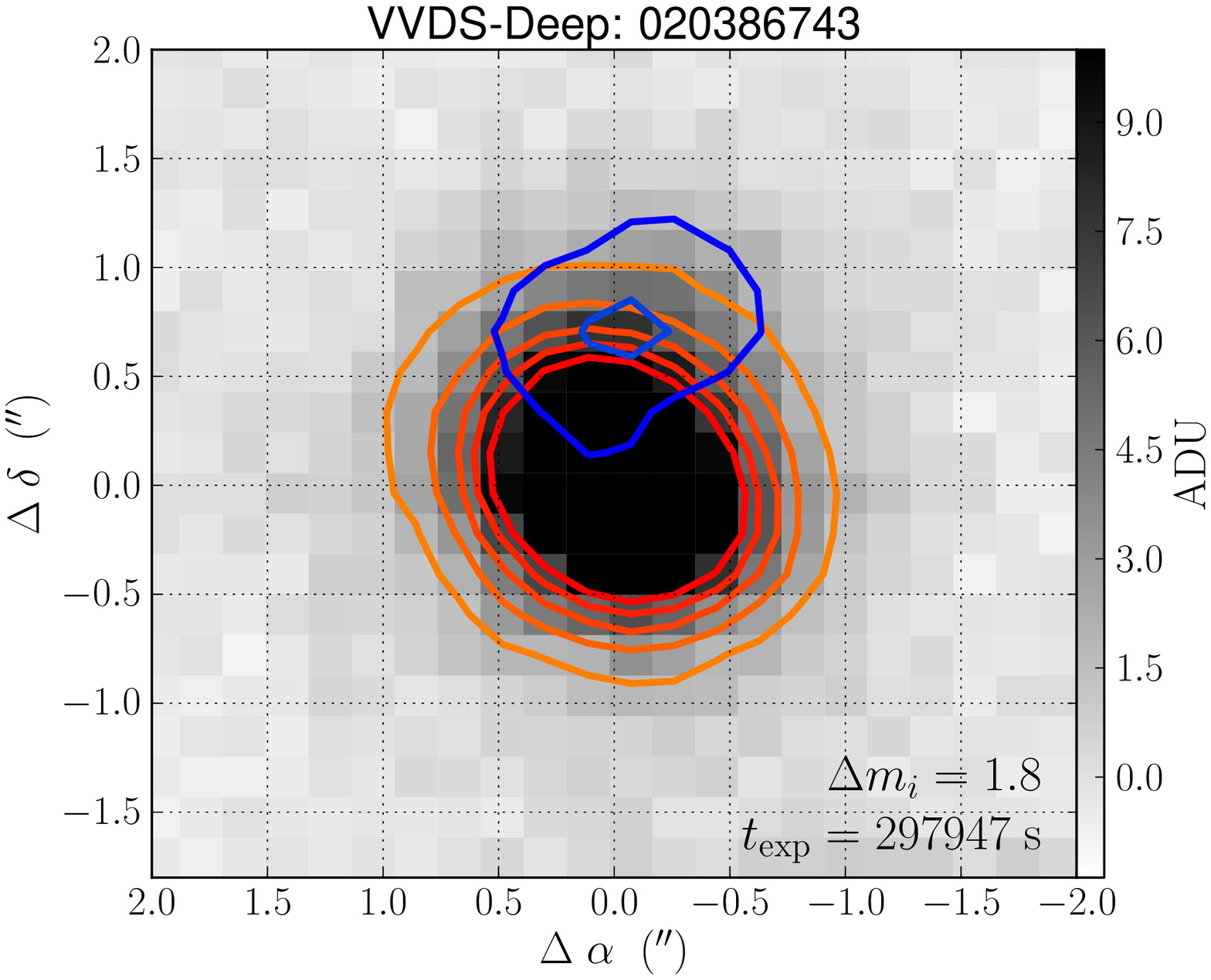}}
	\resizebox{0.32\hsize}{!}{\includegraphics{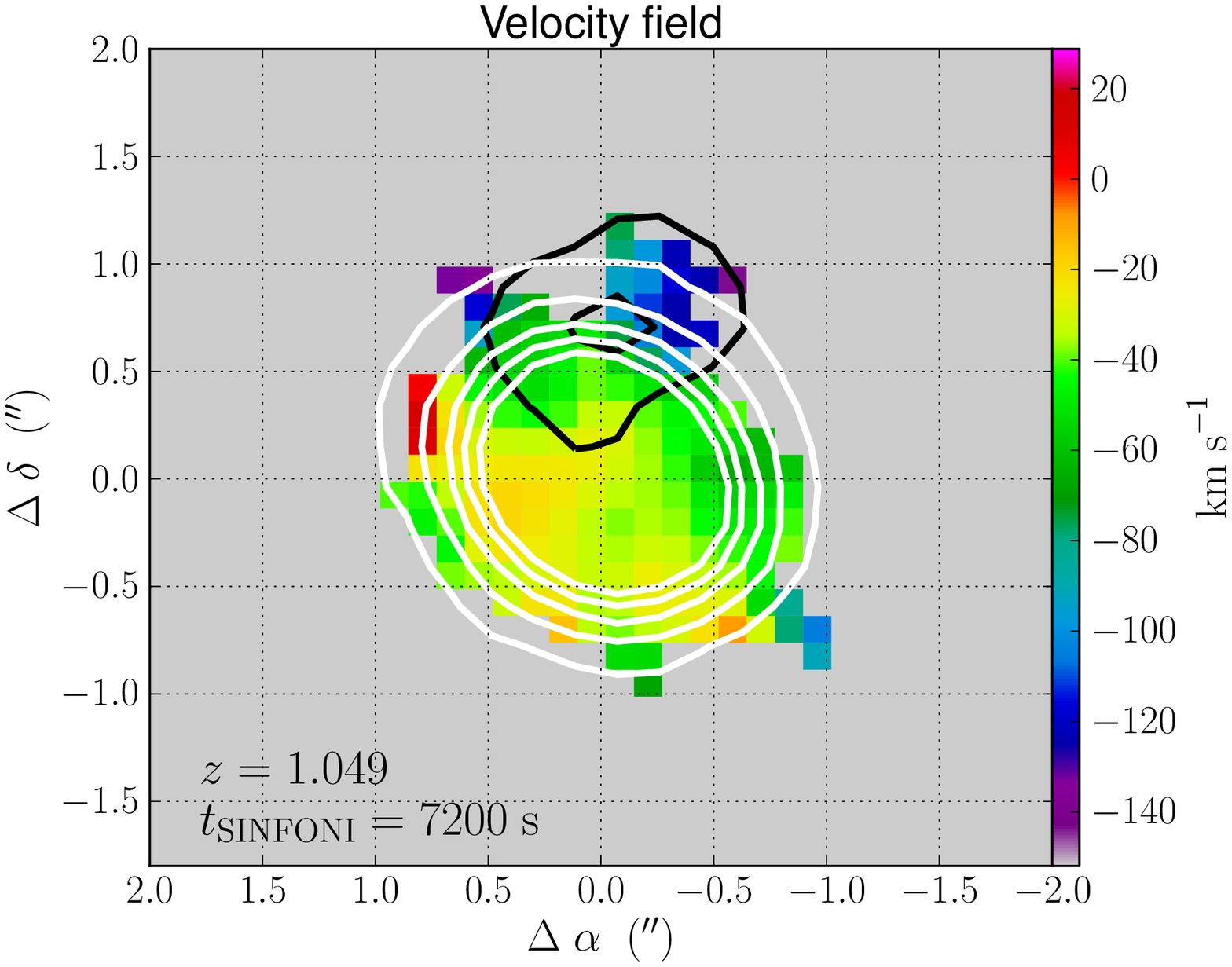}}
	\resizebox{0.32\hsize}{!}{\includegraphics{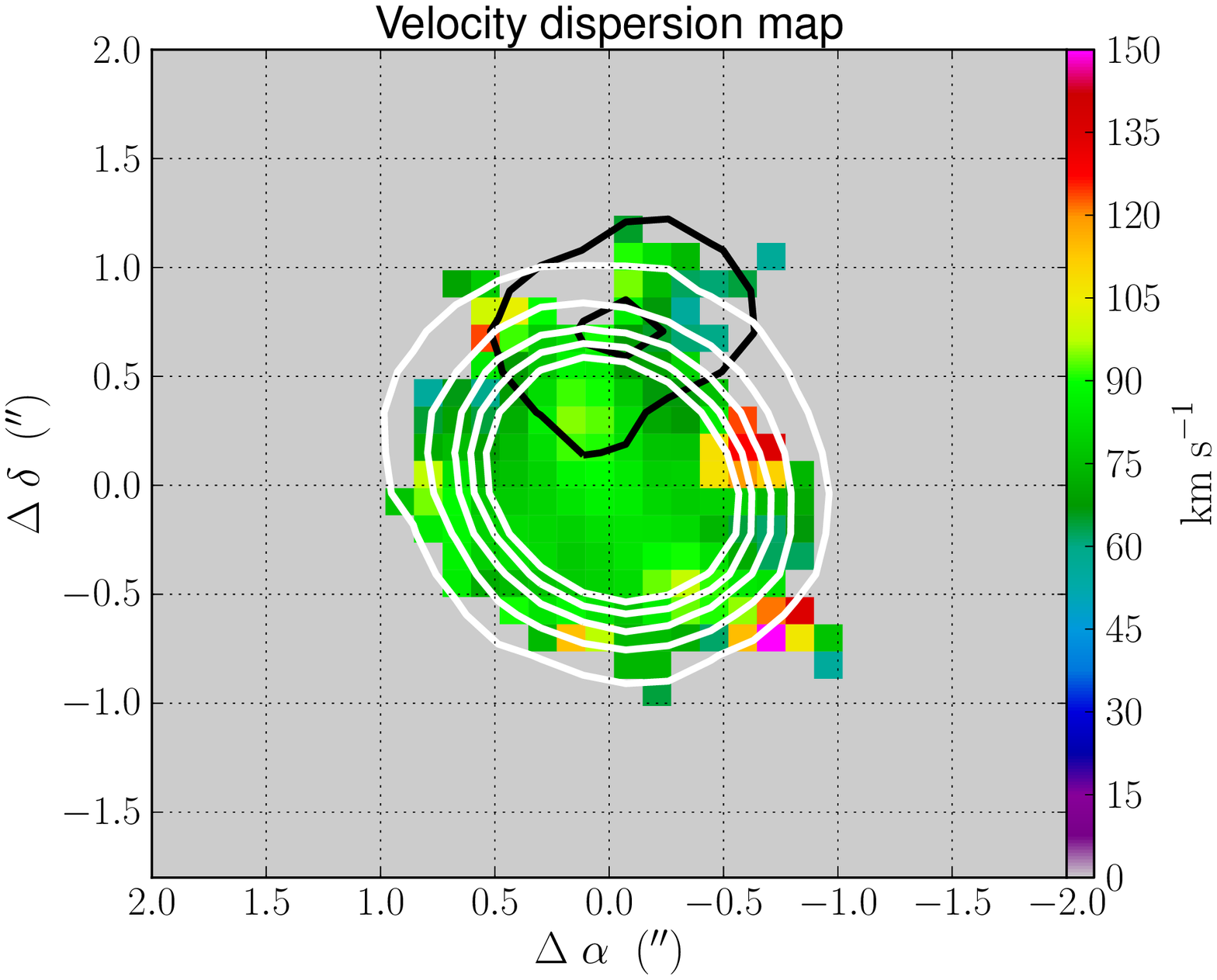}}
	\caption{ The same as Fig.~\ref{src_020294045_i}, but for the MASSIV source 020386743 (no major merger). 
		The outer contour marks the 1.41 ADU (2$\sigma_{\rm sky}$) isophote, while brighter isophotes increase 
		in 1.76 ADU (2.5$\sigma_{\rm sky}$) steps up to 10 ADU to avoid crowded figures.
		[{\it A colour version of this plot is available at the electronic edition}].}
	\label{src_020386743}
	\end{figure*}

	
	\begin{figure*}[t!]
	\resizebox{0.32\hsize}{!}{\includegraphics{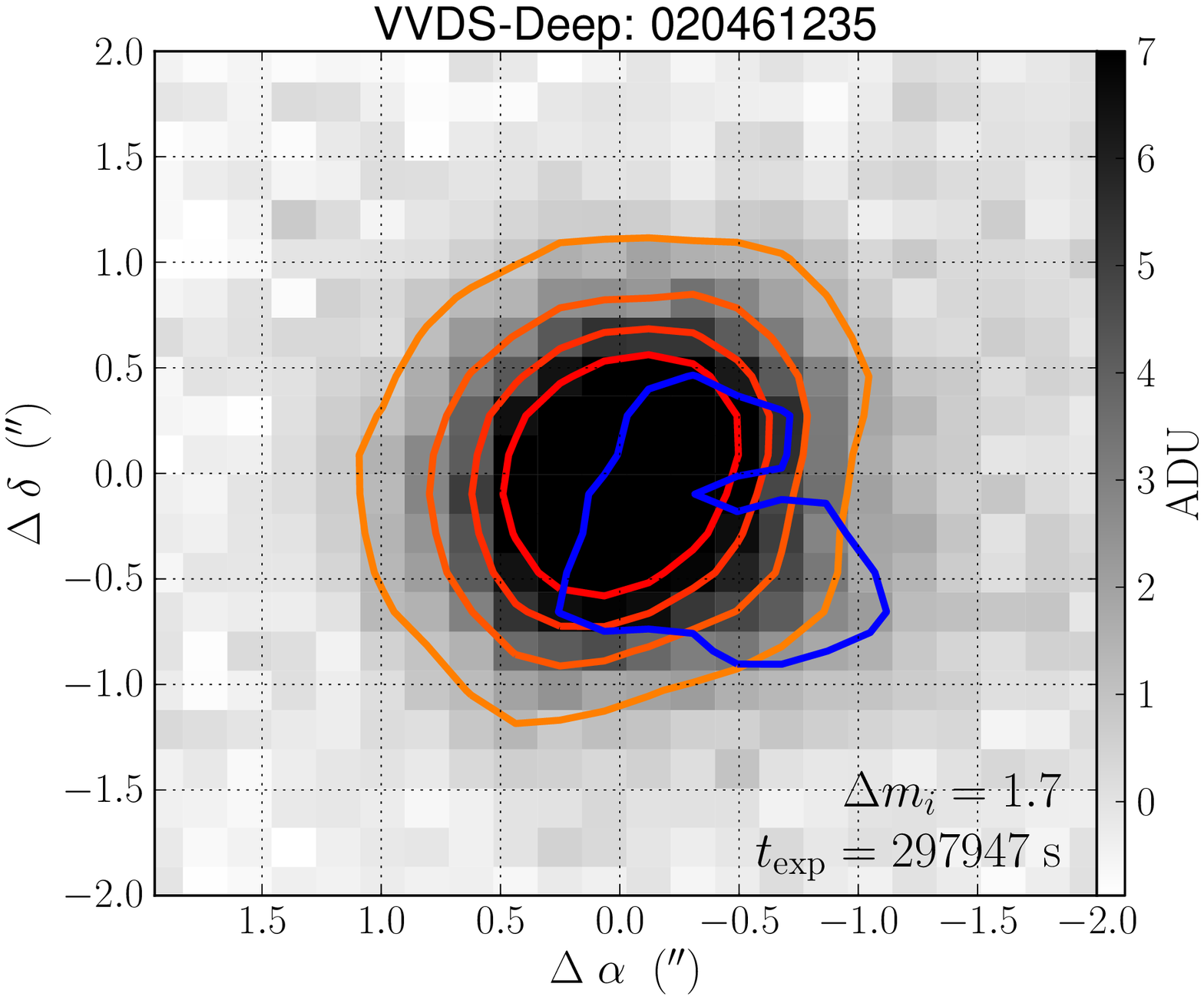}}
	\resizebox{0.32\hsize}{!}{\includegraphics{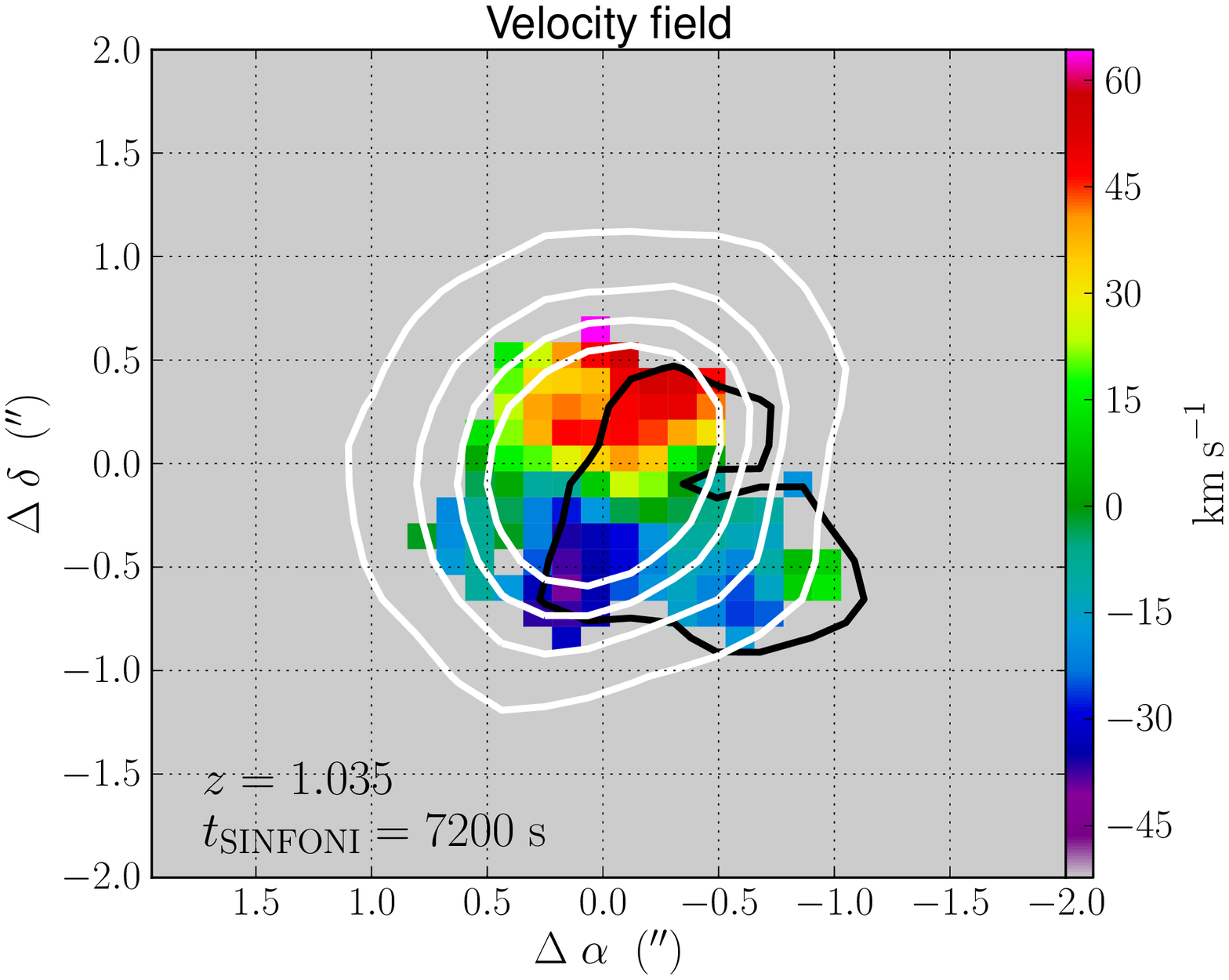}}
	\resizebox{0.32\hsize}{!}{\includegraphics{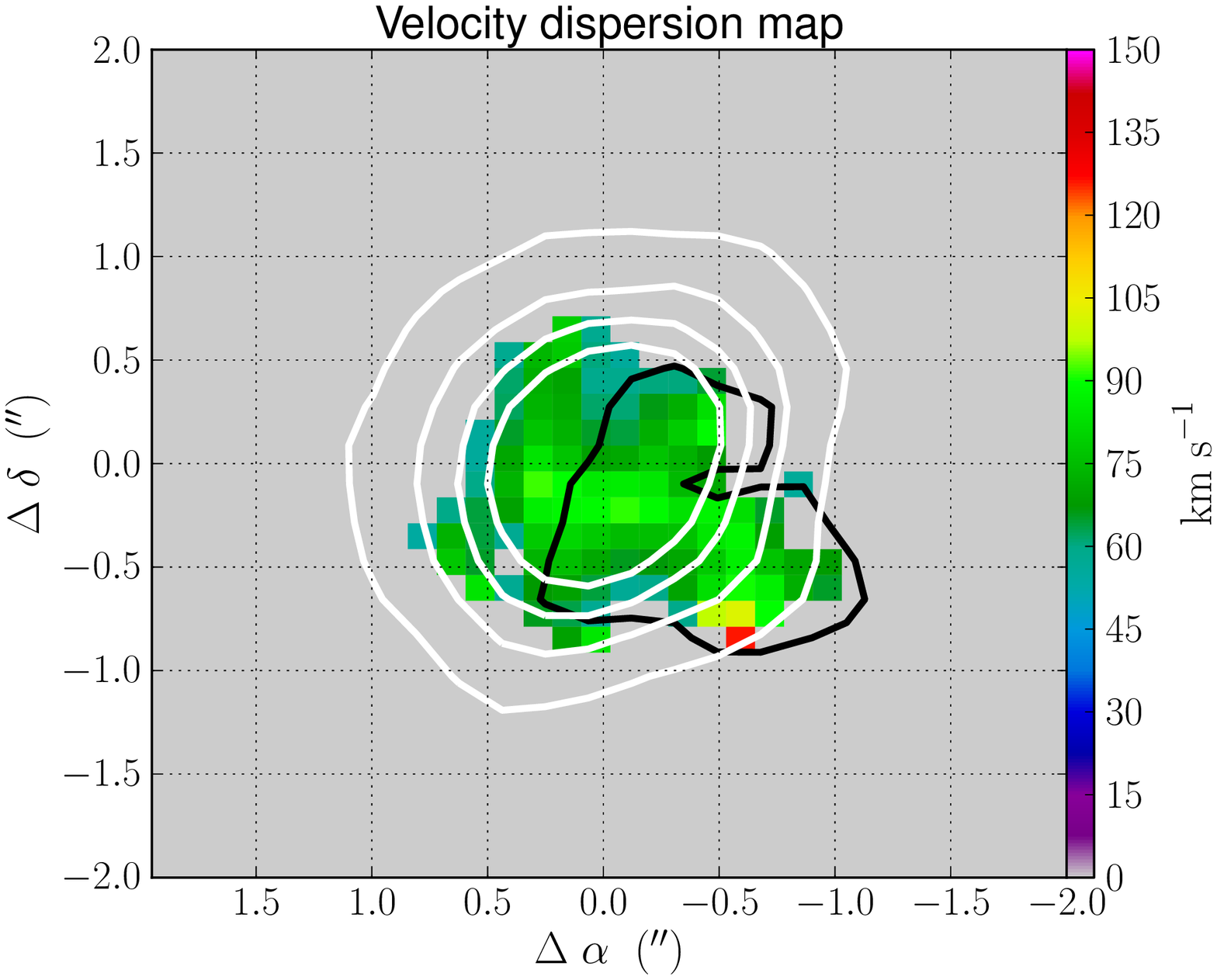}}
	\caption{The same as Fig.~\ref{src_020294045_i}, but for the MASSIV source 020461235 (no major merger). The outer contour 			marks the 1.31 ADU (3$\sigma_{\rm sky}$) isophote, while
		brighter isophotes increase in 1.75 ADU (4$\sigma_{\rm sky}$) steps 
		up to 7 ADU to avoid crowded figures. [{\it A colour version of this plot 
		is available at the electronic edition}].}
	\label{src_020461235}
	\end{figure*}

	
	\begin{figure*}[t!]
	\resizebox{0.32\hsize}{!}{\includegraphics{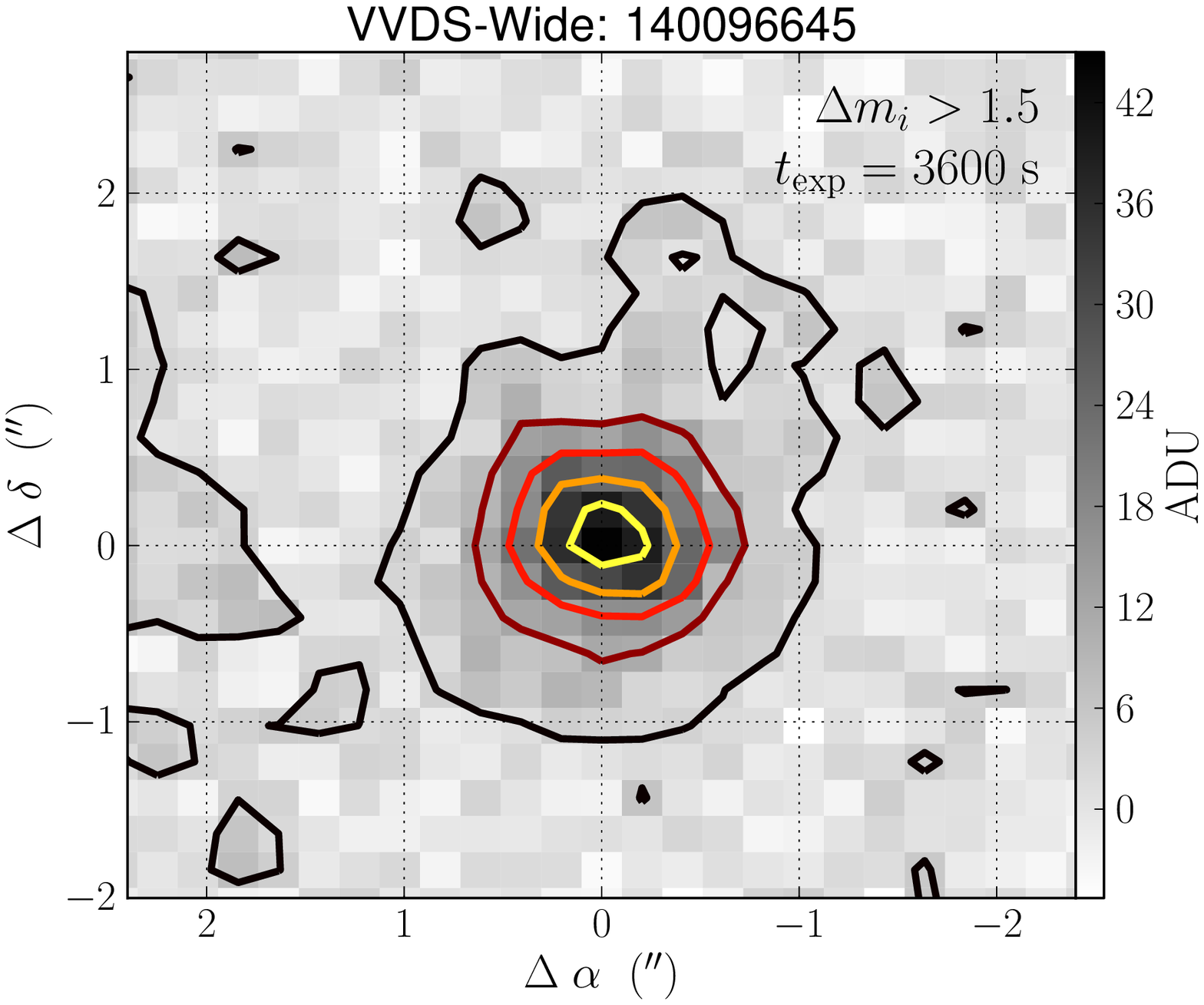}}
	\resizebox{0.32\hsize}{!}{\includegraphics{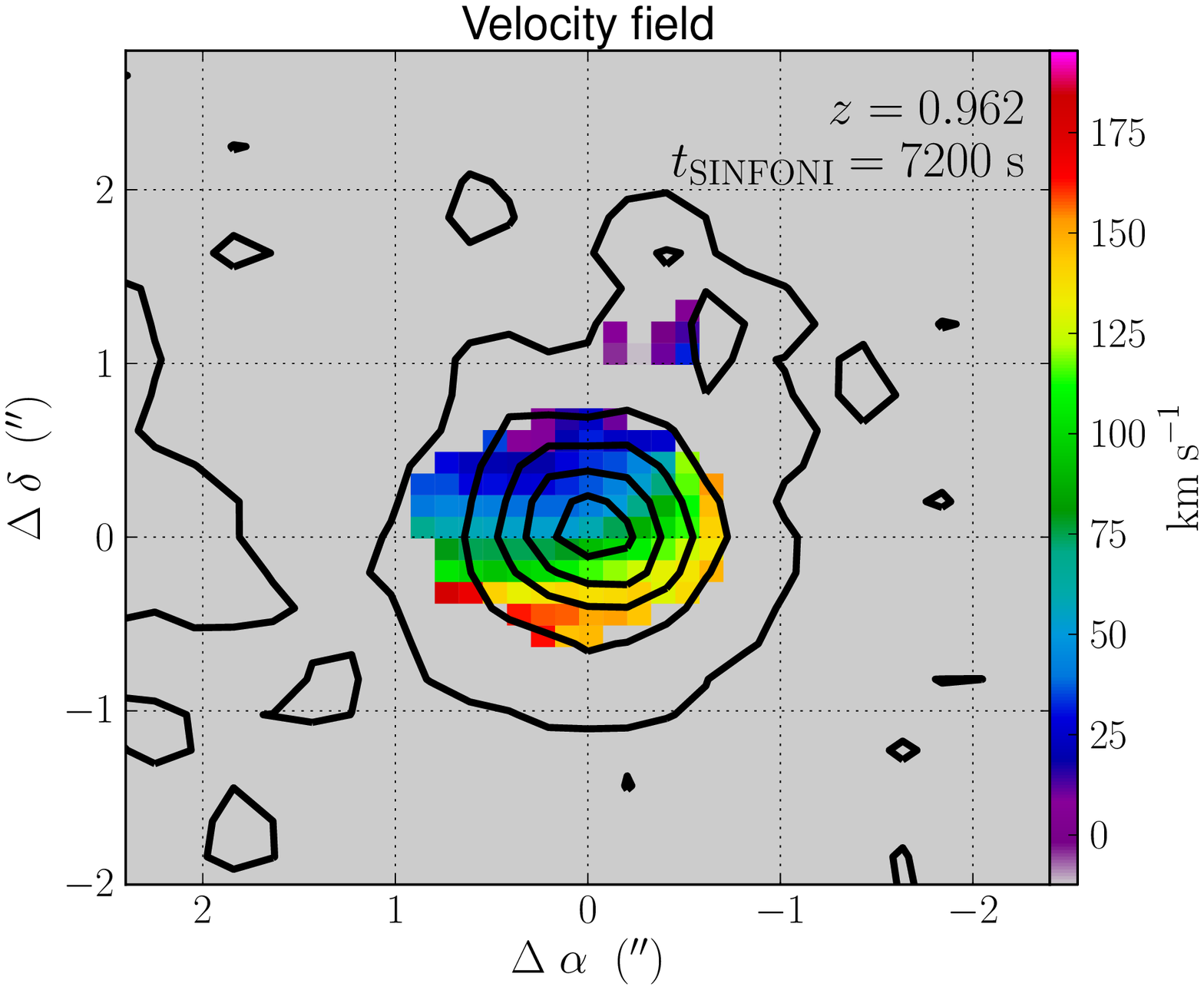}}
	\resizebox{0.32\hsize}{!}{\includegraphics{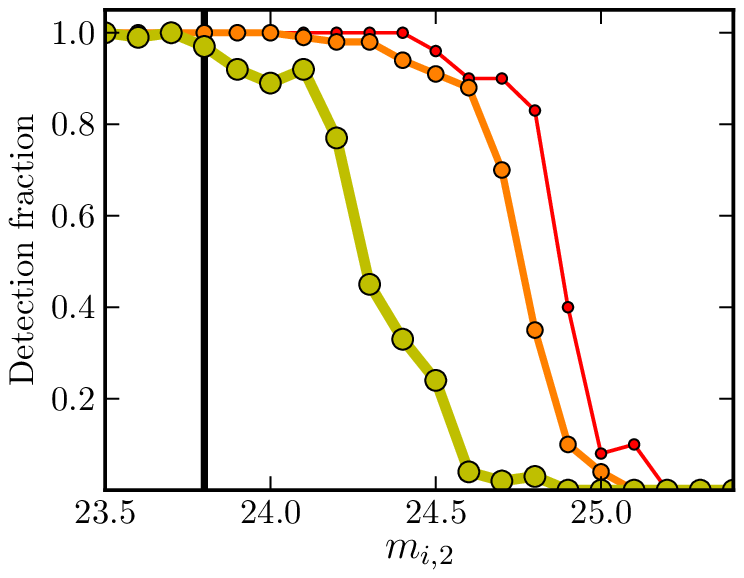}}
	\caption{{\it Left} and {\it central panels} are the same as in Fig.~\ref{src_020294045_i}, 
		but for the MASSIV source 140096645 (no major merger). The level contours mark the isophotes of the original $i$-band image. 
		The H$\alpha$ companion source is not detected in the $i$ band. 
		The outer contour marks the 2.64 ADU (1$\sigma_{\rm sky}$) isophote, while brighter isophotes increase in 
		7.94 ADU (3$\sigma_{\rm sky}$) steps up to 45 ADU to avoid crowded figures. 
		The {\it right panel} shows the detection fraction of fake galaxies injected in the $i-$band image as a 
		function of the magnitude of the companion source $m_{i,2}$. The size of the dots marks the detection curves 
		for extended, normal and compact galaxies (see text for details).
		The vertical line marks the limiting magnitude for major companion ($\Delta m_{i} = 1.5$). 
		[{\it A colour version of this plot is available at the electronic edition}].}
	\label{src_140096645}
	\end{figure*}


\subsection{Close pair candidates at $0.94 \leq z \leq 1.06$}
The weighted mean redshift of the first redshift bin is $\overline{z_{\rm r}}_{,1} = 1.03$.
We have identified 6 close pairs over 18 sources in this redshift
range:

\begin{itemize}


	\item 020294045 (Fig.~\ref{src_020294045_i}). {\bf Major merger}. The velocity map suggests two
		projected components. The companion is toward the north
		and presents a steep velocity gradient compared with the
		principal galaxy. The GALFIT model with two components
		in the $i$ band reproduces both the position of the two kinematical 
		components in the velocity map and the high velocity dispersion 
		in the overlapping region between both components. 
		The luminosity difference is $\Delta m_{i} = 0.4$. The separation 
		between the sources is 2.9$h^{-1}$ kpc and their relative velocity is 
		$\Delta v \sim 180$ km s$^{-1}$. Despite the close separation lead into significant
		overlap, this system also gives a satisfactory fit in the WIRDS 
		$K_{\rm s}-$band image (Fig.\ref{src_020294045_ks}). 
		In this case $\Delta m_{K_{\rm s}} = 0.3$, confirming that 
		this system is a major merger.


	\item 020386743 (Fig.~\ref{src_020386743}). {\it No major merger}. The velocity map suggests two
		projected components. The companion is toward the north
		and presents a different velocity than the northern part of 
		the principal galaxy. The GALFIT model with two components suggests that
		we are only detecting the western part of the companion
		galaxy, with the eastern part being too faint. This
		is also consistent with the velocity dispersion map, which
		shows a regular pattern in the overlapping region. 
		The separation between the sources is $3.7h^{-1}$ kpc and 
		the luminosity difference is $\Delta m_{i} = 1.8$. 
		Hence, we do not classify the system as a major merger.


	\item 020461235 (Fig.~\ref{src_020461235}). {\it No major merger}. The velocity map suggests two
		projected components. The companion is toward the south-west and presents 
		a nearly constant velocity,
		in contrast with the velocity gradient of the principal galaxy.
		The GALFIT model with two components suggests that the
		companion is highly distorted, perhaps because the system is
		in an advanced merger stage (i.e., after first pericenter passage 
		or pre-coalescence). The luminosity difference from the GALFIT model is
		$\Delta m_{i} = 1.7$, while from the residual maps is $\Delta m_{i} = 1.9$. 
		This difference strongly suggests 
		that this system is not a major merger. The separation between the 
		components is 2.8$h^{-1}$ kpc.	


	\begin{figure*}[t!]
	\resizebox{0.32\hsize}{!}{\includegraphics{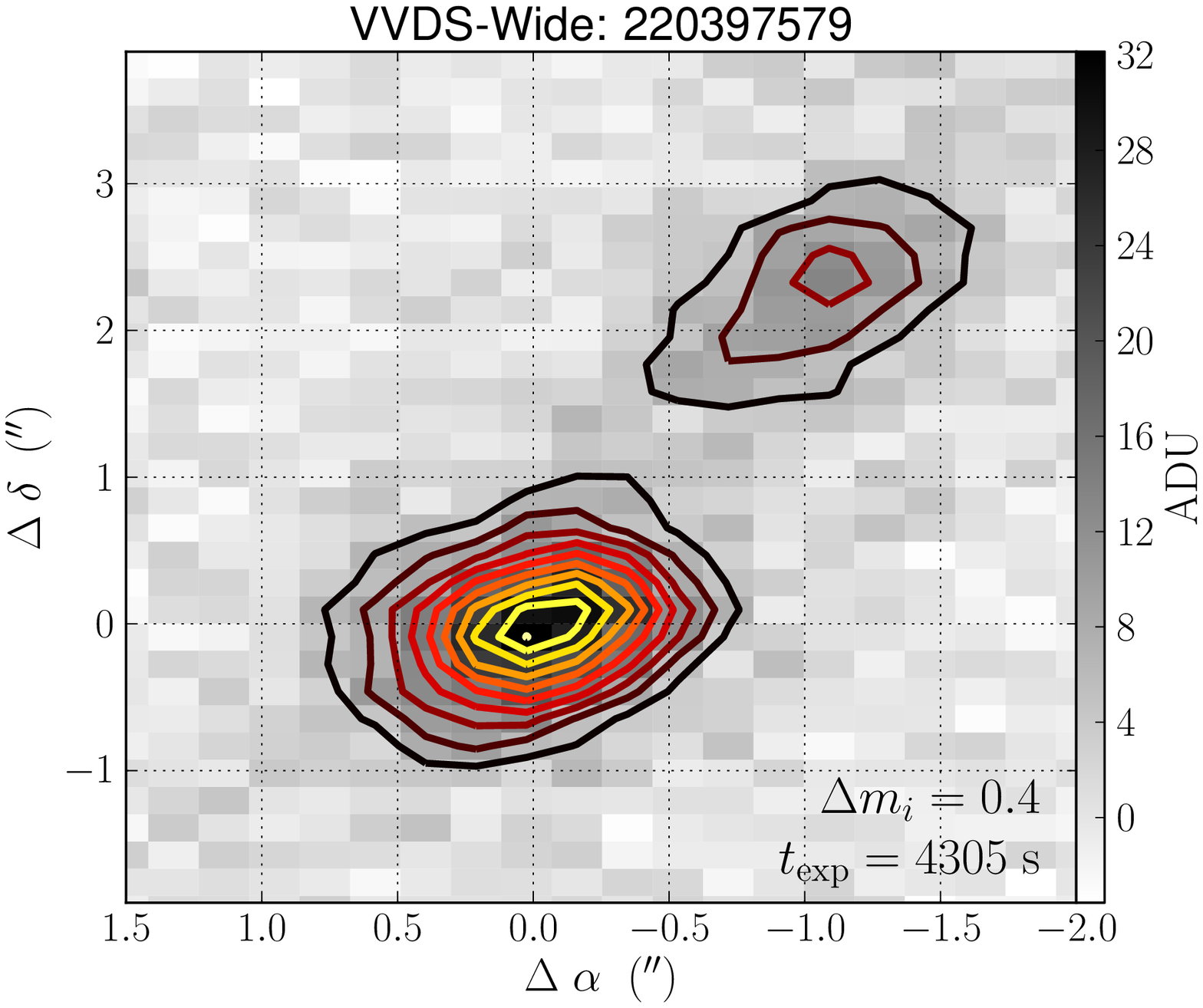}}
	\resizebox{0.32\hsize}{!}{\includegraphics{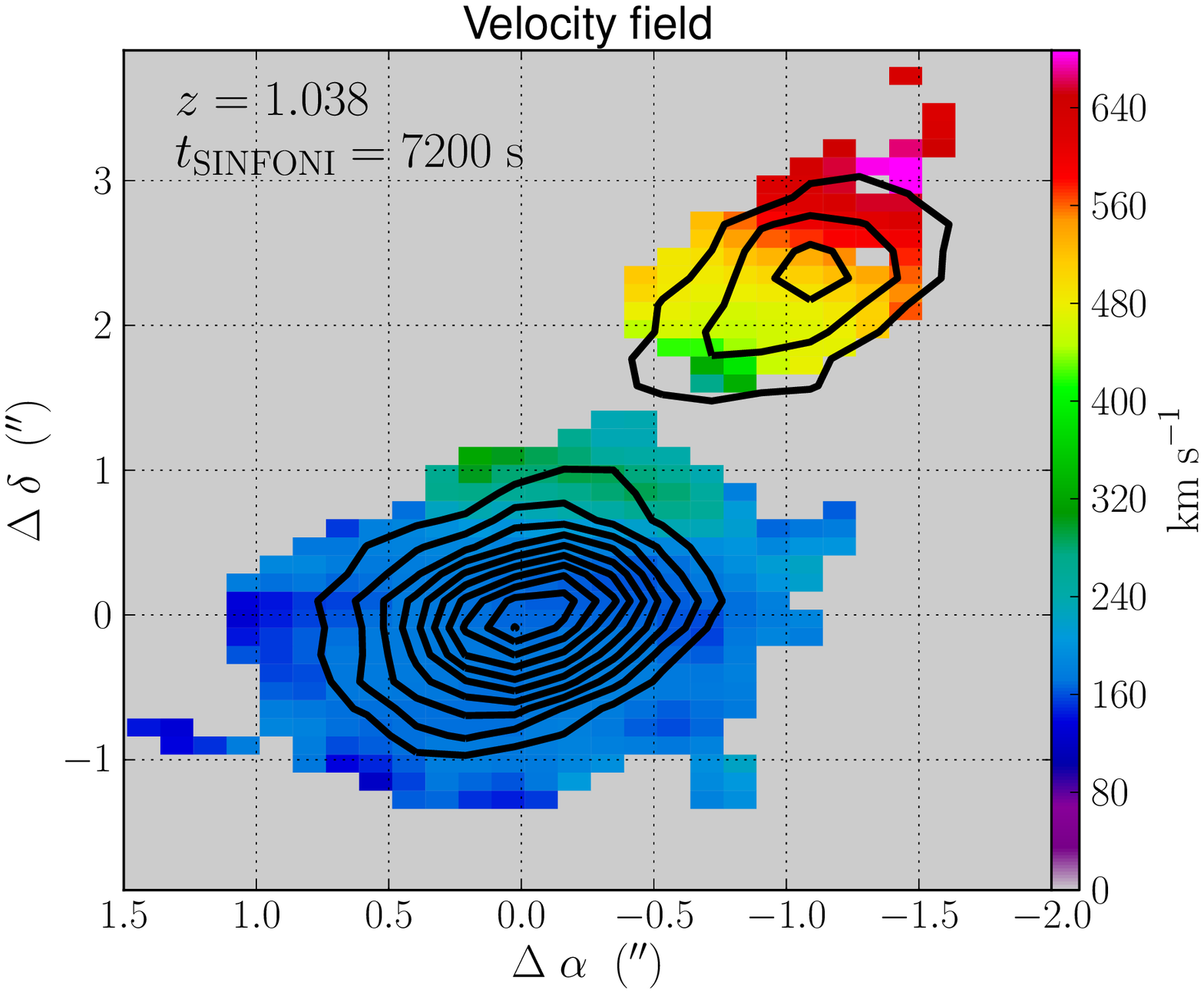}}
	\resizebox{0.32\hsize}{!}{\includegraphics{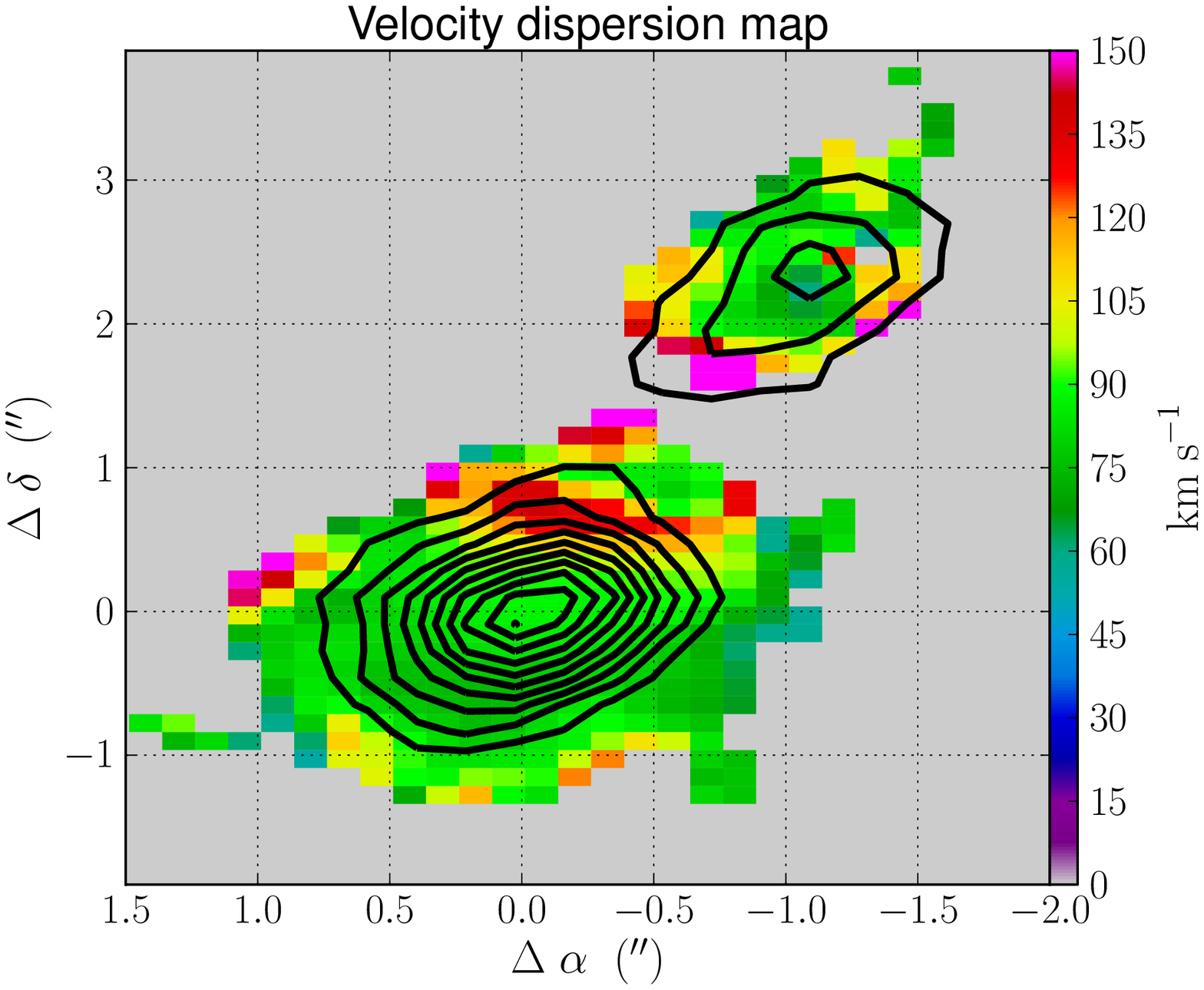}}
	\caption{The same as Fig.~\ref{src_020294045_i}, but for the MASSIV source 220397579 (major merger). 
		The level contours mark the isophotes of the original $i$-band image. The outer contour marks 
		the 5.36 ADU (3$\sigma_{\rm sky}$) isophote, while brighter isophotes increase 
		in 2.68 ADU (1.5$\sigma_{\rm sky}$) steps. 
		[{\it A colour version of this plot is available at the electronic edition}].}
	\label{src_220397579}
	\end{figure*}


	\begin{figure*}[t!]
	\resizebox{0.32\hsize}{!}{\includegraphics{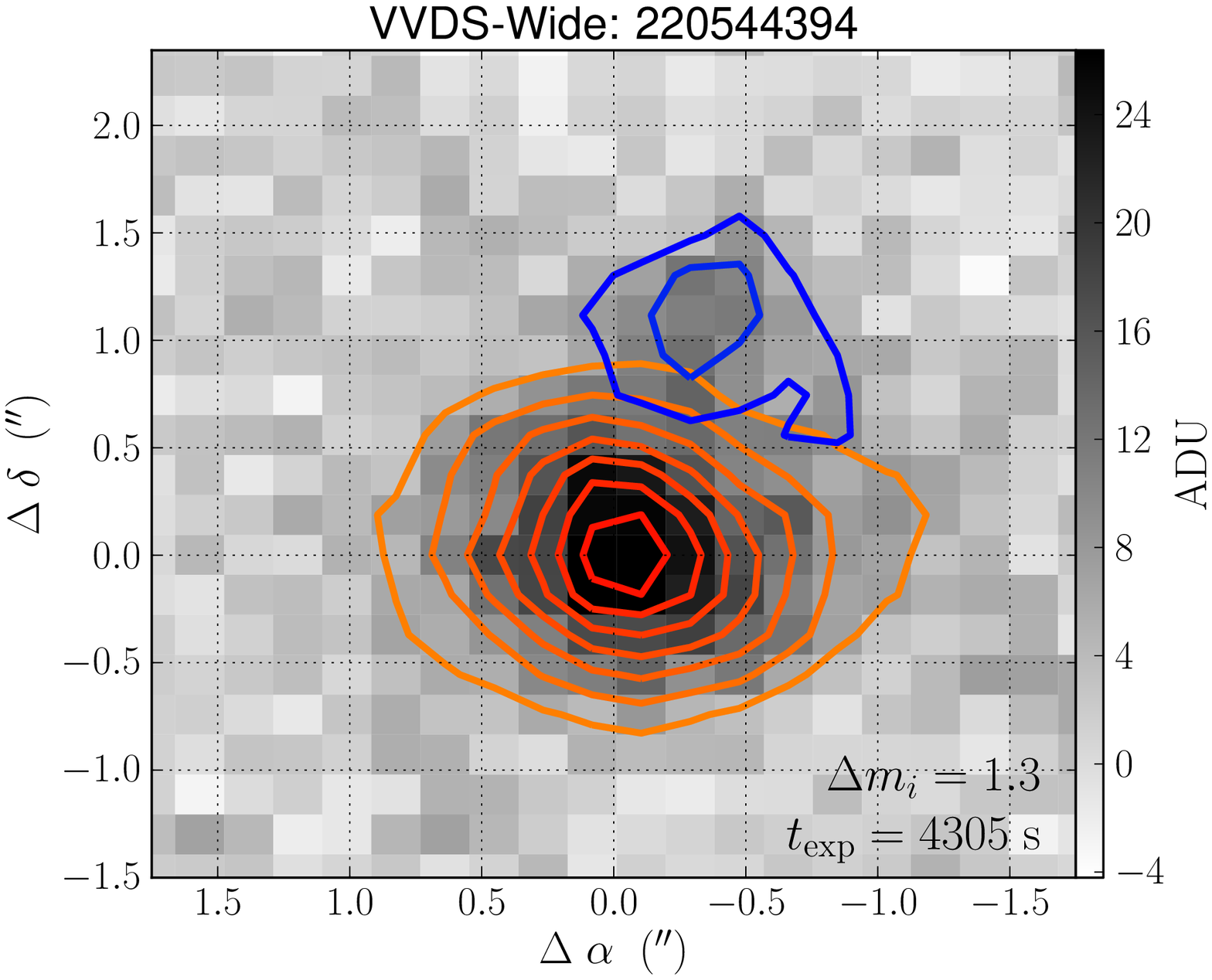}}
	\resizebox{0.32\hsize}{!}{\includegraphics{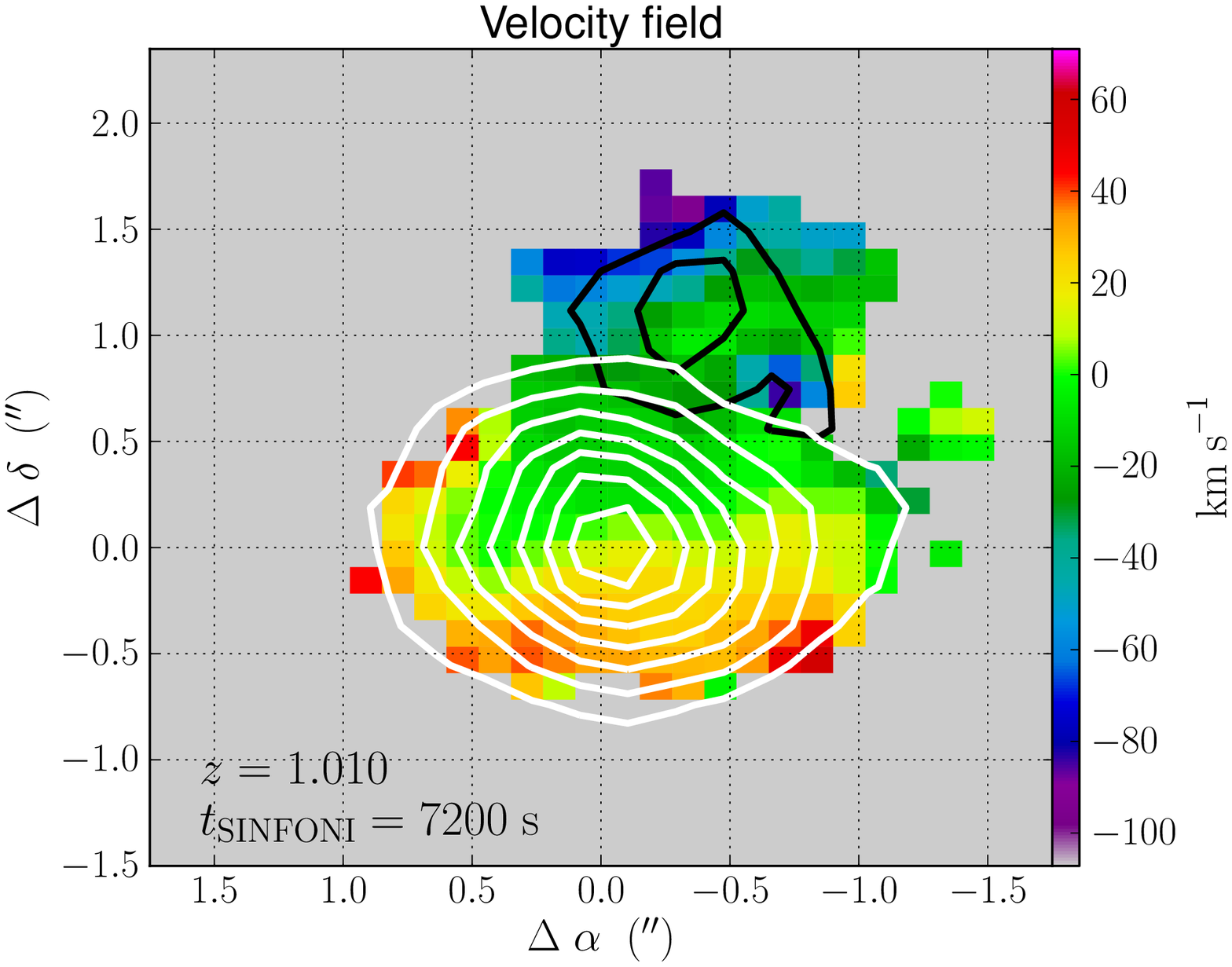}}
	\resizebox{0.32\hsize}{!}{\includegraphics{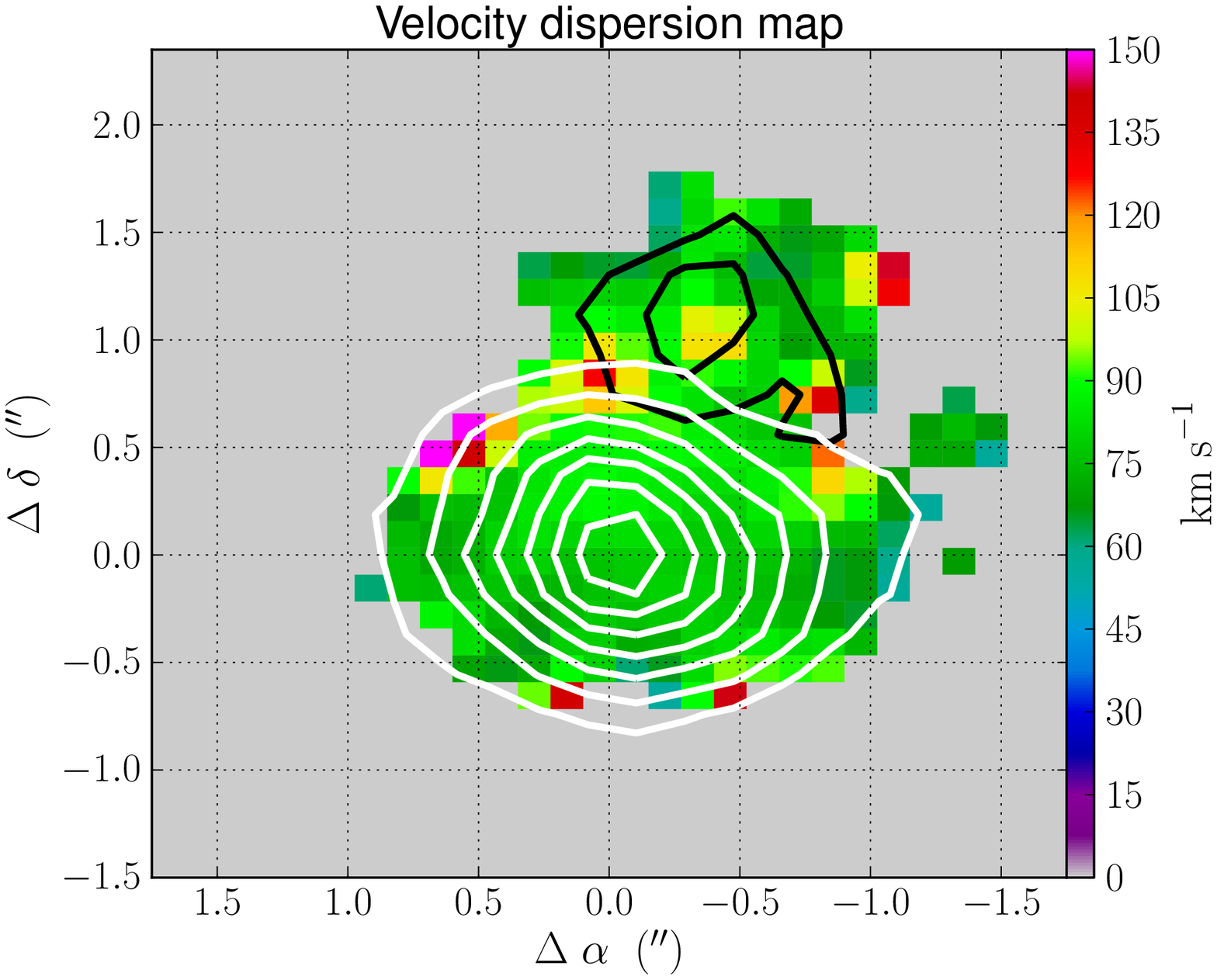}}
	\caption{ The same as Fig.~\ref{src_020294045_i}, but for the MASSIV source 220544394 (major merger). The outer contour 		marks the 4.31 ADU (2$\sigma_{\rm sky}$) isophote, while brighter isophotes 
		increase in 3.23 ADU (1.5$\sigma_{\rm sky}$) steps. 
		[{\it A colour version of this plot is available at the electronic edition}].}
	\label{src_220544394}
	\end{figure*}


	\begin{figure*}[t!]
	\resizebox{0.32\hsize}{!}{\includegraphics{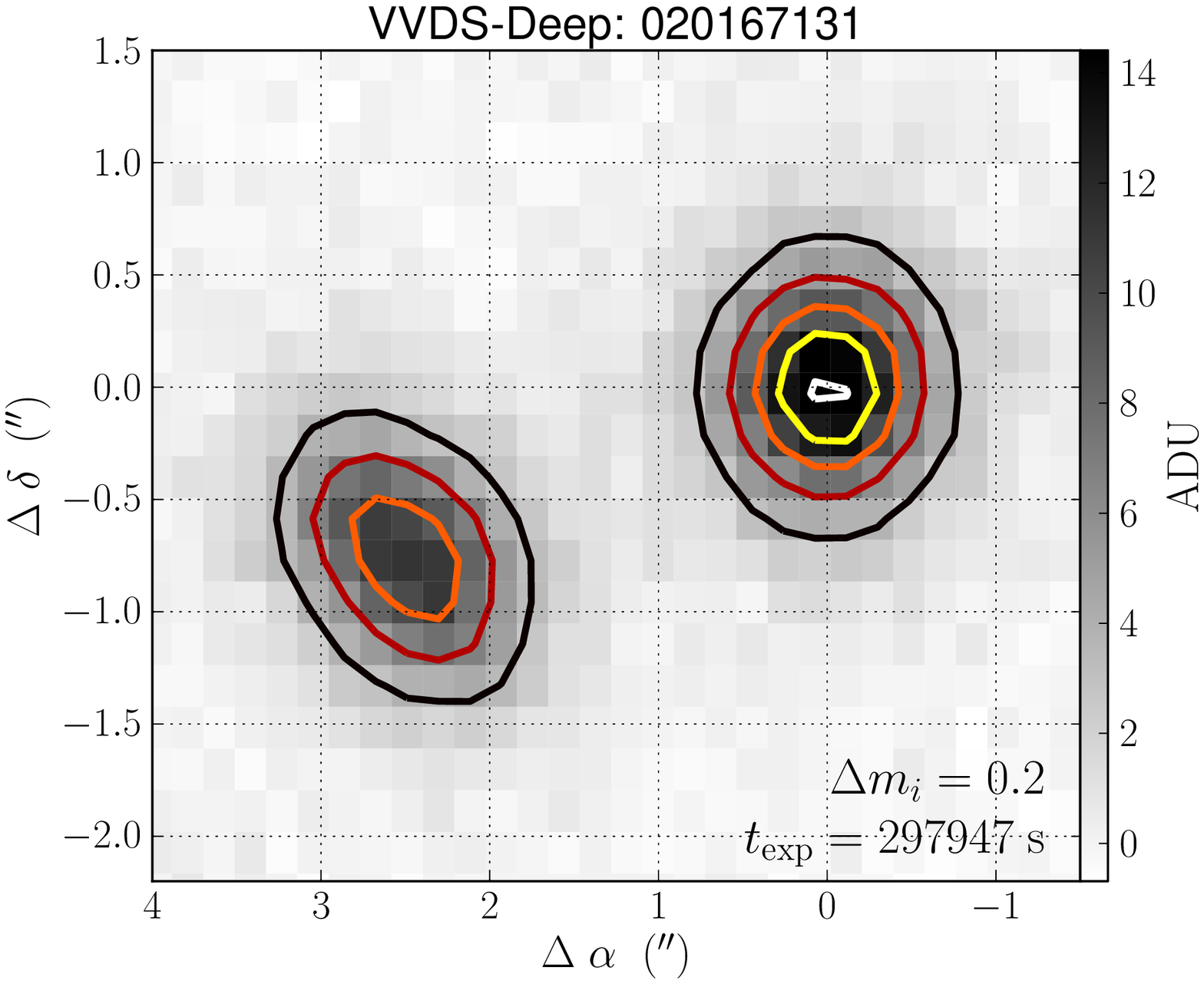}}
	\resizebox{0.32\hsize}{!}{\includegraphics{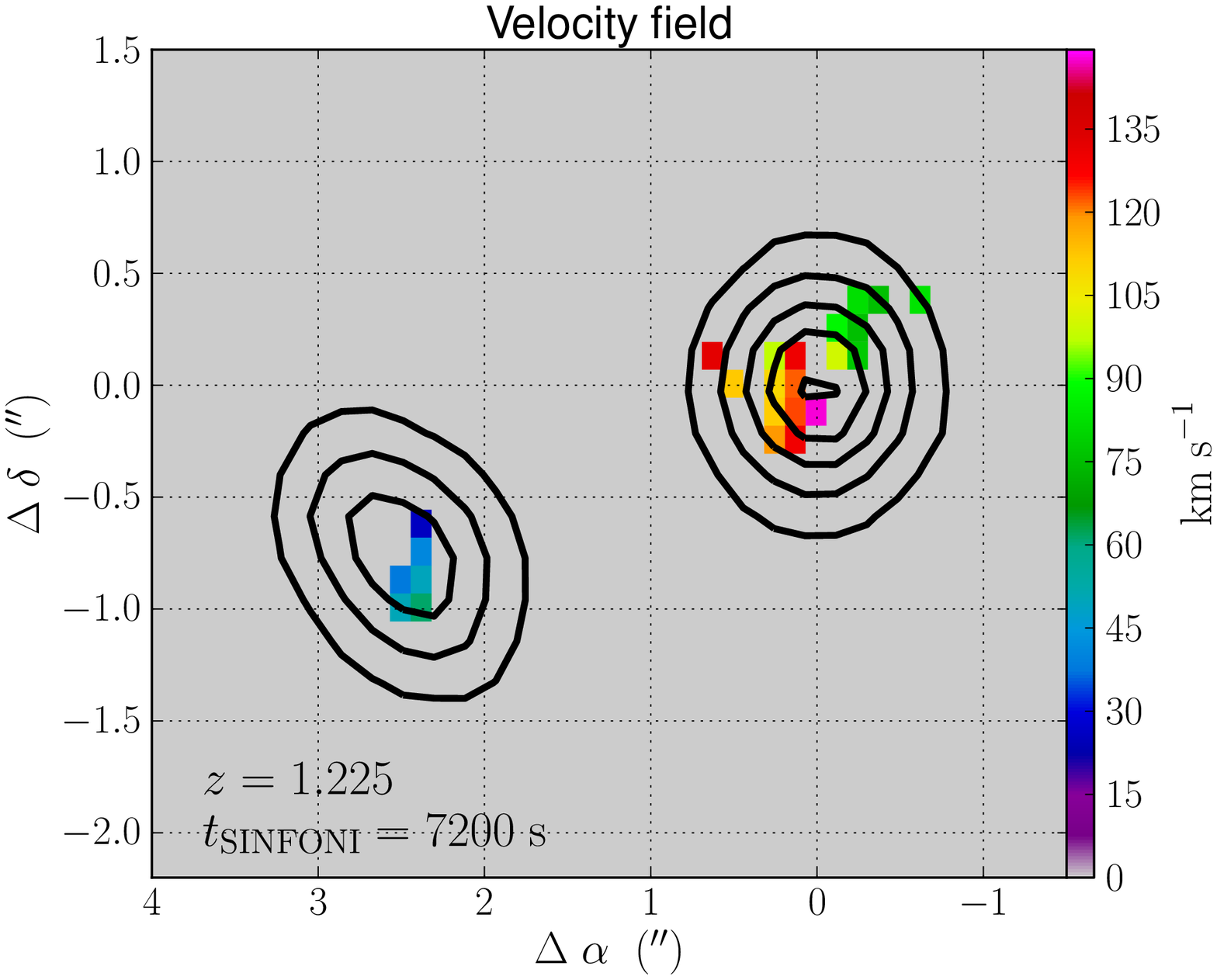}}
	\resizebox{0.32\hsize}{!}{\includegraphics{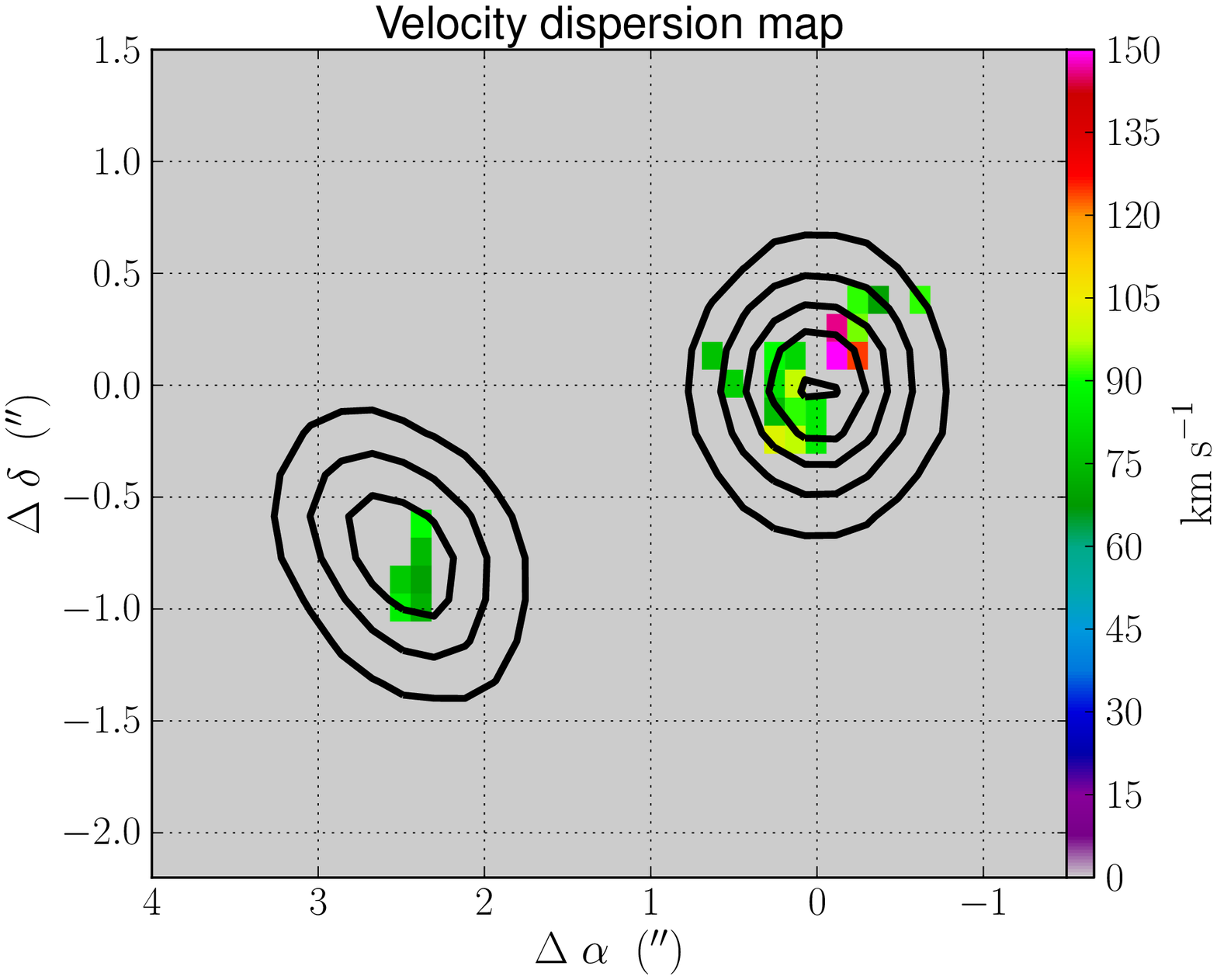}}
	\caption{The same as Fig.~\ref{src_220397579}, but for the MASSIV source 020167131 (major merger). 
		The outer contour marks the 3.43 ADU (10$\sigma_{\rm sky}$ 
		isophote, while brighter isophotes increase in 2.74 ADU (8$\sigma_{\rm sky}$) steps. 
		[{\it A colour version of this plot is available at the electronic edition}].}
	\label{src_020167131}
	\end{figure*}


	\begin{figure*}[t!]
	\resizebox{0.32\hsize}{!}{\includegraphics{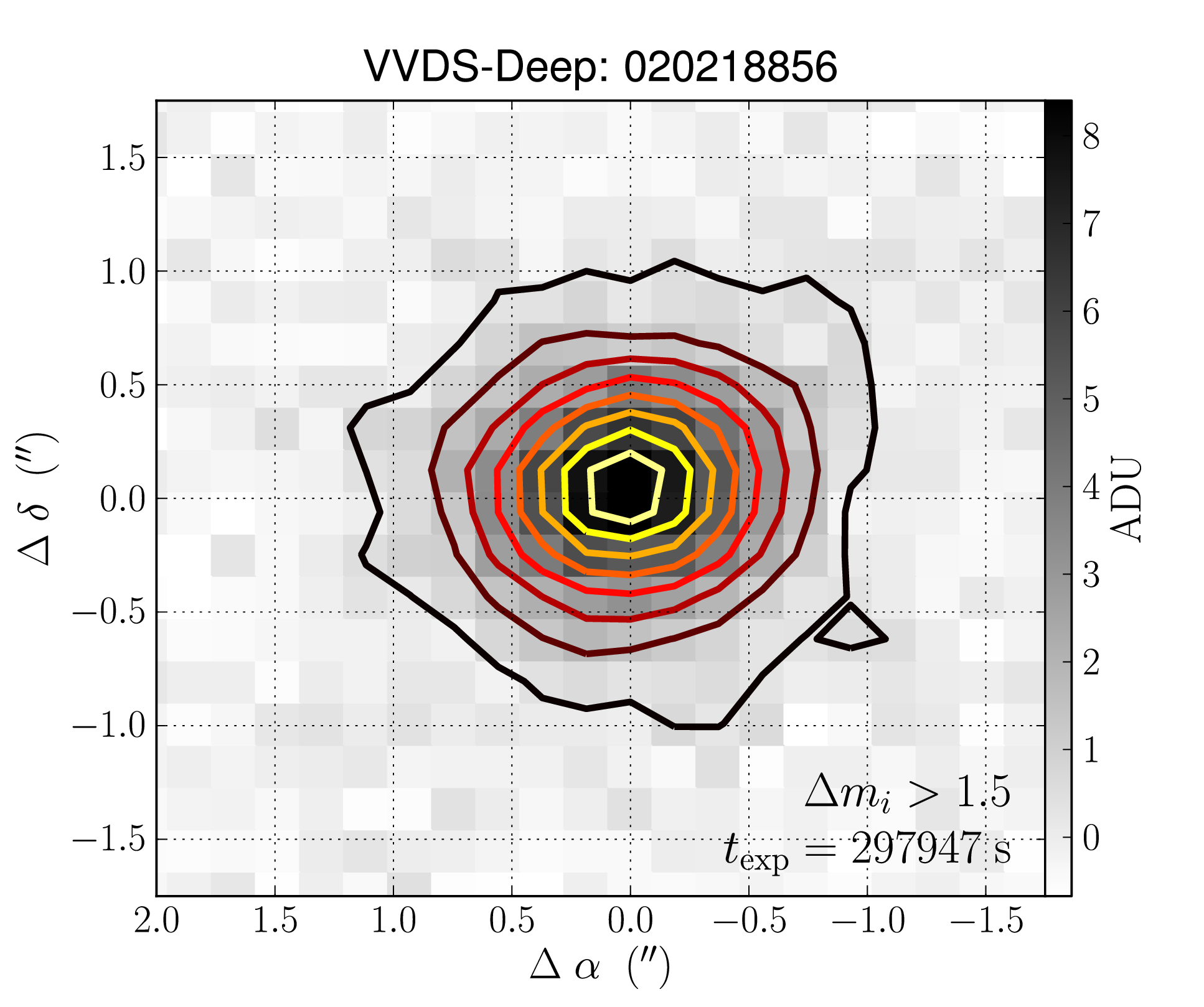}}
	\resizebox{0.32\hsize}{!}{\includegraphics{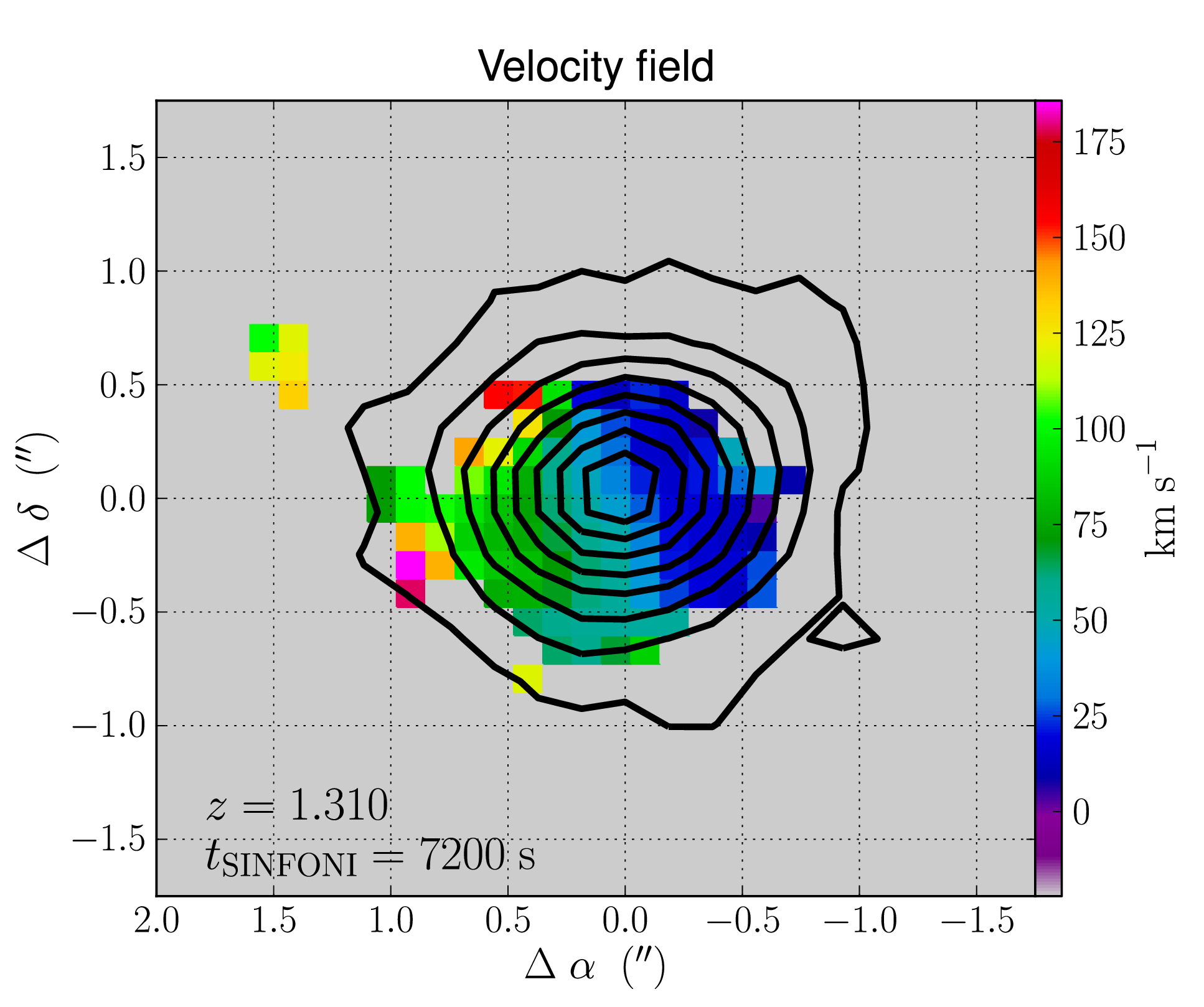}}
	\resizebox{0.32\hsize}{!}{\includegraphics{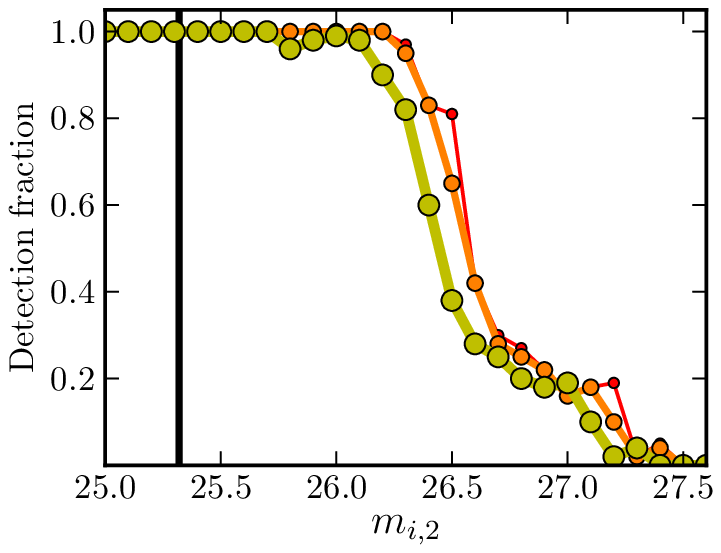}}
	\caption{The same as Fig.~\ref{src_140096645}, but for the MASSIV source 020218856 (no major merger). 
		The H$\alpha$ companion galaxy is not detected in the $i$ band. The outer contour marks the 0.34 ADU 
		(1$\sigma_{\rm sky}$) isophote, while brighter isophotes increase in 1.01 ADU (3$\sigma_{\rm sky}$) steps. 
		[{\it A colour version of this plot is available at the electronic edition}].}
	\label{src_020218856}
	\end{figure*}



	\begin{figure*}[t!]
	\resizebox{0.32\hsize}{!}{\includegraphics{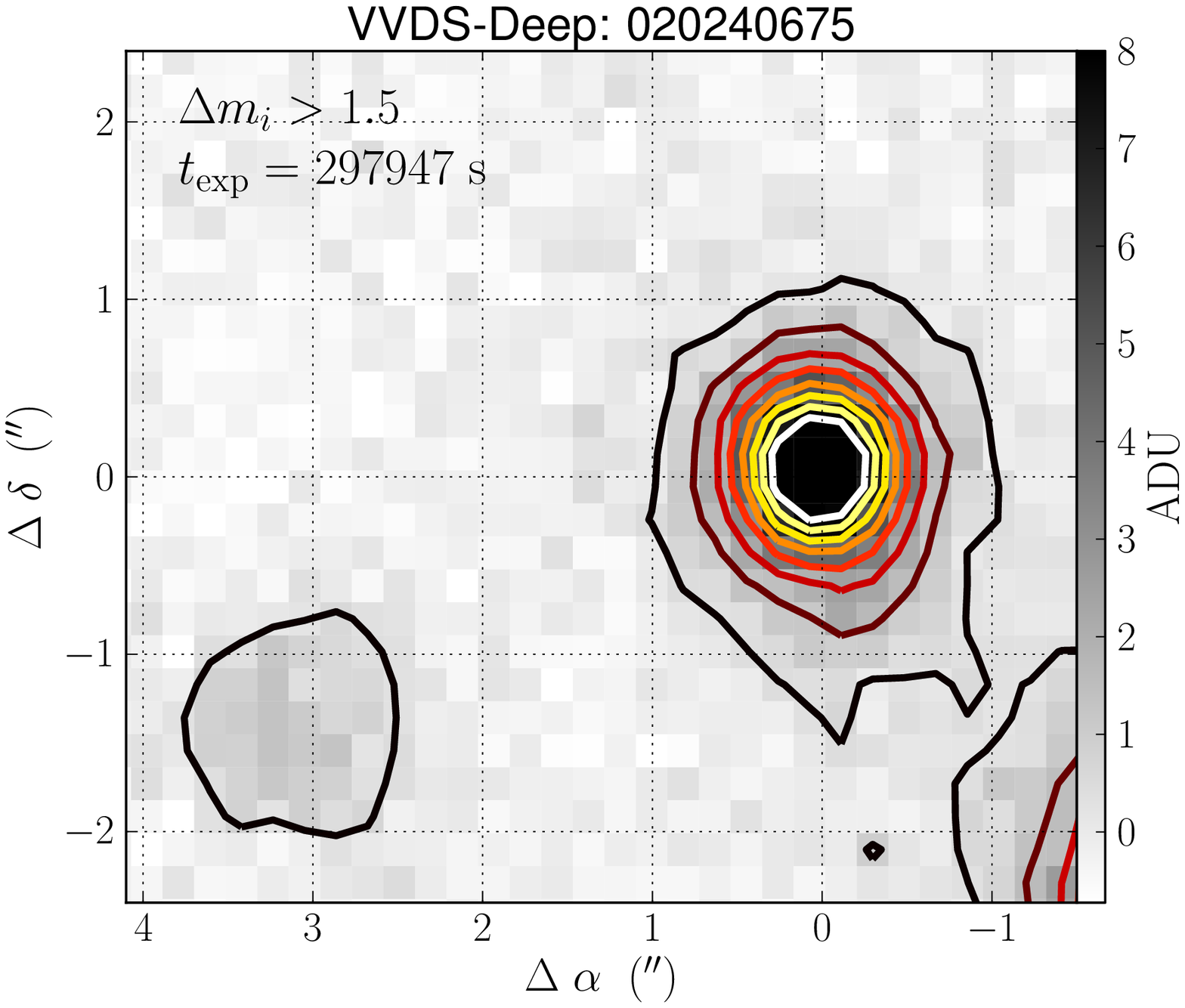}}
	\resizebox{0.32\hsize}{!}{\includegraphics{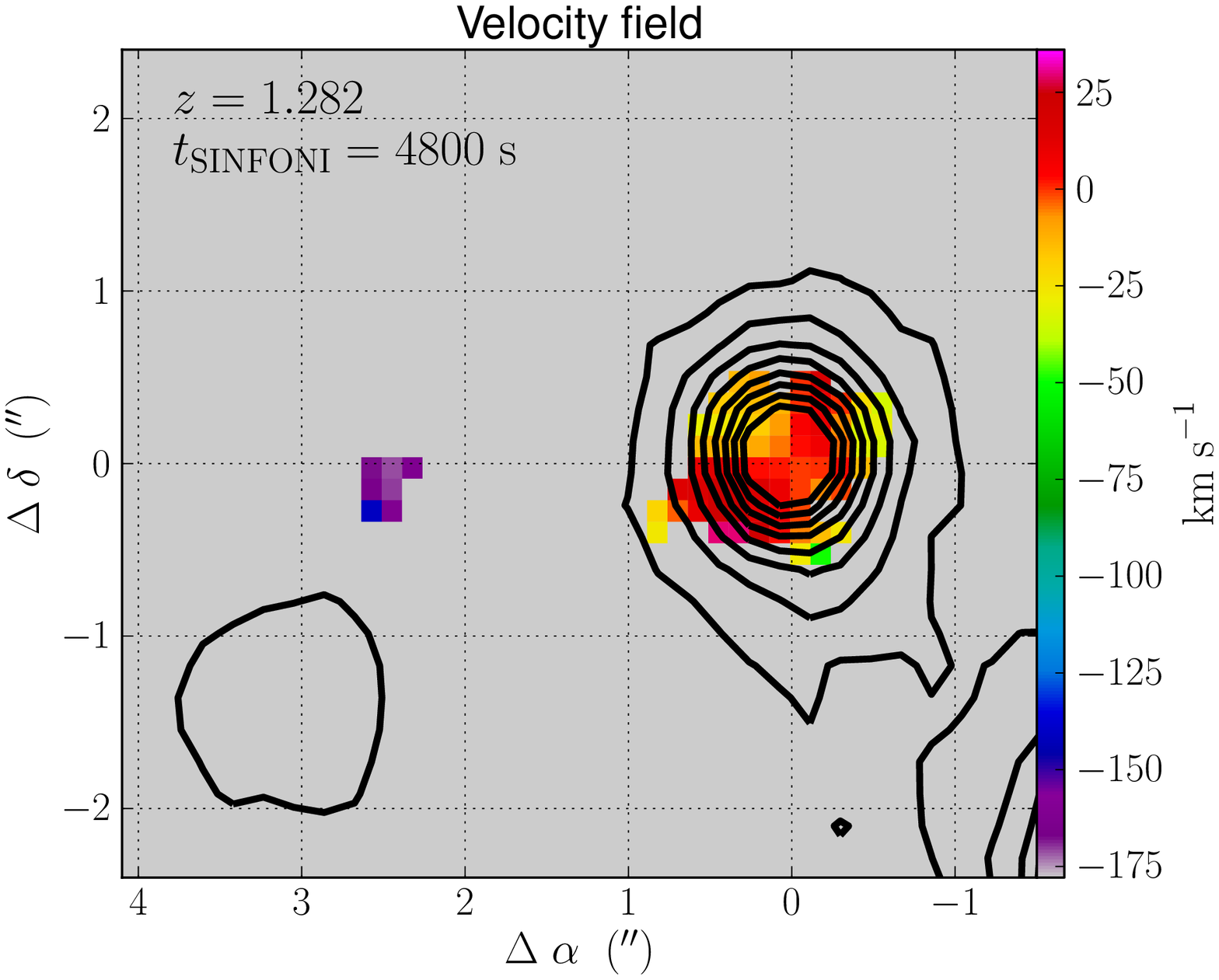}}
	\resizebox{0.32\hsize}{!}{\includegraphics{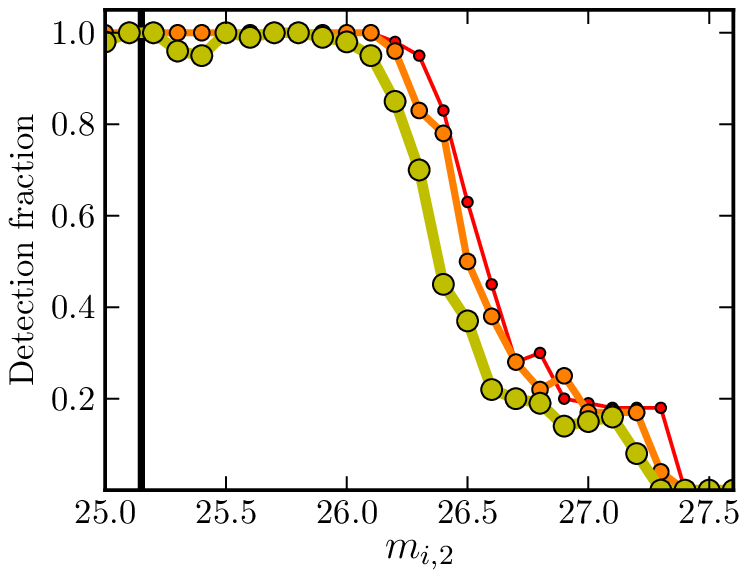}}
	\caption{The same as Fig.~\ref{src_140096645}, but for the MASSIV source 020240675 (no major merger). 
		The H$\alpha$ companion galaxy is not detected in the $i$ band. The outer contour marks the 0.36 ADU (1$\sigma_{\rm sky}$)
		isophote, while brighter isophotes increase in 1.08 ADU (3$\sigma_{\rm sky}$) 
		steps up to 8 ADU to avoid crowded figures. 
		[{\it A colour version of this plot is available at the electronic edition}].}
	\label{src_020240675}
	\end{figure*}


	\begin{figure*}[t!]
	\resizebox{0.32\hsize}{!}{\includegraphics{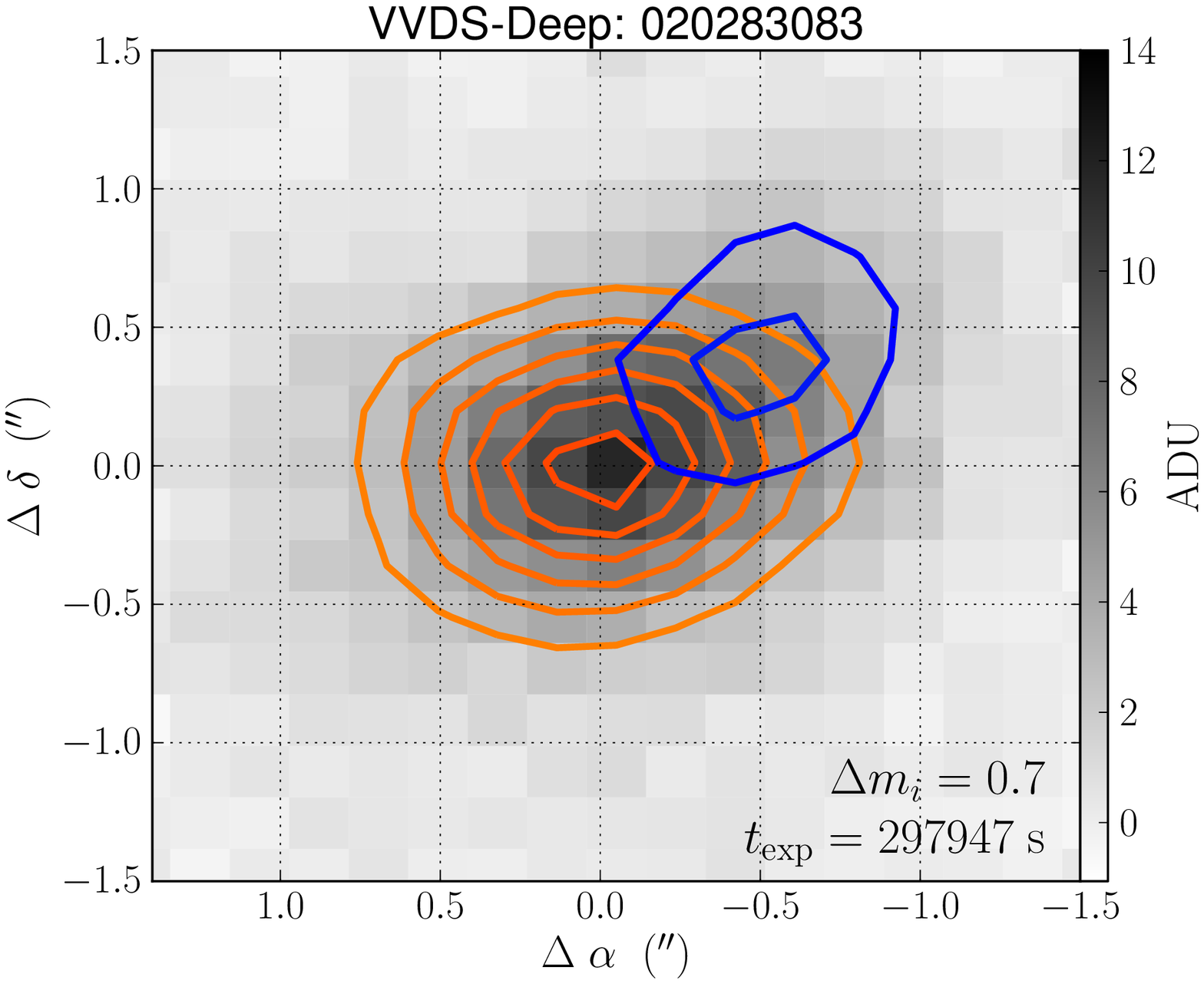}}
	\resizebox{0.32\hsize}{!}{\includegraphics{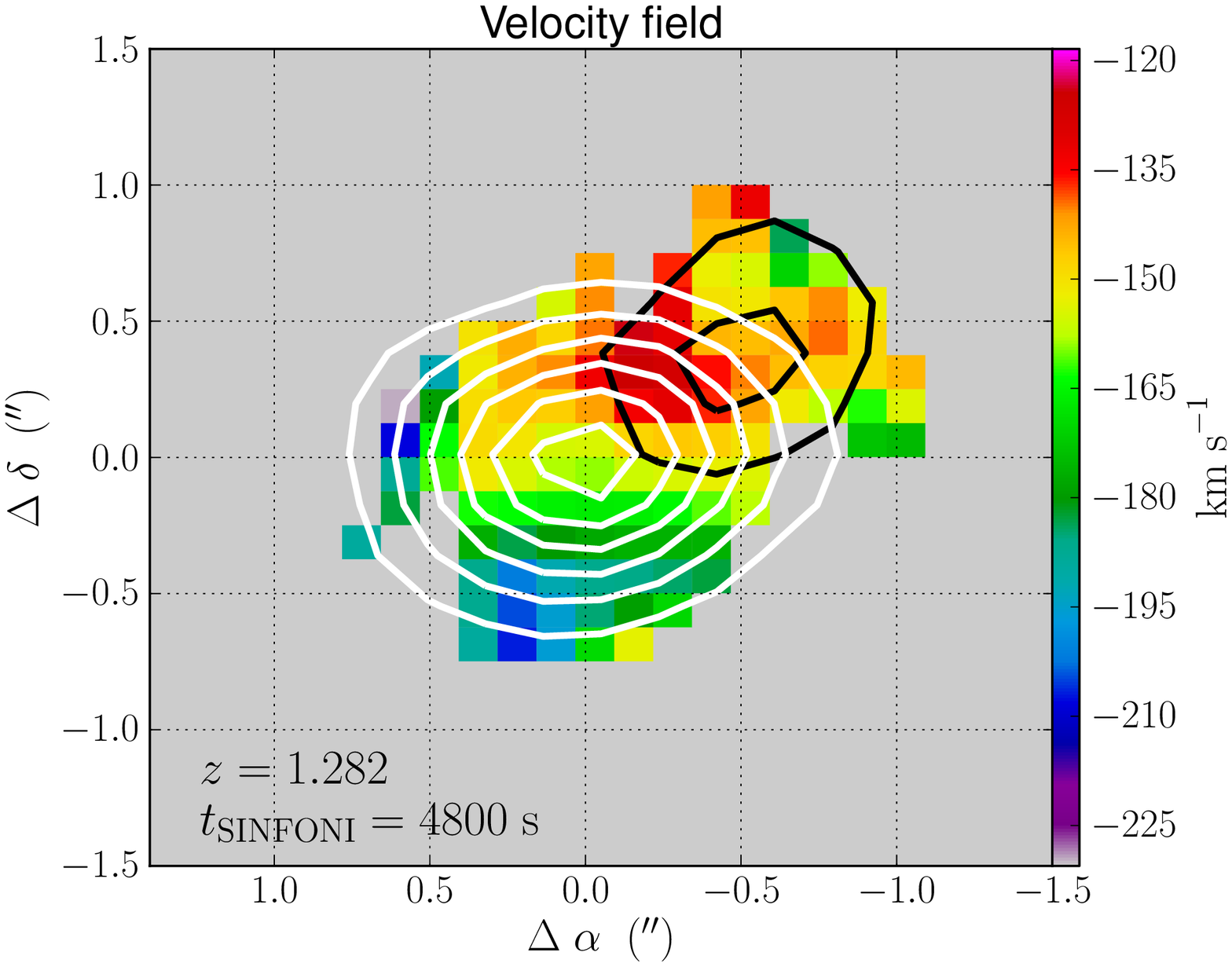}}
	\resizebox{0.32\hsize}{!}{\includegraphics{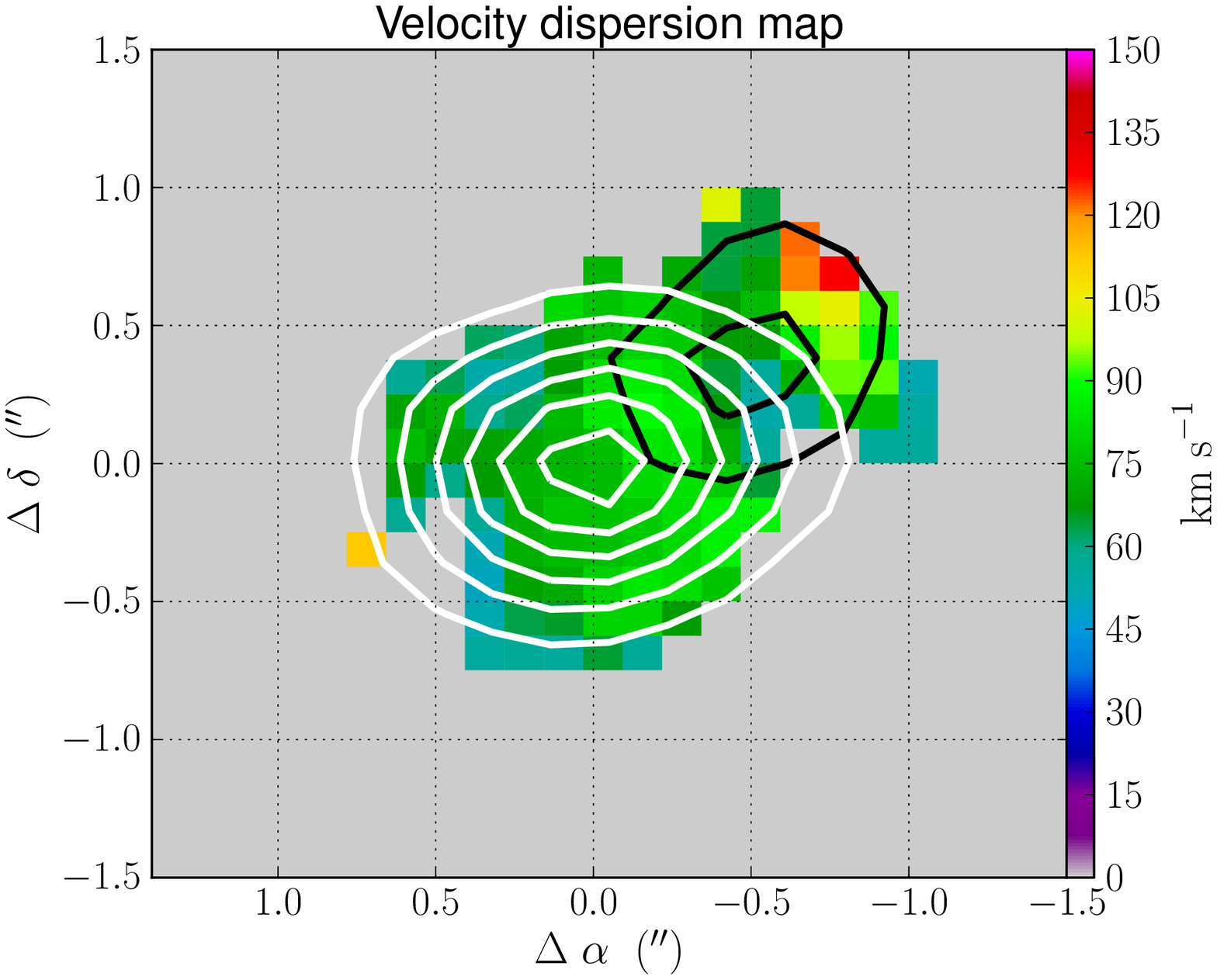}}
	\caption{The same as Fig.~\ref{src_020294045_i}, but for the MASSIV source 020283083 (major merger). 
		The outer contour marks the 2.12 ADU (4$\sigma_{\rm sky}$) isophote, while brighter 
		isophotes increase in 1.06 ADU (2$\sigma_{\rm sky}$) steps. 
		[{\it A colour version of this plot is available at the electronic edition}].}
	\label{src_020283083}
	\end{figure*}


	\begin{figure*}[t!]
	\resizebox{0.32\hsize}{!}{\includegraphics{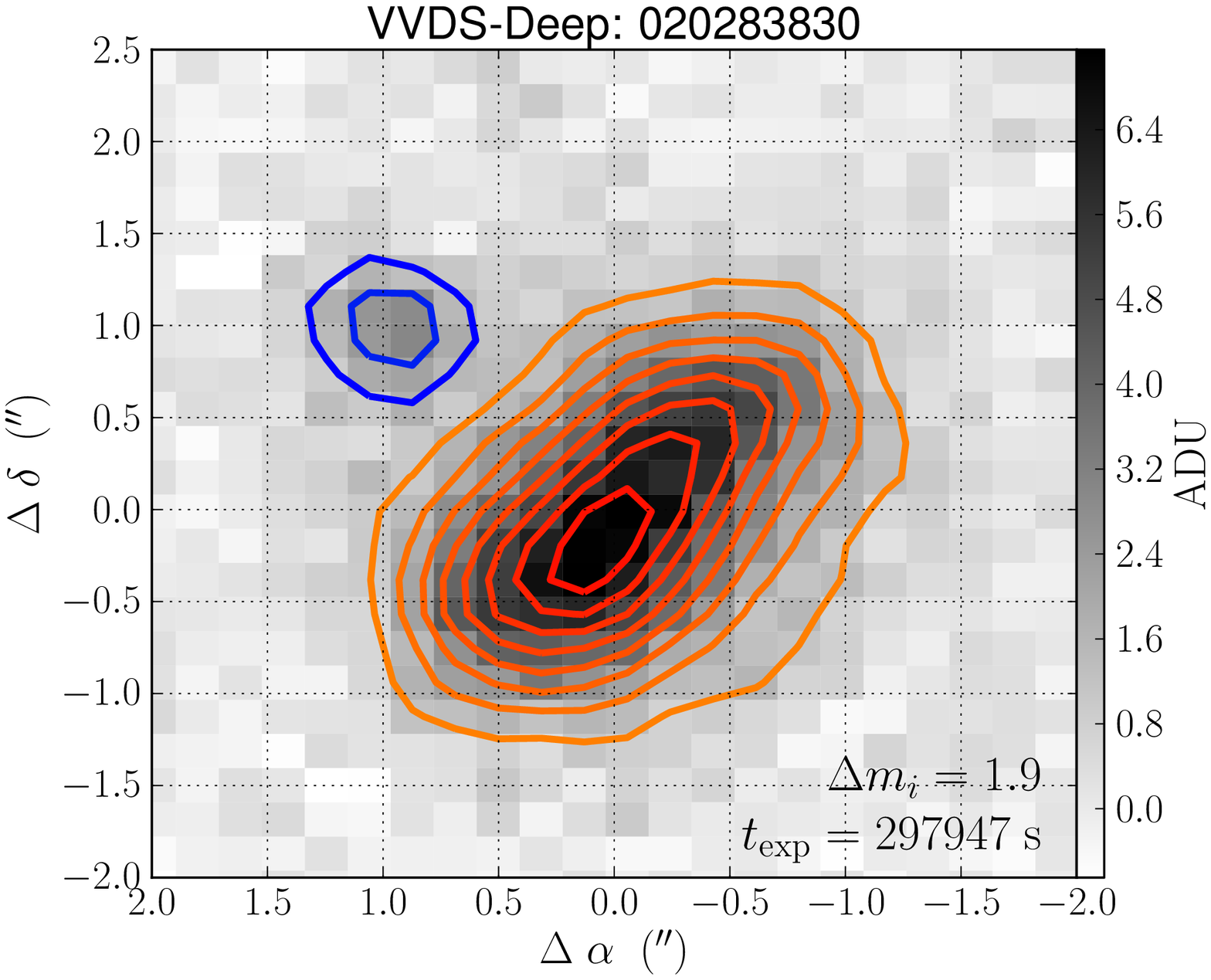}}
	\resizebox{0.32\hsize}{!}{\includegraphics{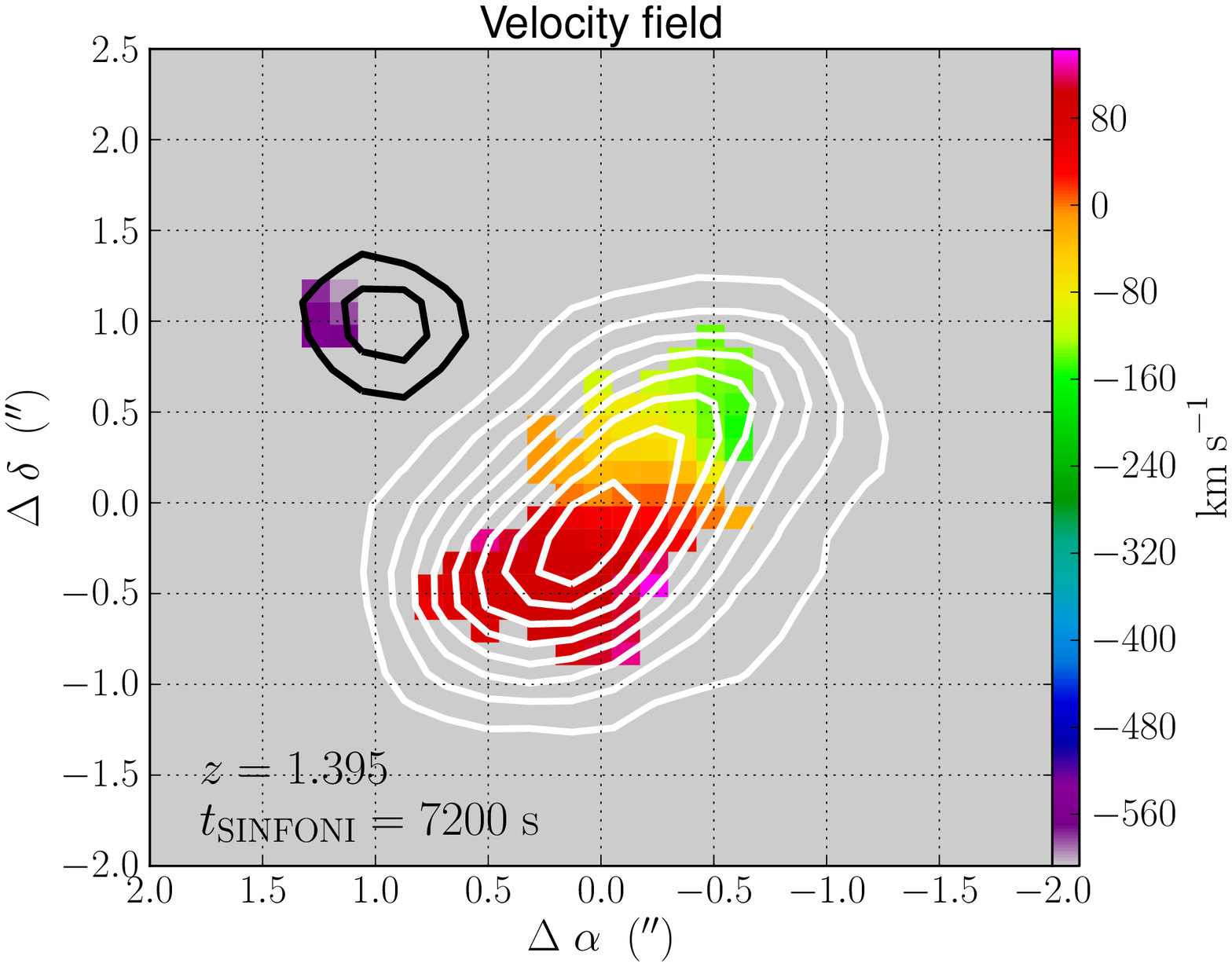}}
	\resizebox{0.32\hsize}{!}{\includegraphics{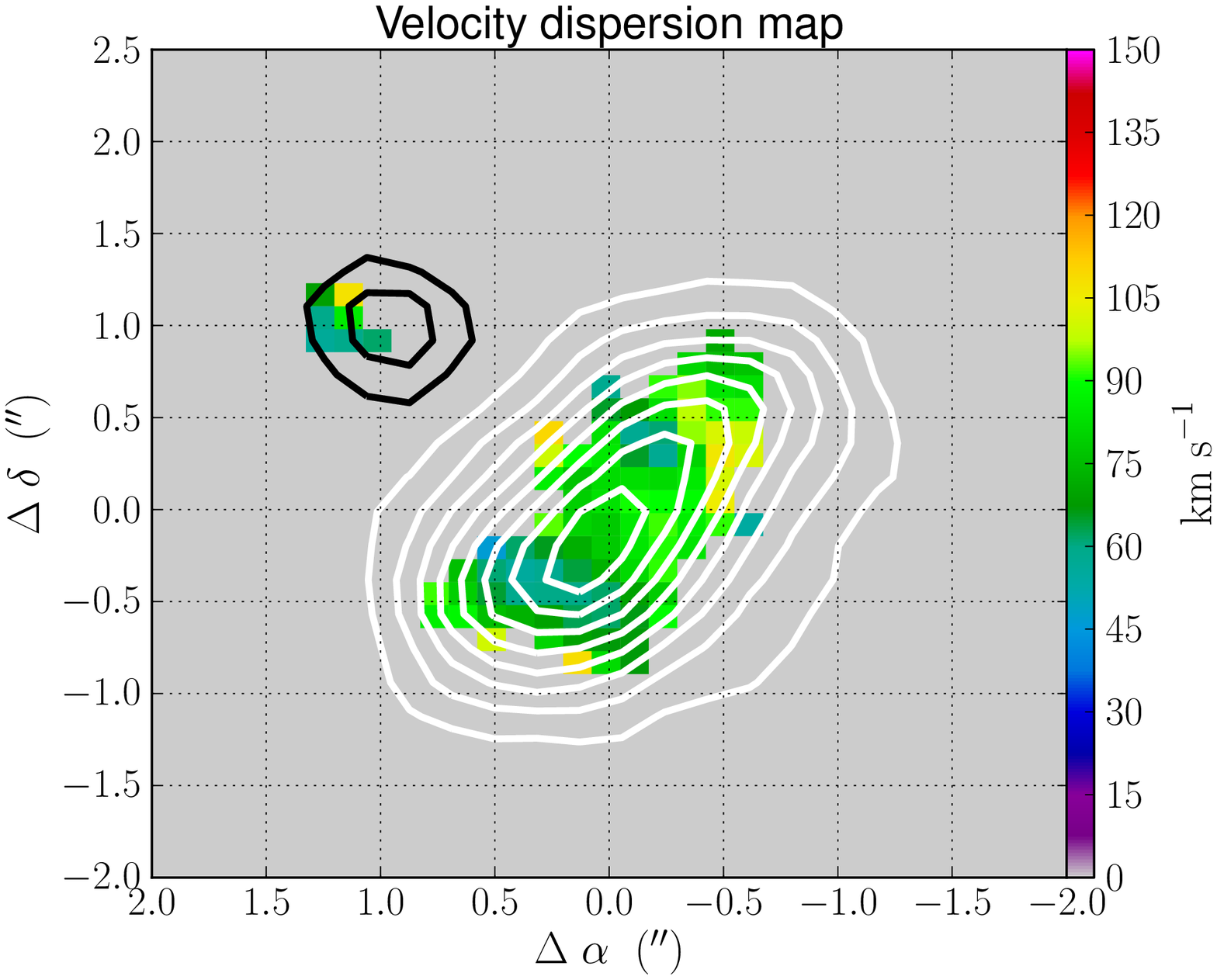}}
	\caption{The same as Fig.~\ref{src_020294045_i}, but for the MASSIV source 020283830 (no major merger). 
		The outer contour marks the 1.14 ADU (3.5$\sigma_{\rm sky}$) isophote, while brighter isophotes increase in 
		0.75 ADU (2.3$\sigma_{\rm sky}$) steps. 
		[{\it A colour version of this plot is available at the electronic edition}].}
	\label{src_020283830}
	\end{figure*}


	\begin{figure*}[t!]
	\resizebox{0.32\hsize}{!}{\includegraphics{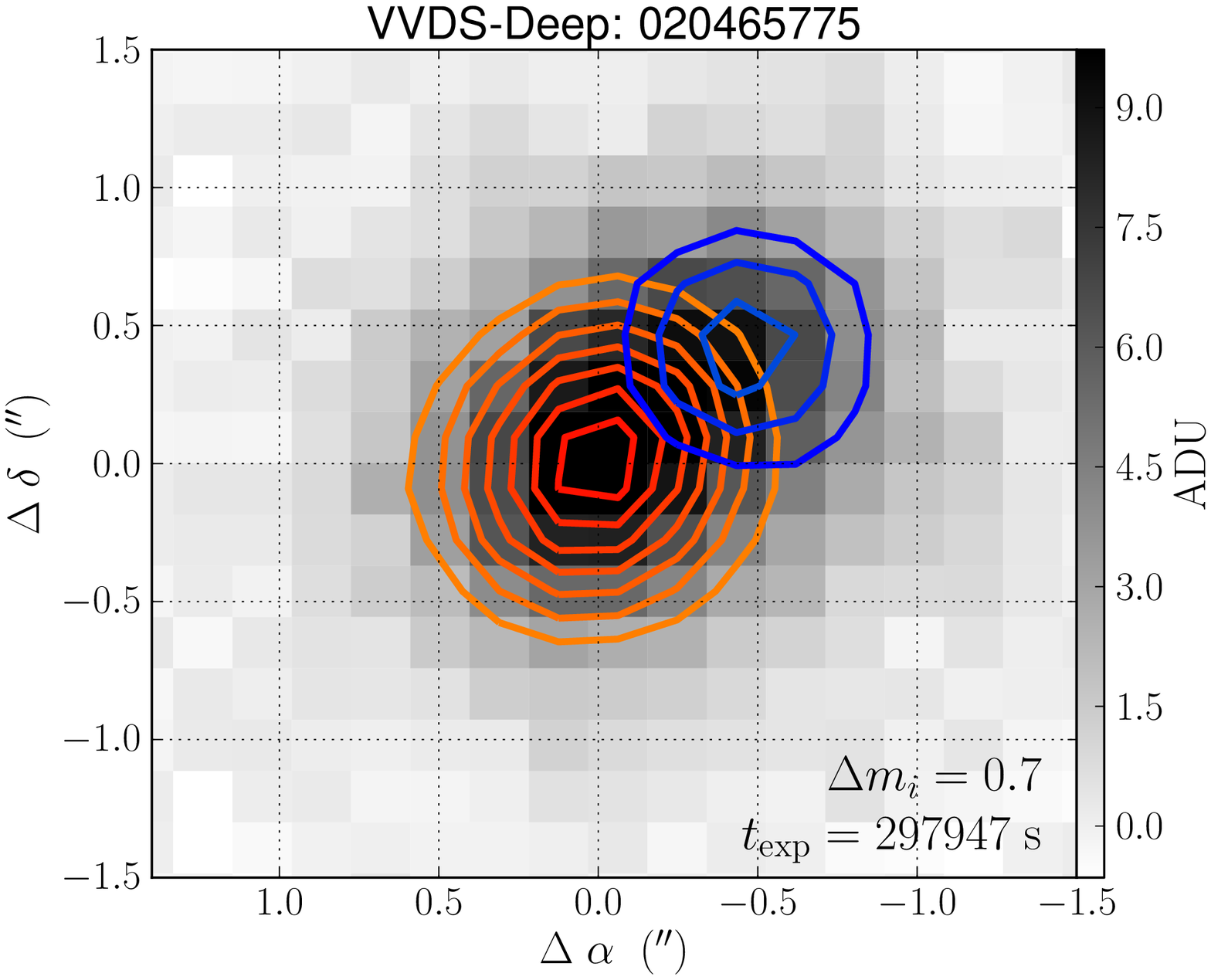}}
	\resizebox{0.32\hsize}{!}{\includegraphics{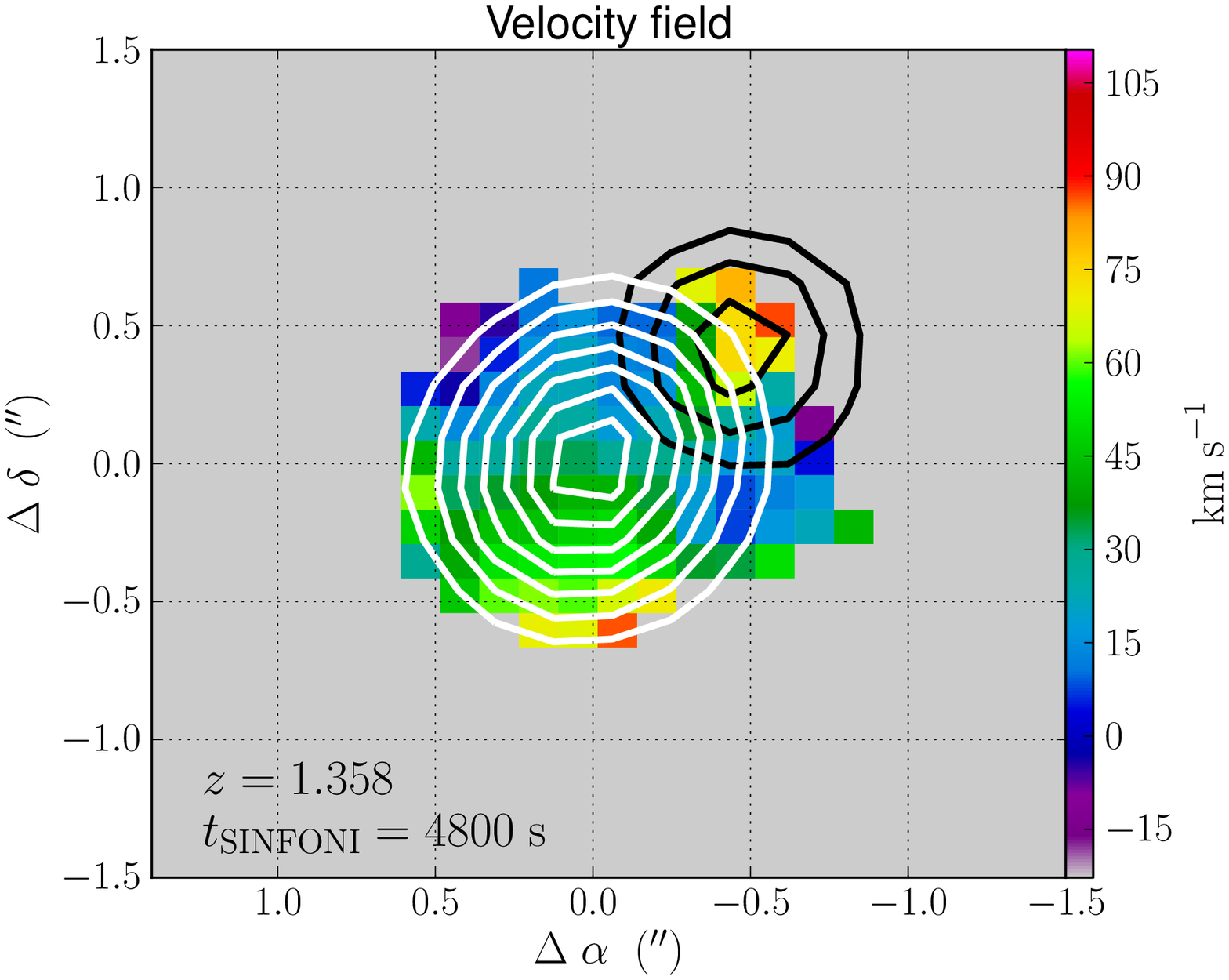}}
	\resizebox{0.32\hsize}{!}{\includegraphics{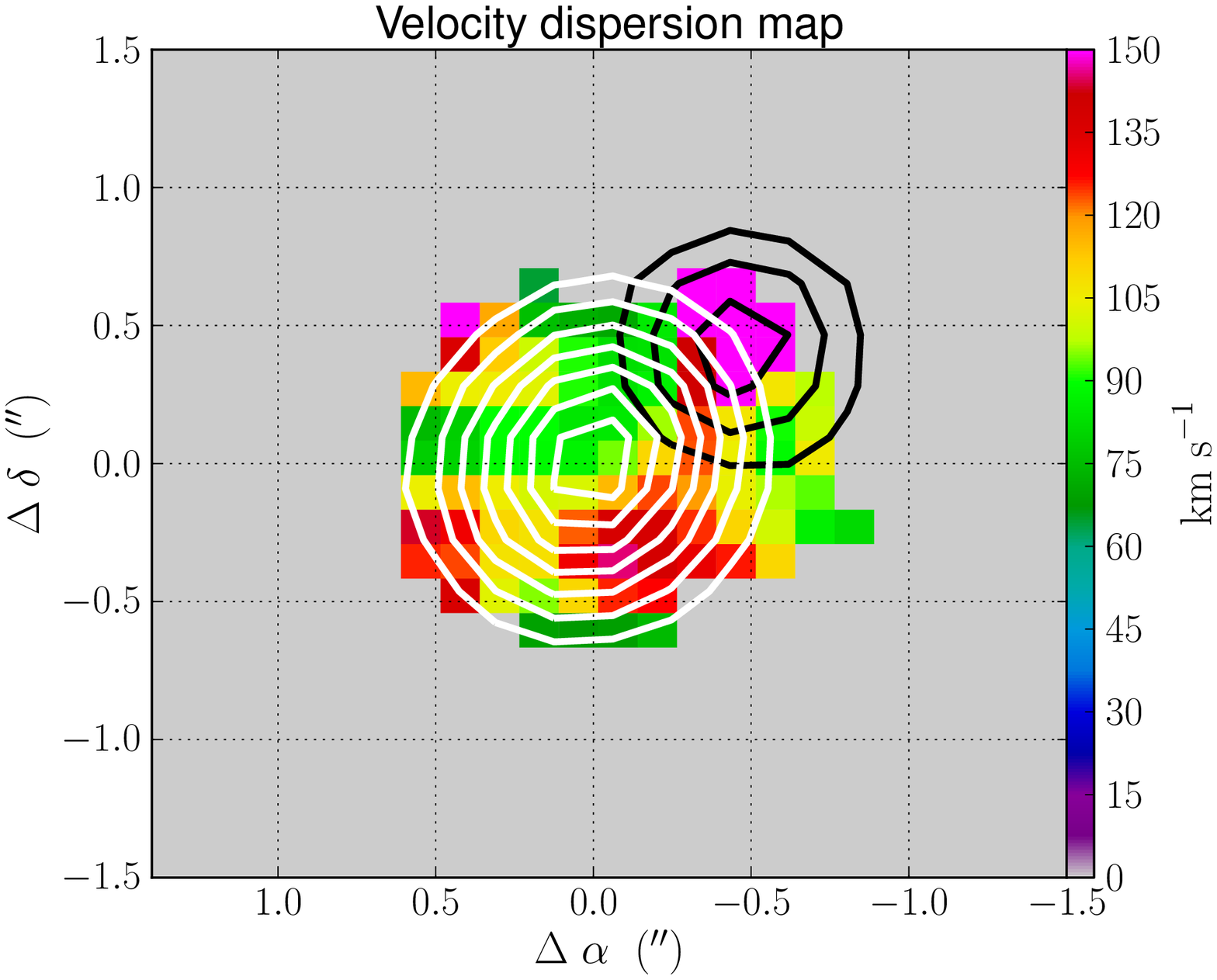}}
	\caption{The same as Fig.~\ref{src_020294045_i}, but for the MASSIV source 020465775 (major merger). 
		The outer contour marks the 2.92 ADU (9$\sigma_{\rm sky}$) isophote, while brighter 
		isophotes increase in 0.97 ADU (3$\sigma_{\rm sky}$) steps. 
		[{\it A colour version of this plot is available at the electronic edition}].}
	\label{src_020465775}
	\end{figure*}


	\begin{figure*}[t!]
	\resizebox{0.32\hsize}{!}{\includegraphics{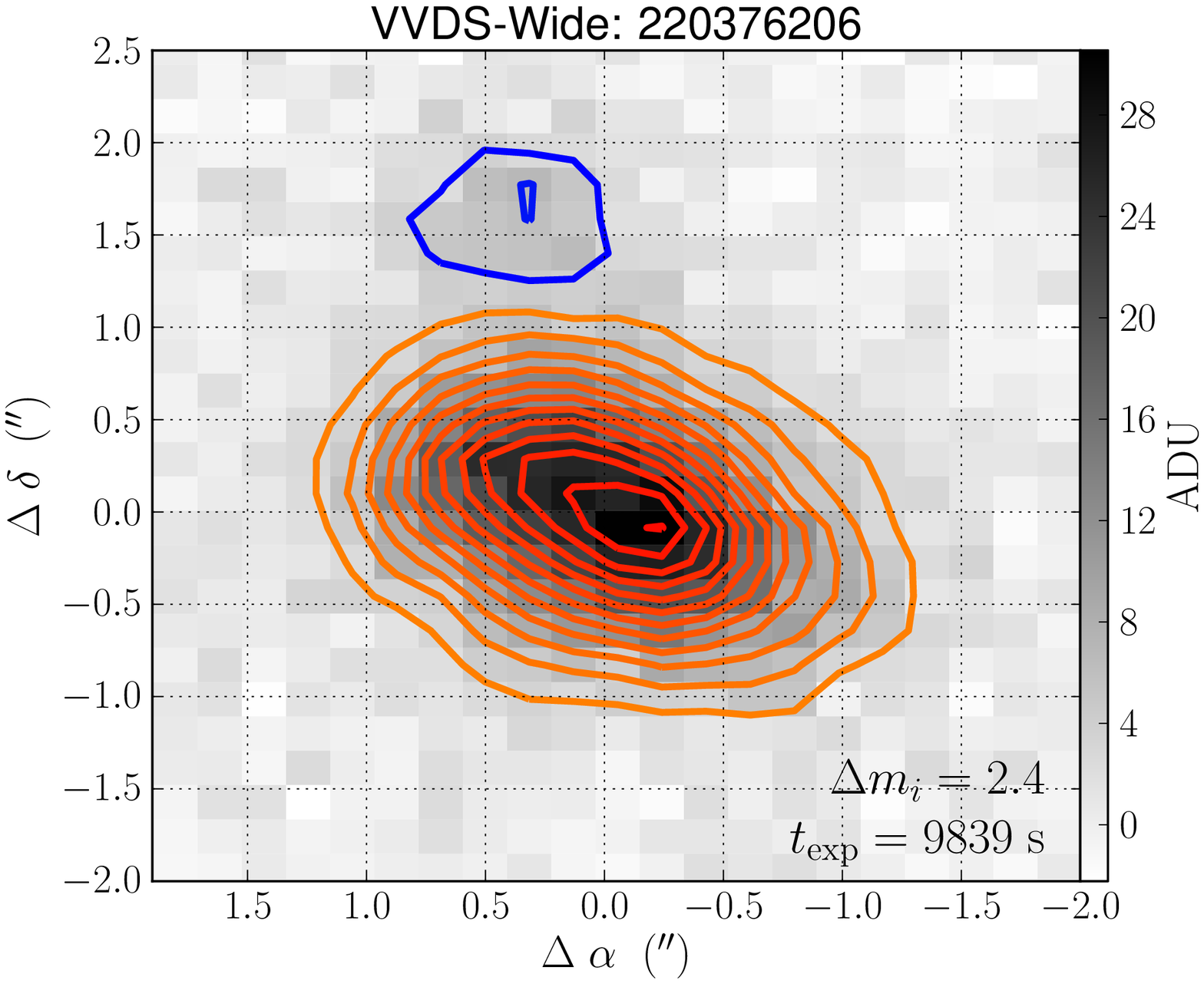}}
	\resizebox{0.32\hsize}{!}{\includegraphics{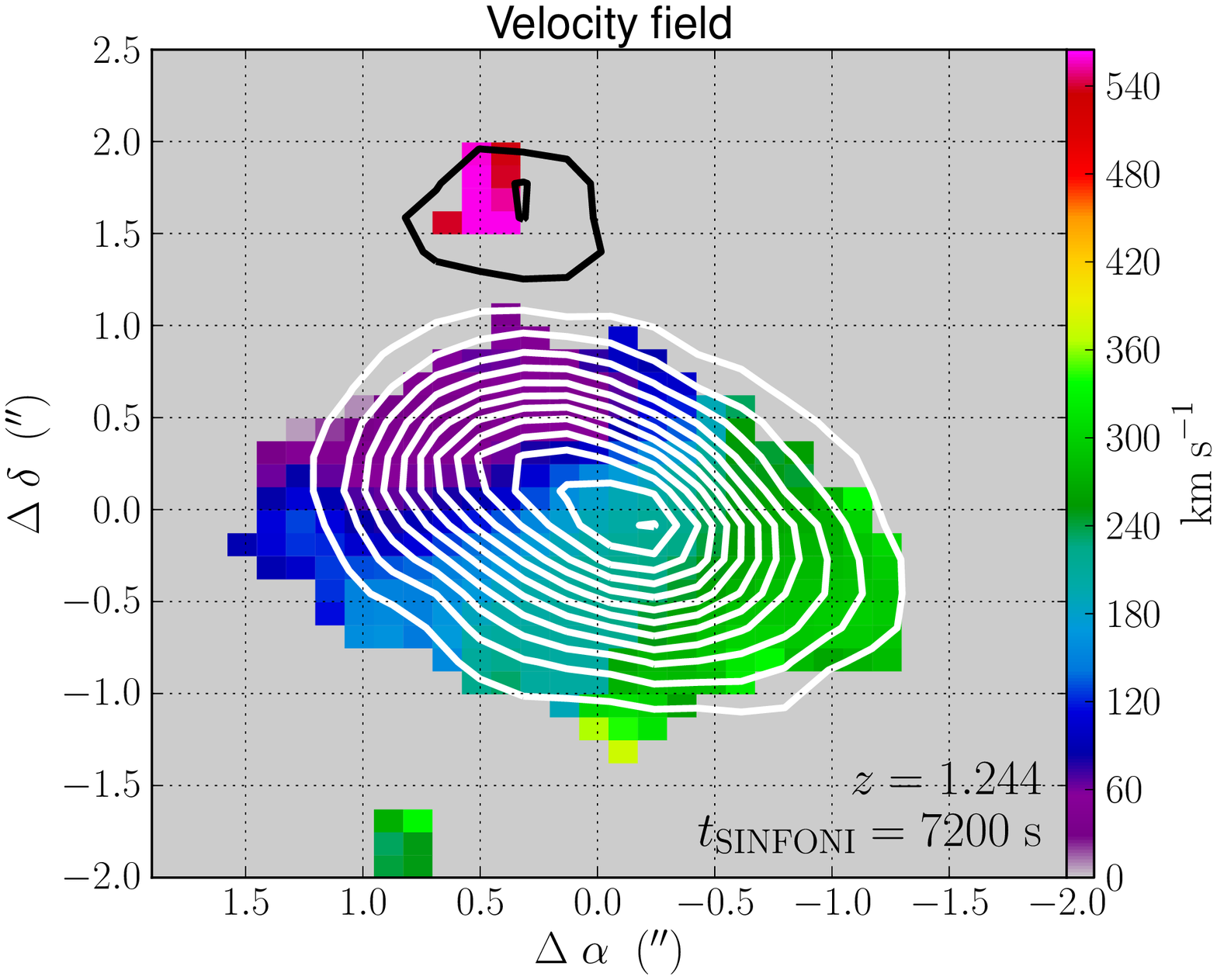}}
	\resizebox{0.32\hsize}{!}{\includegraphics{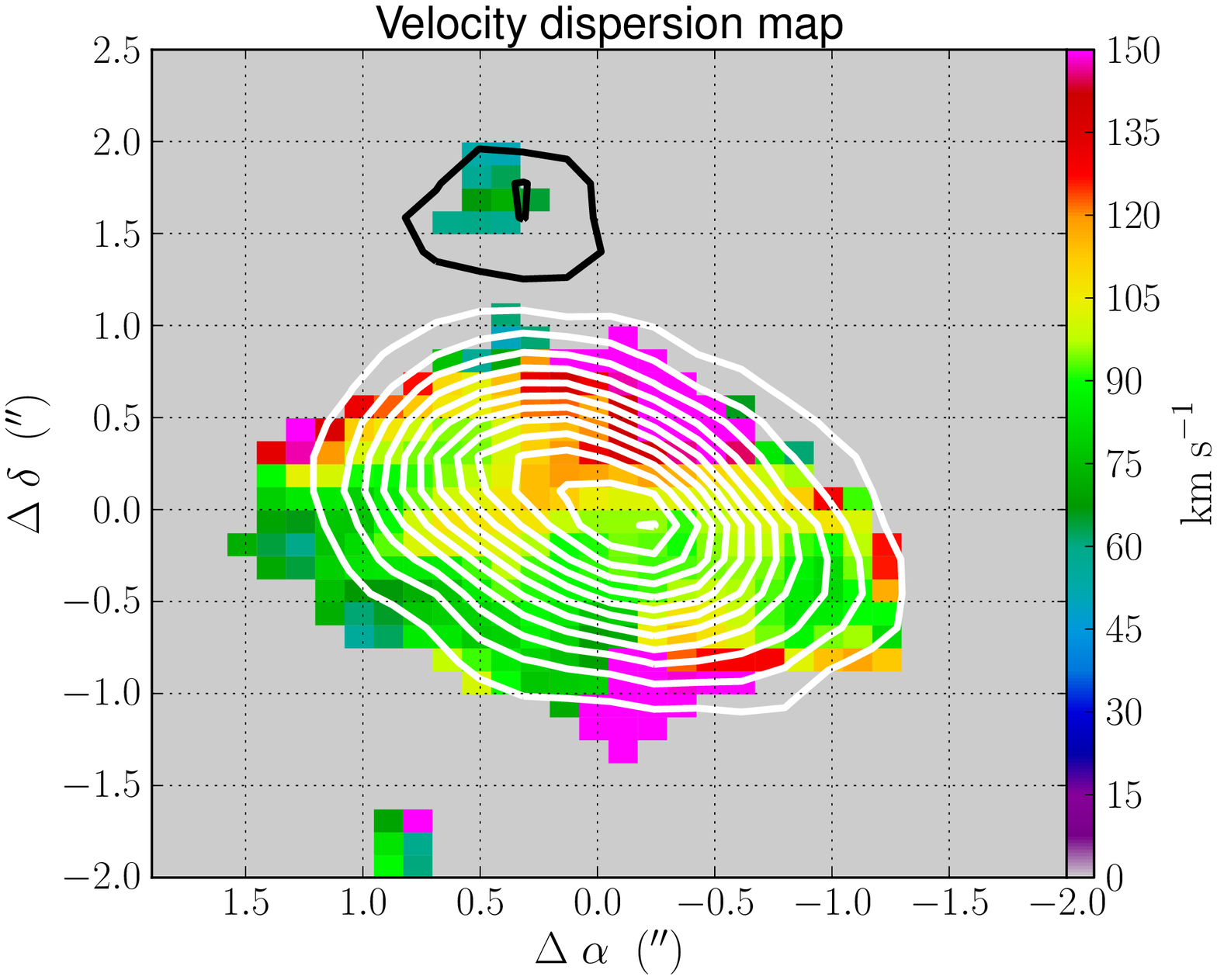}}
	\caption{The same as Fig.~\ref{src_020294045_i}, but for the MASSIV source 220376206 (no major merger). 
		The outer contour marks the 3.39 ADU (3$\sigma_{\rm sky}$) isophote, while brighter 
		isophotes increase in 2.26 ADU (2$\sigma_{\rm sky}$) steps. 
		[{\it A colour version of this plot is available at the electronic edition}].}
	\label{src_220376206}
	\end{figure*}


	\begin{figure*}[t!]
	\resizebox{0.32\hsize}{!}{\includegraphics{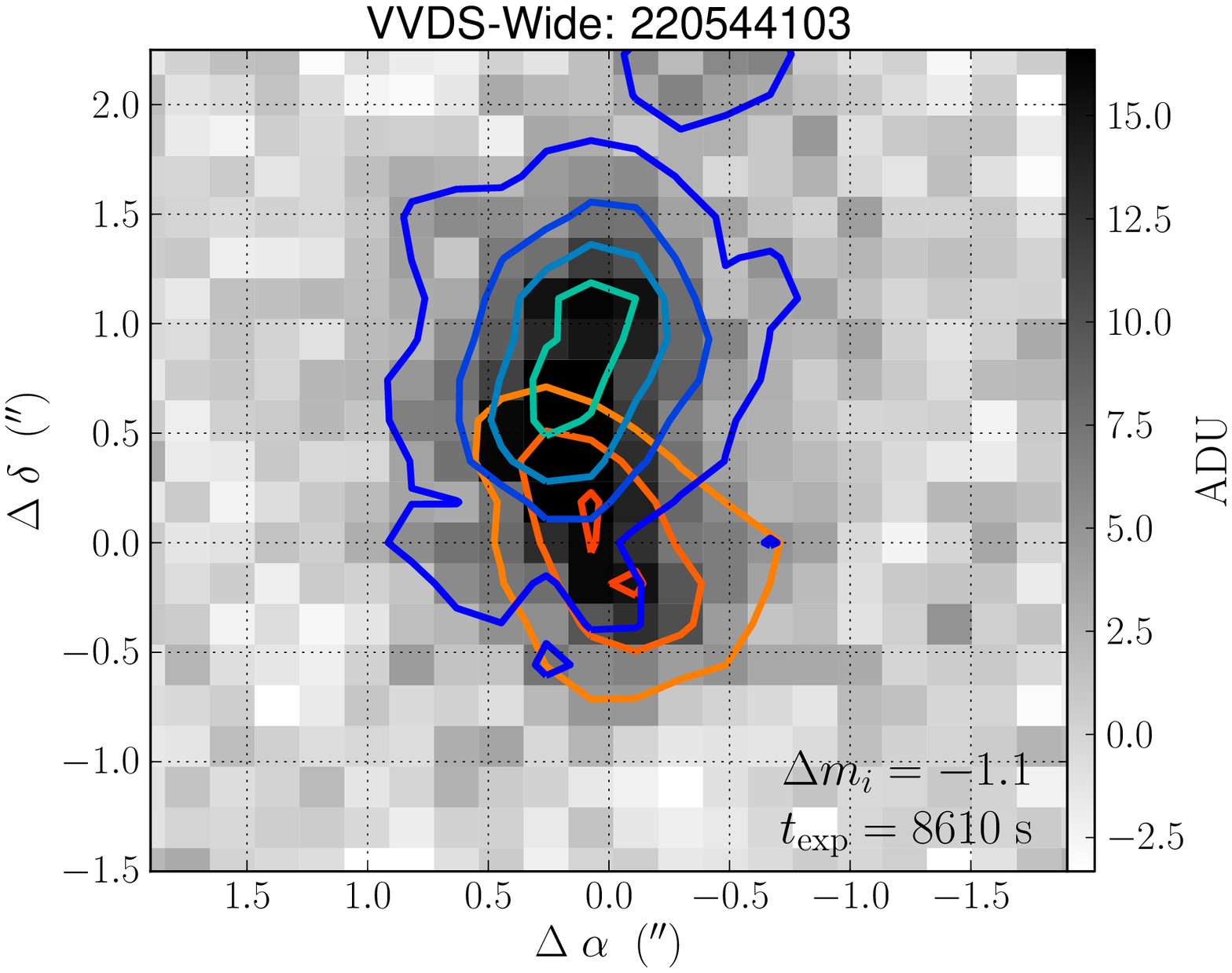}}
	\resizebox{0.32\hsize}{!}{\includegraphics{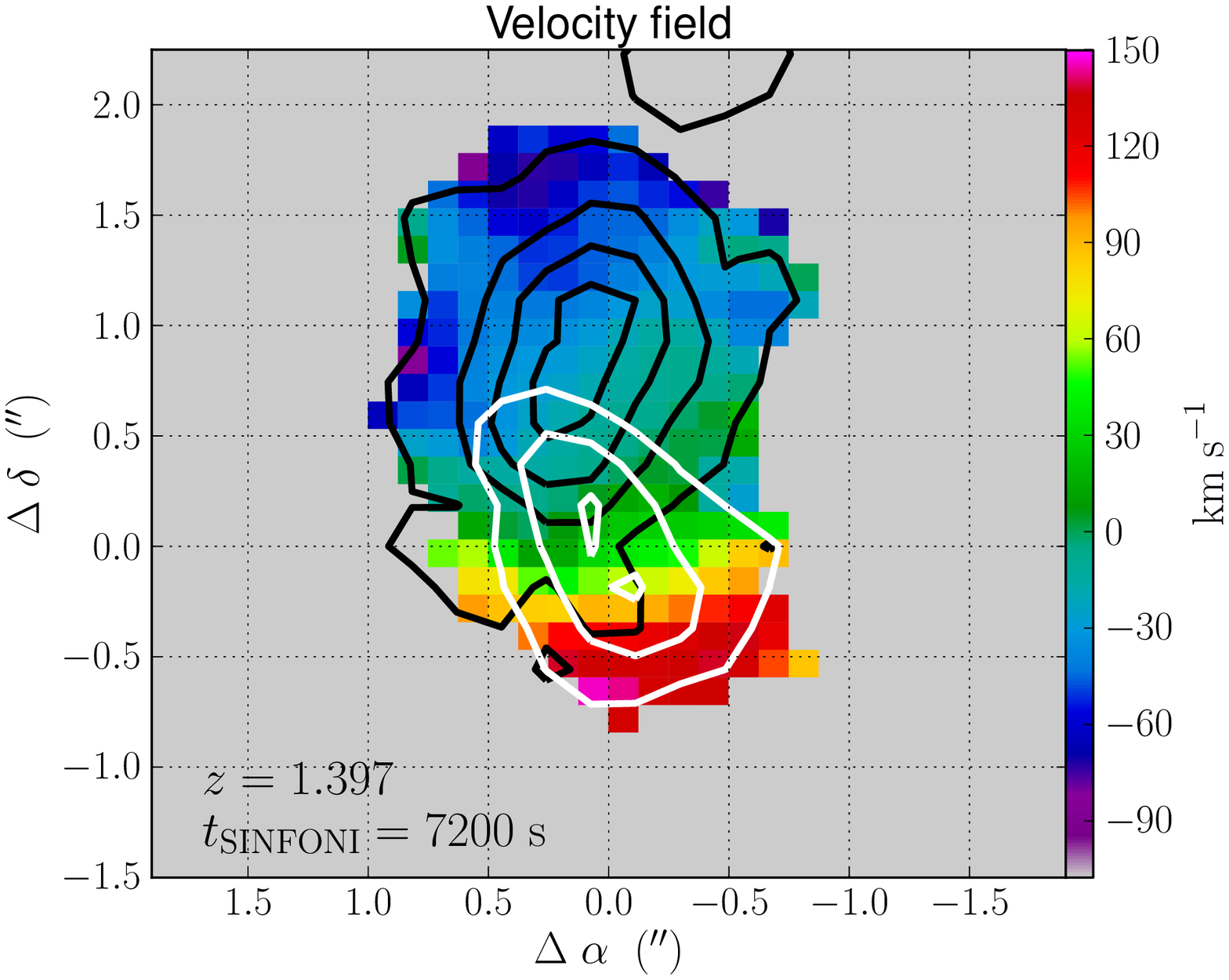}}
	\resizebox{0.32\hsize}{!}{\includegraphics{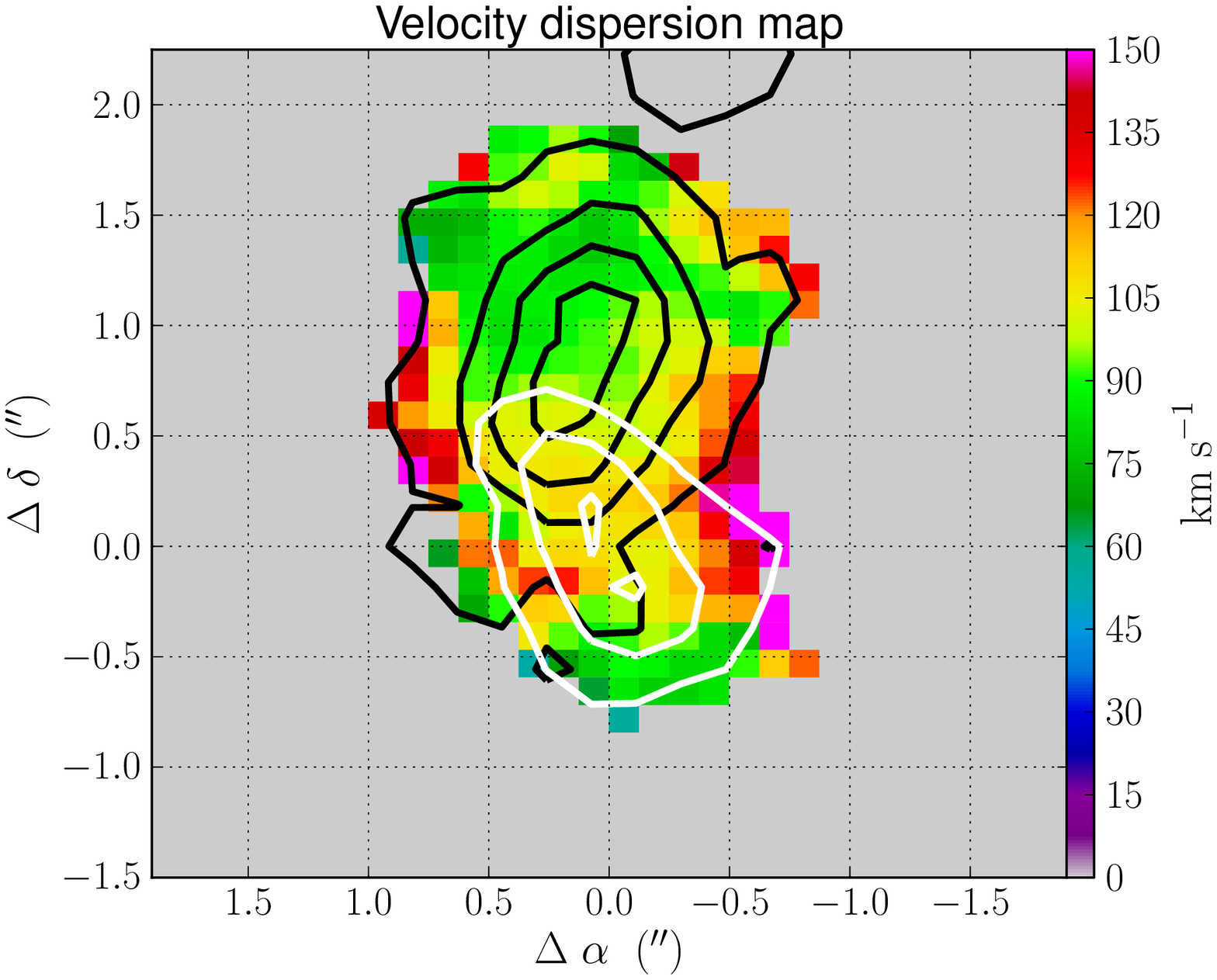}}
	\caption{The same as Fig.~\ref{src_020294045_i}, but for the MASSIV source 220544103 (major merger). 
		The outer contour marks the 3.32 ADU (2$\sigma_{\rm sky}$) isophote, while brighter 
		isophotes increase in 3.32 ADU (2$\sigma_{\rm sky}$) steps. 
		[{\it A colour version of this plot is available at the electronic edition}].}
	\label{src_220544103}
	\end{figure*}


	\begin{figure*}[t!]
	\resizebox{0.32\hsize}{!}{\includegraphics{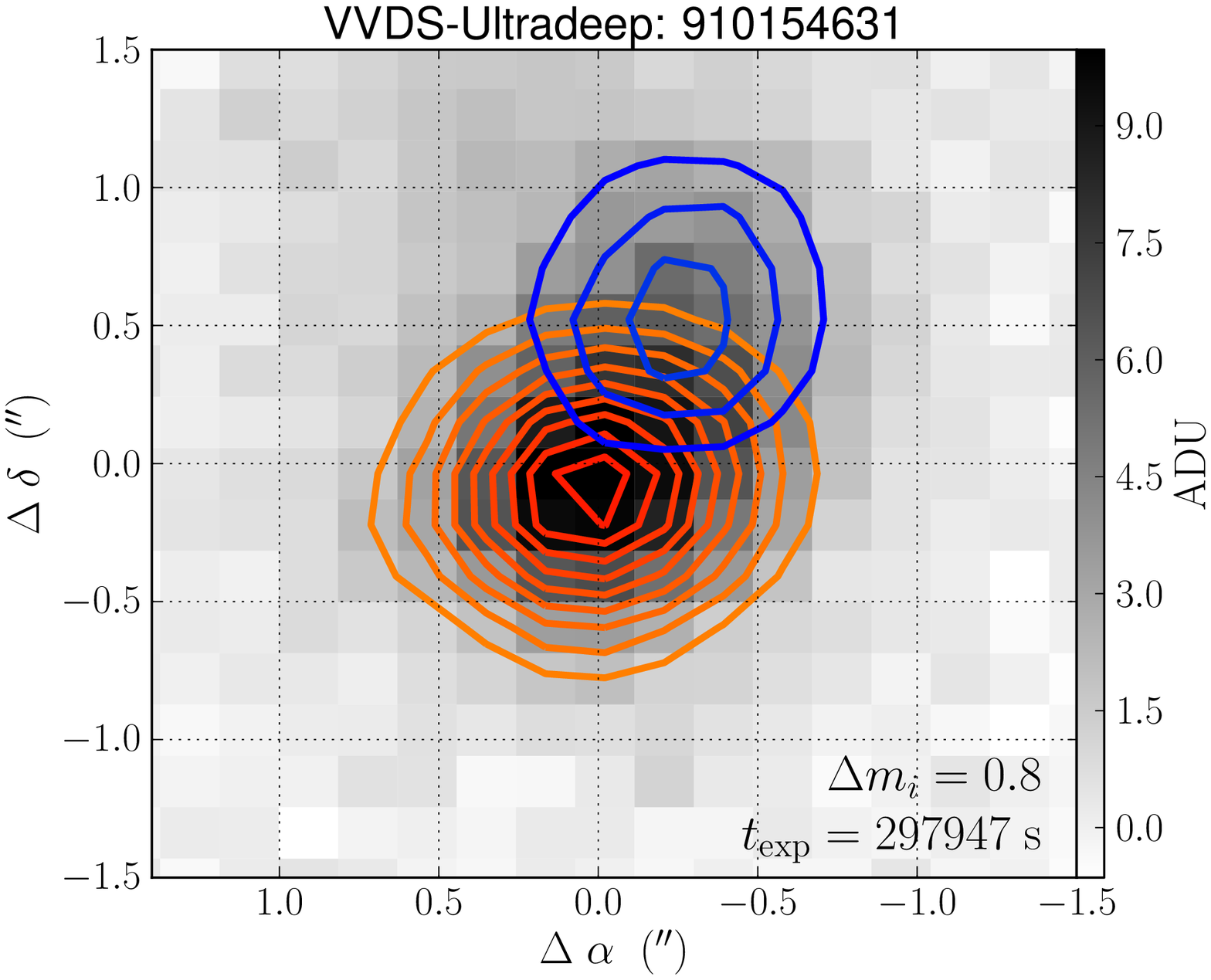}}
	\resizebox{0.32\hsize}{!}{\includegraphics{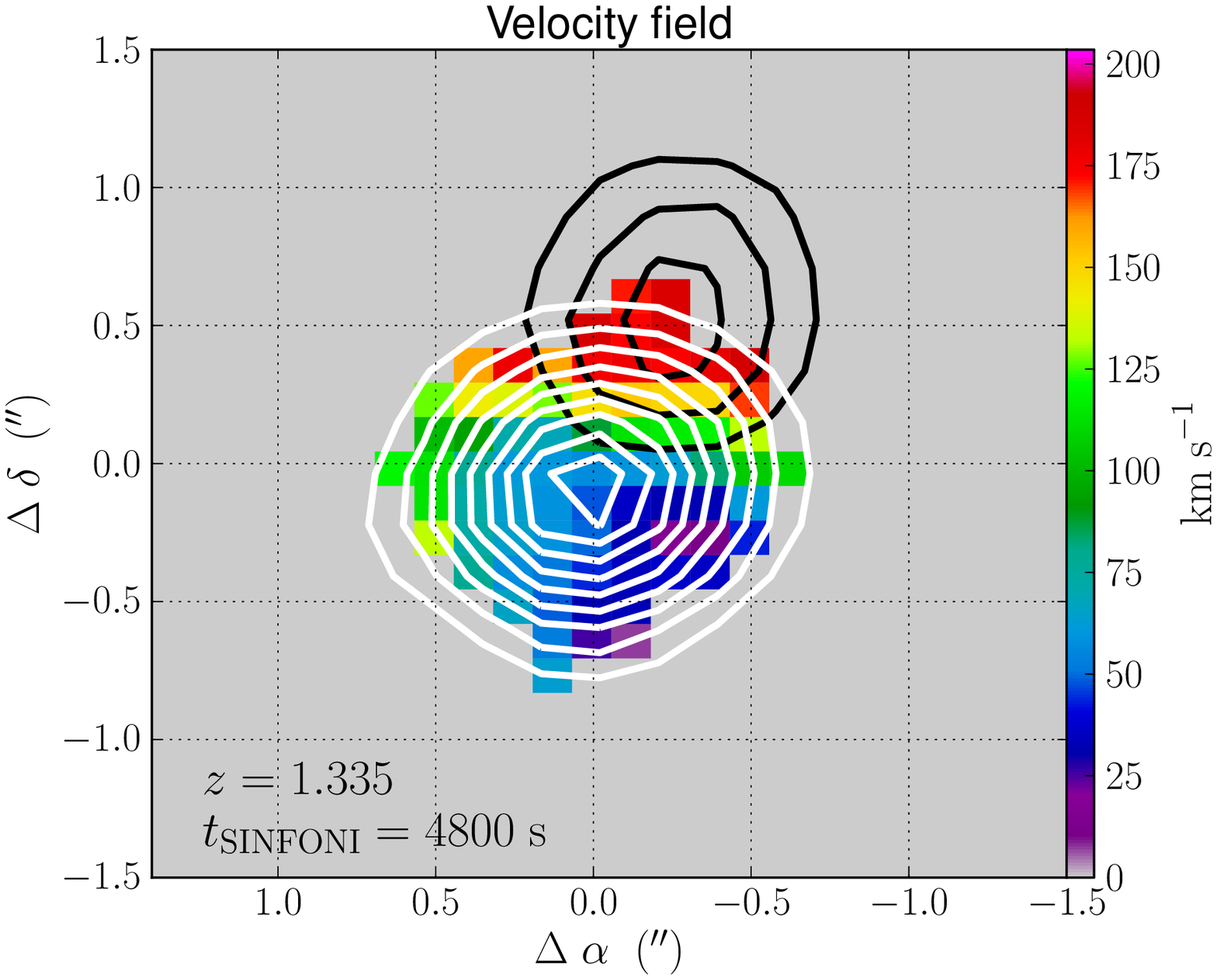}}
	\resizebox{0.32\hsize}{!}{\includegraphics{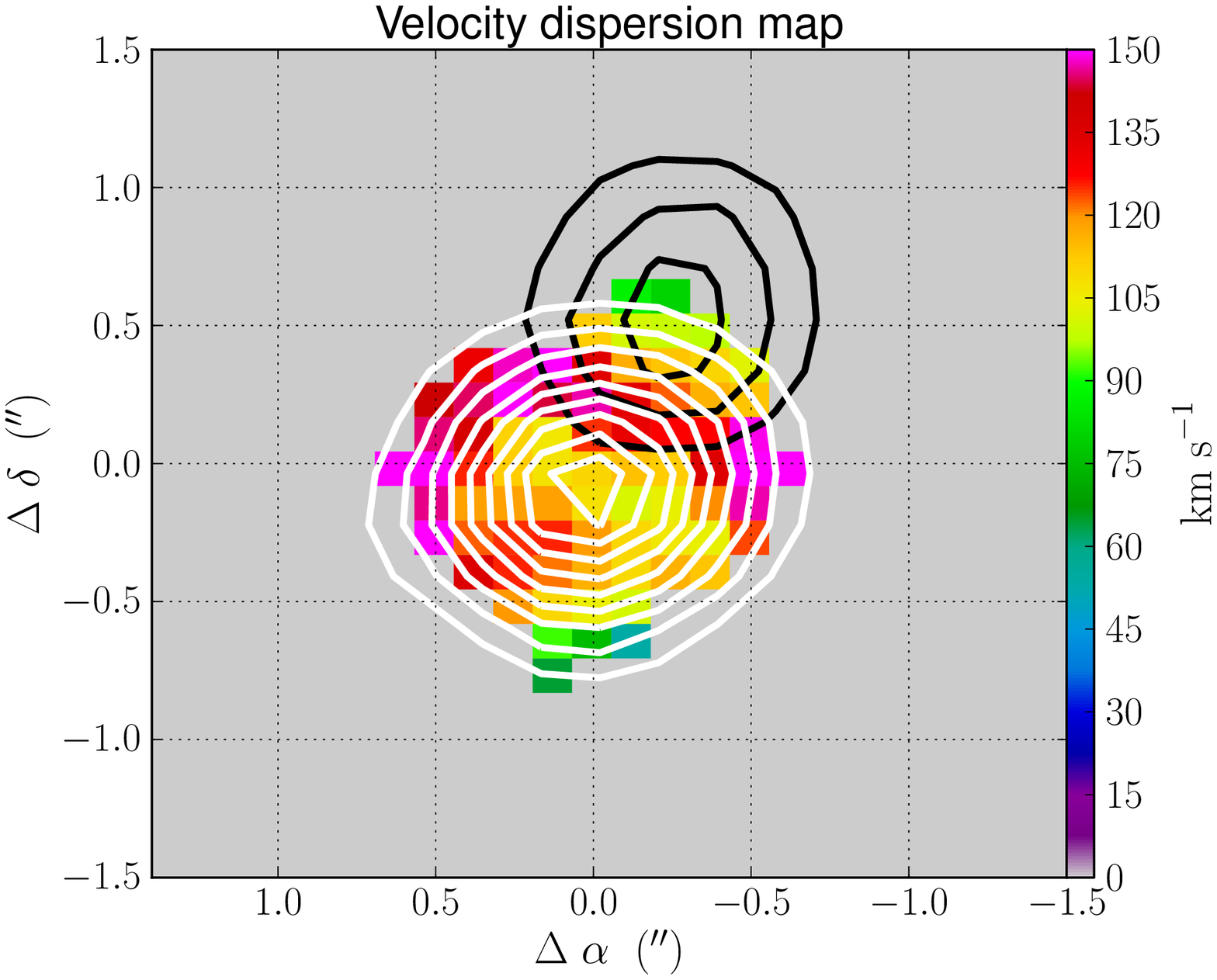}}
	\caption{The same as Fig.~\ref{src_020294045_i}, but for the MASSIV source 910154631 (major merger). 
		The outer contour marks the 1.9 ADU (6$\sigma_{\rm sky}$) isophote, while brighter 
		isophotes increase in 0.8 ADU (2.5$\sigma_{\rm sky}$) steps. 
		[{\it A colour version of this plot is available at the electronic edition}].}
	\label{src_910154631}
	\end{figure*}

	
	\item 140096645 (Fig.~\ref{src_140096645}). {\it No major merger}.  A small companion is identified
		in the velocity map about 1.2 arcsec away
		from the principal galaxy. There is no continuum detection
		in the $i-$band image in the position of the H$\alpha$ companion,
		although there is a $\sim 1\sigma_{\rm sky}$ excess emission. 
		To estimate if the non-detected companion could be bright/massive
		enough to lead to a major merger, we injected fake companion sources 
		in the $i-$band image and studied their detection fraction as a function 
		of the fake companion luminosity $m_{i,2}$.
		The fake sources were modelled with a S\'ersic function and
		convolved with a typical PSF of the 14h field using GALFIT.
		For each fake source we assumed a random inclination and
		position angle, a S\'ersic index $n_{\rm s} = 1$, i.e., an exponential disc
		(we checked that the detection curve is similar assuming either $n_{\rm s} = 0.5$ or 
		$n_{\rm s} = 2$), and an effective radius $r_{\rm e}$ given by
		the $r_{\rm e} - M_{\star}$ relation in MASSIV, 
		$\log r_{\rm e} = 0.36 + 0.37[\log(M_{\star}/M_{\odot}) - 10]$ (see also \citealt{massiv4}). 
		To estimate the stellar mass of the fake companion, we took $\Delta m_{i}$
		as a proxy of the mass ratio between the principal galaxy, for
		which the stellar mass is known, and the companion. Then,
		we applied Poissonian noise to the model and injected it in
		the expected position of the possible companion. We measured, 
		for different luminosities $m_{i,2}$, which fraction of the
		500 injected fake sources were detected. We repeated the
		previous steps for fake sources with $2r_{\rm e}$ (extended sources) and $0.5r_{\rm e}$
		(compact sources), spanning all the possible sizes of the real sources. 
		We show the result of this experiment in the right panel of Fig.~\ref{src_140096645}.

		We find that even for extended sources, we are $\sim 90$\% complete at $m_{i,2} \sim 24$, 
		while the limiting magnitude for a major companion is $m^{\rm MM}_{i,2} = 23.8$ 
		(vertical line in the right panel of Fig.~\ref{src_140096645}). Because of the small probability 
		of non detection of a major companion, we classify this system as no major merger.


	\item 220397579 (Fig.~\ref{src_220397579}). {\bf Major merger}. The velocity map shows two
		different components with a separation of $r_{\rm p} = 14.4h^{-1}$ kpc
		and a relative velocity of $\Delta v \sim 340$ km s$^{-1}$. The companion 
		is toward the north-west. The luminosity difference is $\Delta m_{i} = 0.4$, 
		suggesting a major merger. The difference in the 
		$K_{\rm s}$ band is $\Delta m_{K_{\rm s}} = -1.4$. The negative sign implies that
		the companion is more luminous than the principal, which is brighter in the 
		NUV rest-frame and in H$\alpha$. This is consistent with the measured integrated 
		metallicity of these sources, that is higher for the companion \citep{massiv3}. 
		The suggested picture is that the system comprises a nearly face-on principal
		galaxy with intense star formation and low dust reddening,
		and a nearly edge-on companion galaxy with either a low
		level of star formation or strong dust reddening. We classify
		the system as a major merger.


	\item 220544394 (Fig.~\ref{src_220544394}). {\bf Major merger}. The velocity map shows two
		different components with a separation of $r_{\rm p} = 7.1h^{-1}$ kpc
		and a relative velocity of $\Delta v \sim 50$ km s$^{-1}$, that are also
		well recovered by the GALFIT model with two components. 
		The companion is toward the north, while the luminosity difference is $\Delta m_{i} = 1.3$. 
		We classify the system as a major merger.

\end{itemize}

In summary for this redshift range, we classify 3 of the 6
close pair candidates as major mergers. This translates to a gas-rich major merger fraction of
$f_{\rm MM} = 0.208^{+0.152}_{-0.068}$ at $\overline{z_{\rm r}}_{,1} = 1.03$.


\subsection{Close pair candidates at $1.2 \leq z < 1.5$}
The weighted mean redshift of the second redshift bin is $\overline{z_{\rm r}}_{,2} = 1.32$. 
This is a redshift range where there is no measurement of the major merger fraction
from spectroscopic close pairs yet. In this framework,
MASSIV provides an unique opportunity to measure the major
merger fraction at this crucial epoch of galaxy evolution. We identify
11 close pair candidates over 30 galaxies in this redshift bin:


	\begin{figure*}[t!]
	\resizebox{0.32\hsize}{!}{\includegraphics{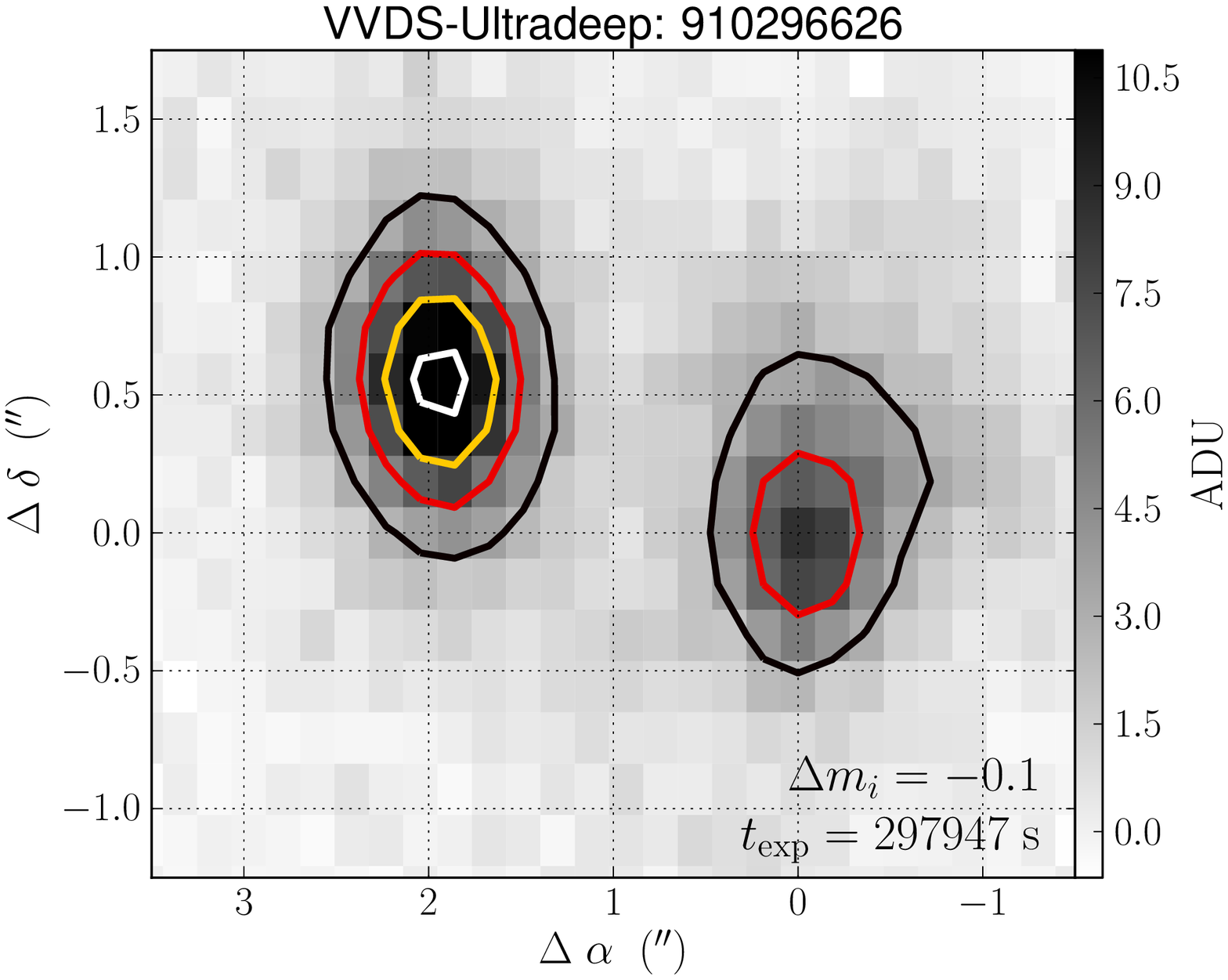}}
	\resizebox{0.32\hsize}{!}{\includegraphics{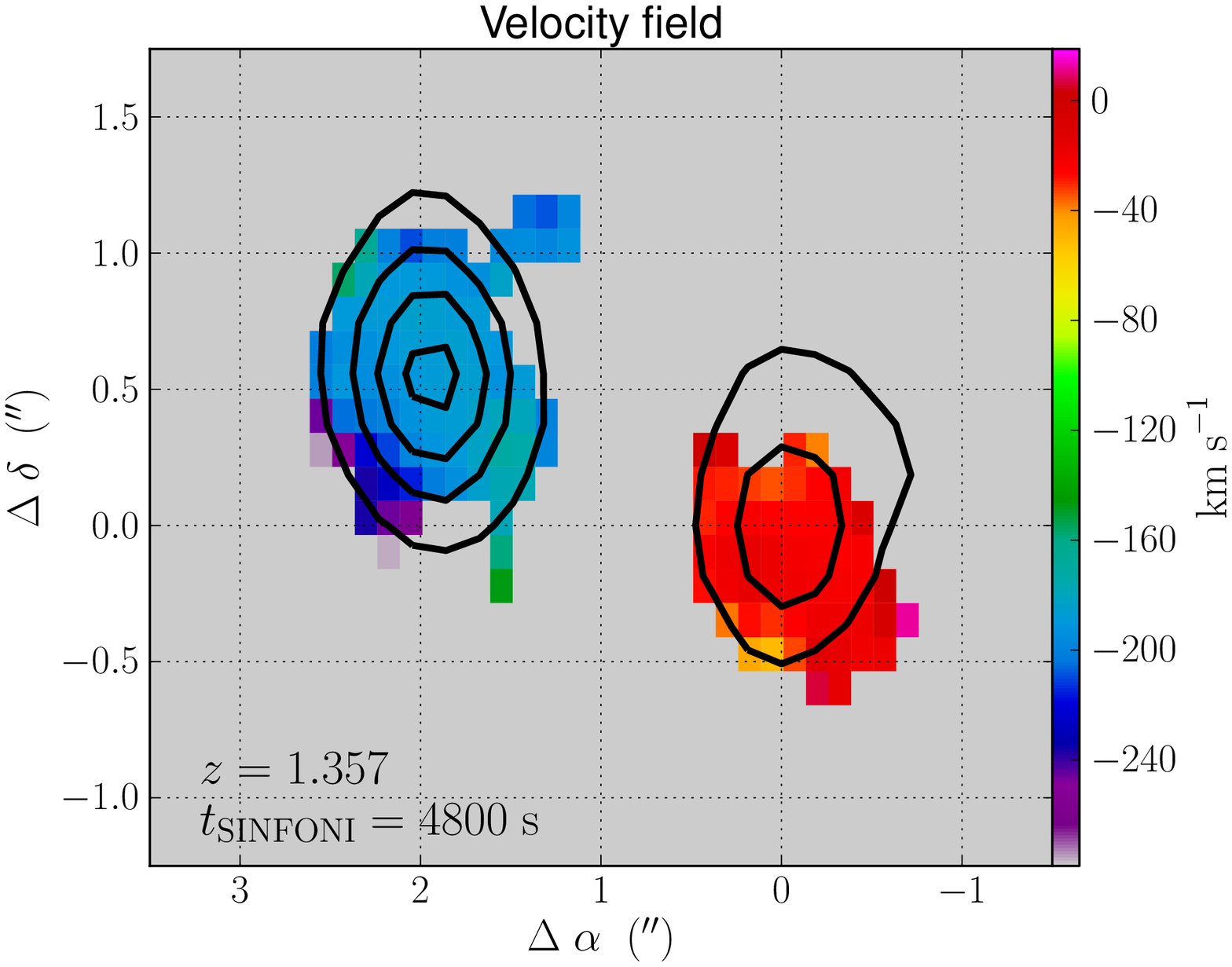}}
	\resizebox{0.32\hsize}{!}{\includegraphics{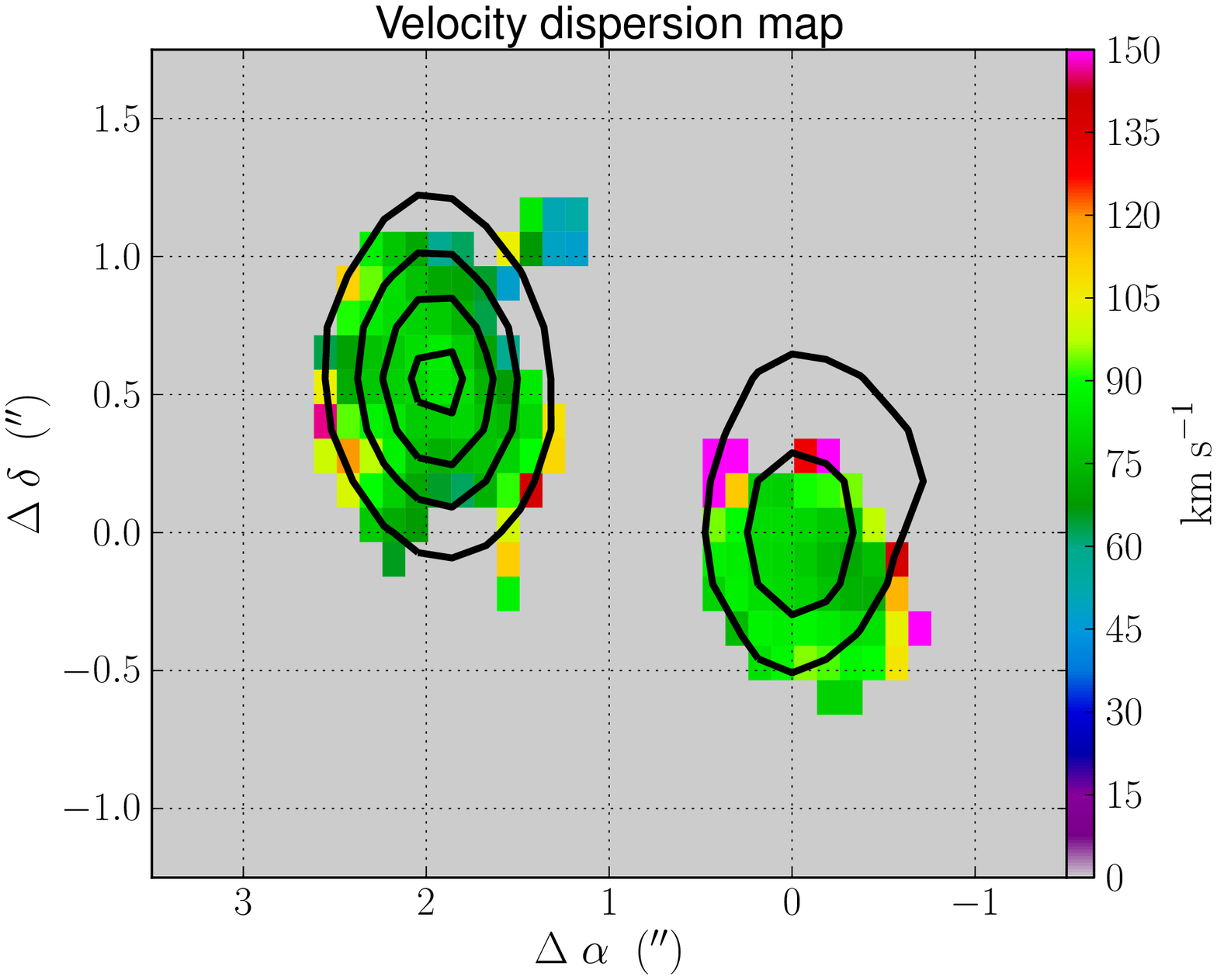}}
	\caption{The same as Fig.~\ref{src_220397579}, but for the MASSIV source 910296626 (major merger). 
		The outer contour marks the 3.32 ADU (10$\sigma_{\rm sky}$) isophote, 
		while brighter isophotes increase in 2.56 ADU (8$\sigma_{\rm sky}$) steps. 
		[{\it A colour version of this plot is available at the electronic edition}].}
	\label{src_910296626}
	\end{figure*}


	\begin{figure*}[t!]
	\resizebox{0.32\hsize}{!}{\includegraphics{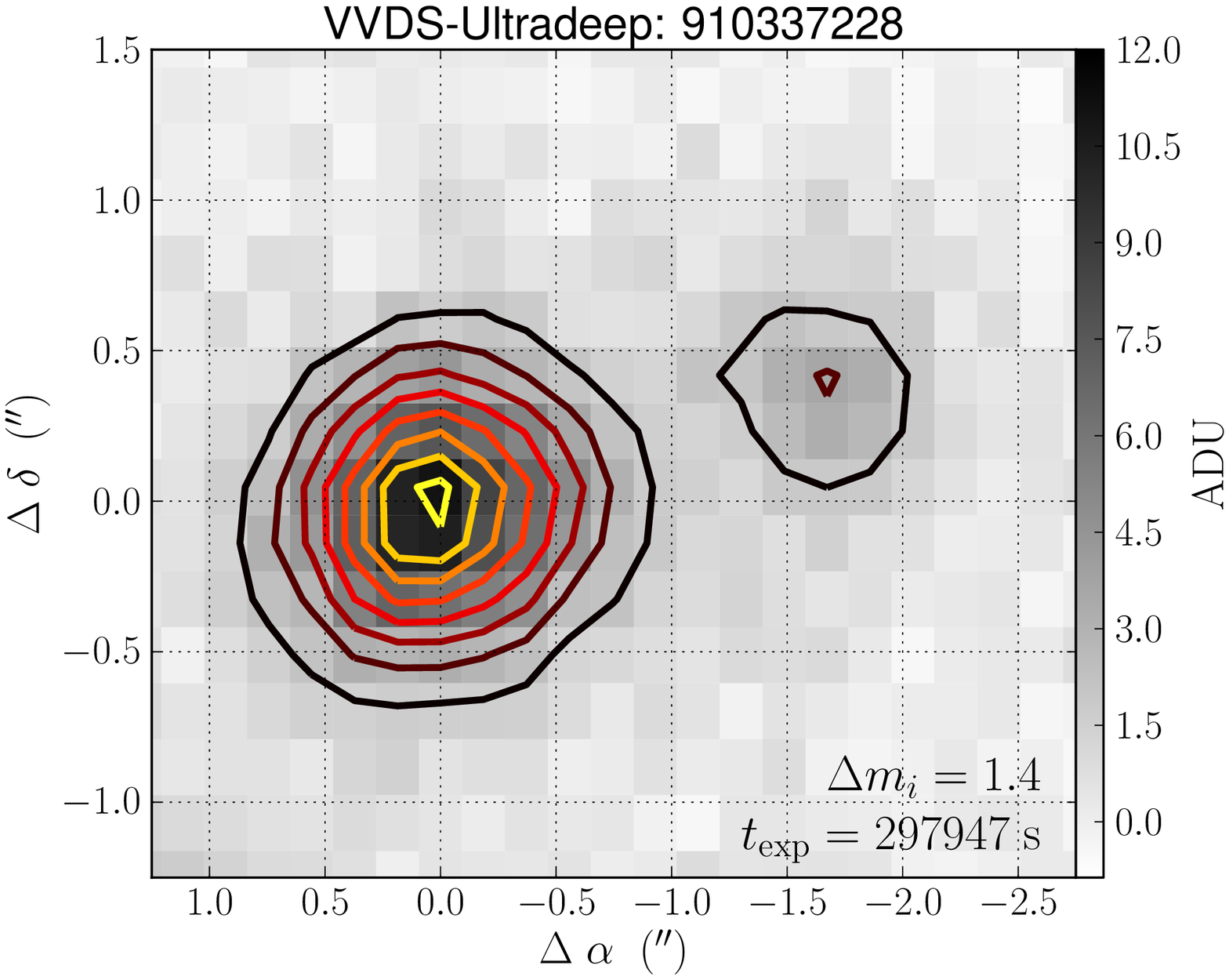}}
	\resizebox{0.32\hsize}{!}{\includegraphics{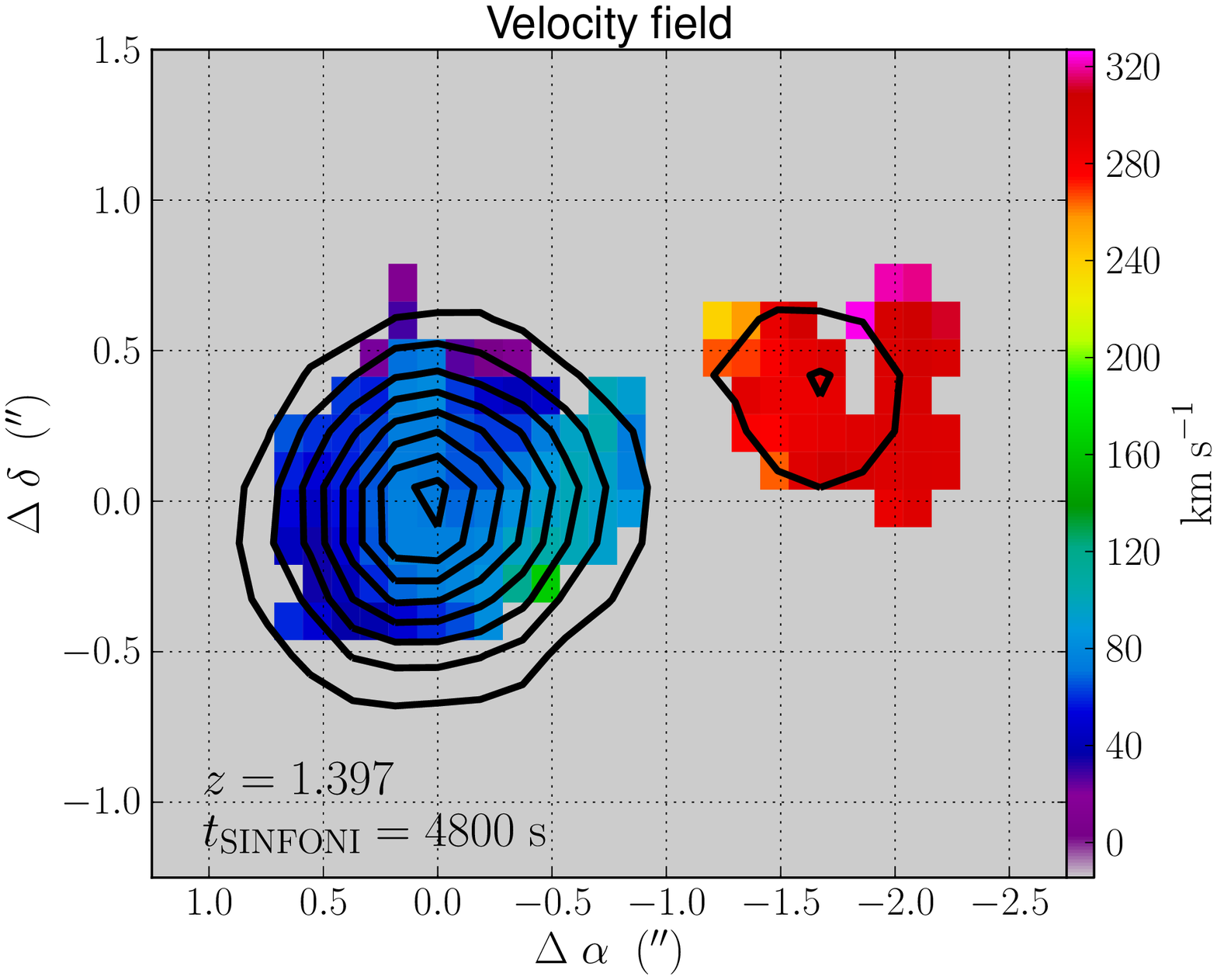}}
	\resizebox{0.32\hsize}{!}{\includegraphics{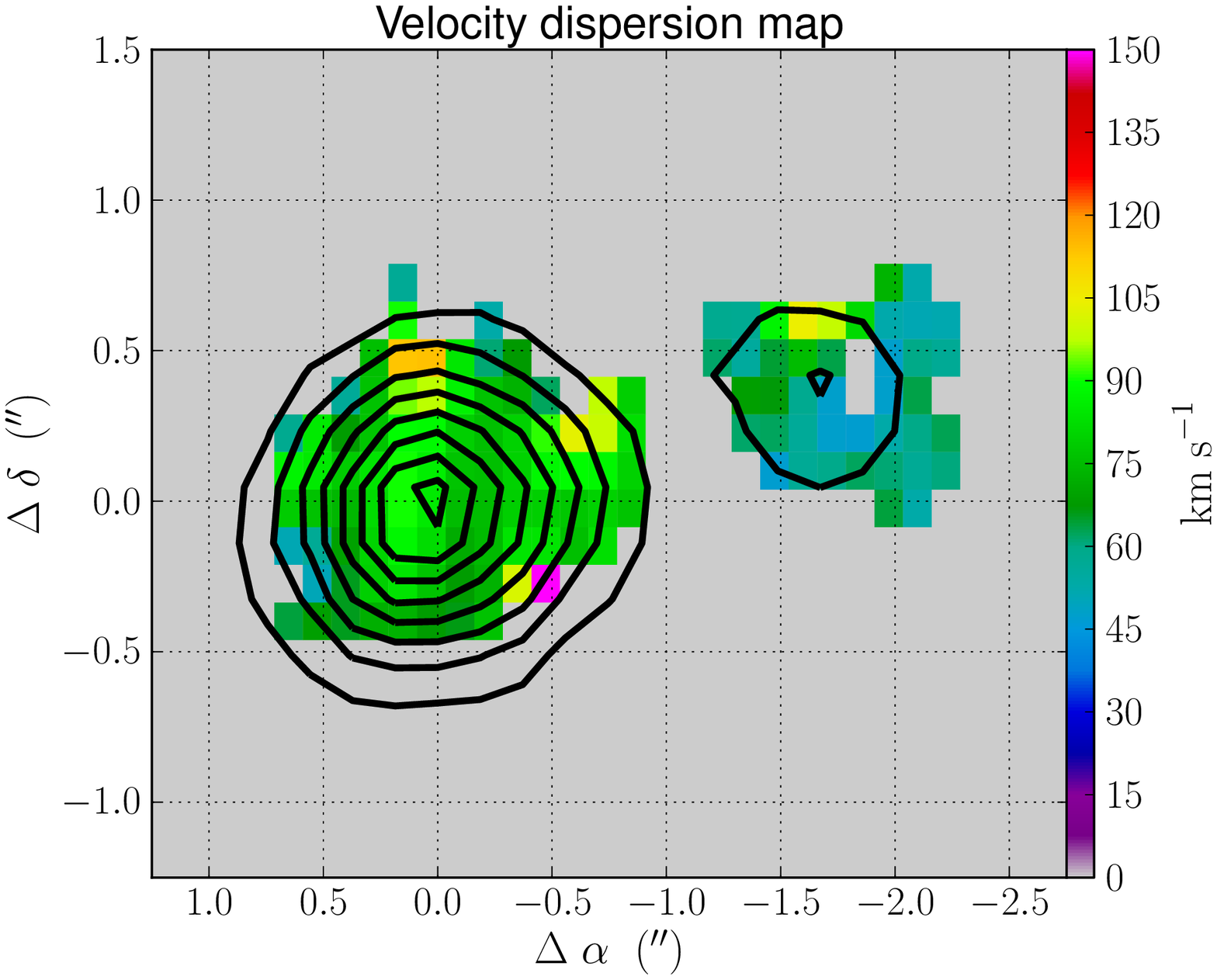}}
	\caption{ The same as Fig.~\ref{src_220397579}, but for the MASSIV source 910337228 (major merger). 
		The outer contour marks the 1.73 ADU (4$\sigma_{\rm sky}$) isophote,
		while brighter isophotes increase in 1.30 ADU (3$\sigma_{\rm sky}$) steps.  
		[{\it A colour version of this plot is available at the electronic edition}].}
	\label{src_910337228}
	\end{figure*}


	\begin{figure*}[t!]
	\resizebox{0.32\hsize}{!}{\includegraphics{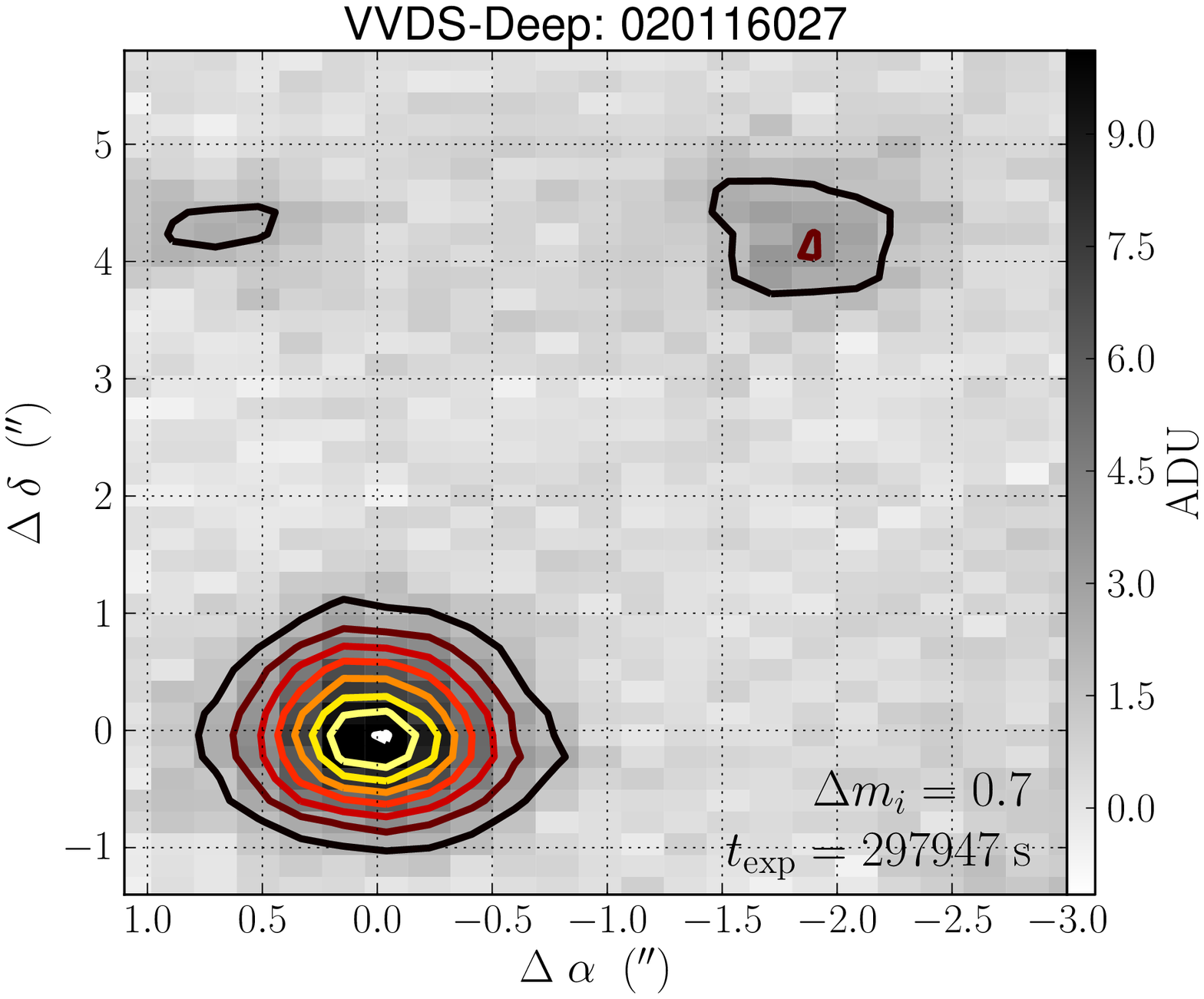}}
	\resizebox{0.32\hsize}{!}{\includegraphics{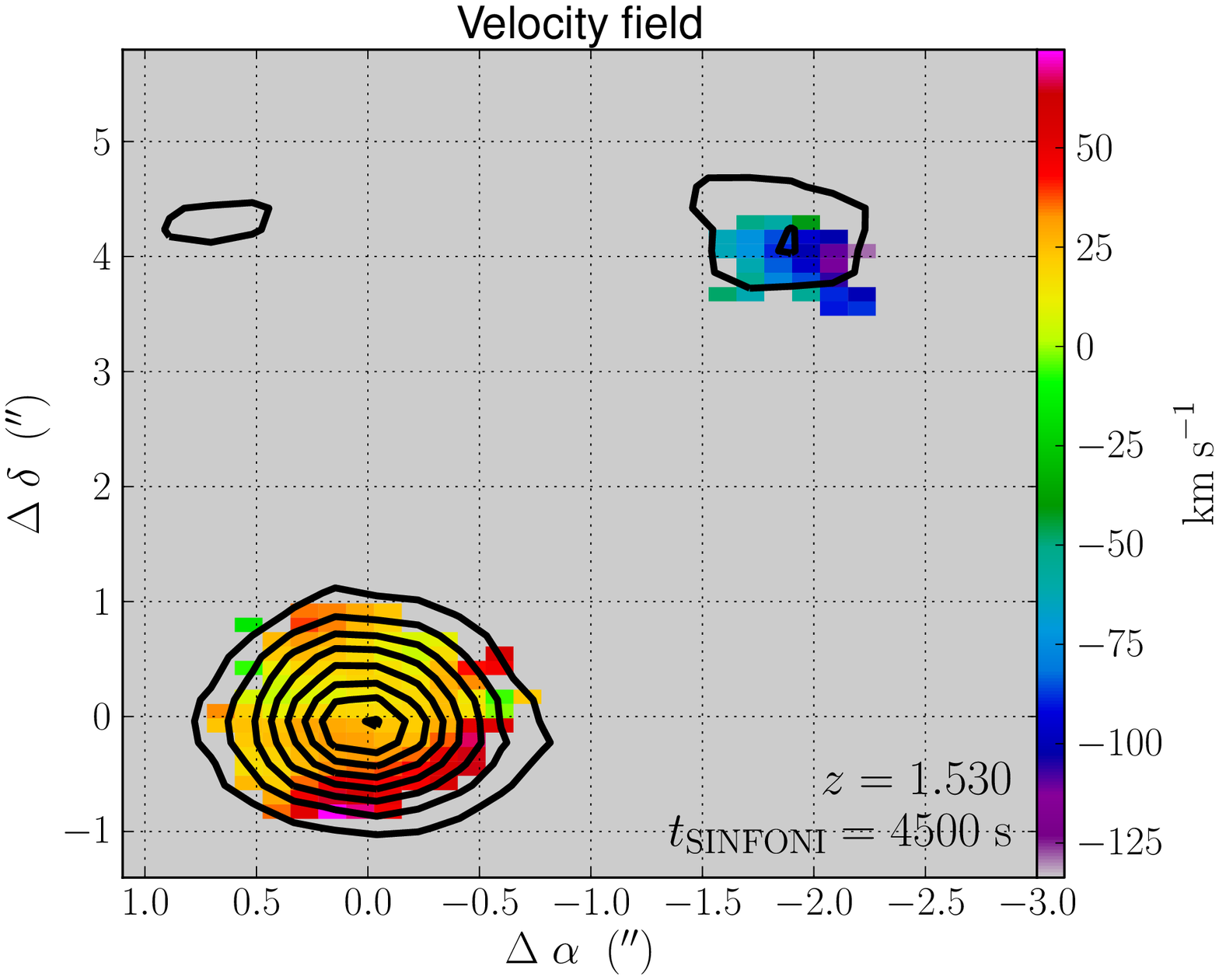}}
	\resizebox{0.32\hsize}{!}{\includegraphics{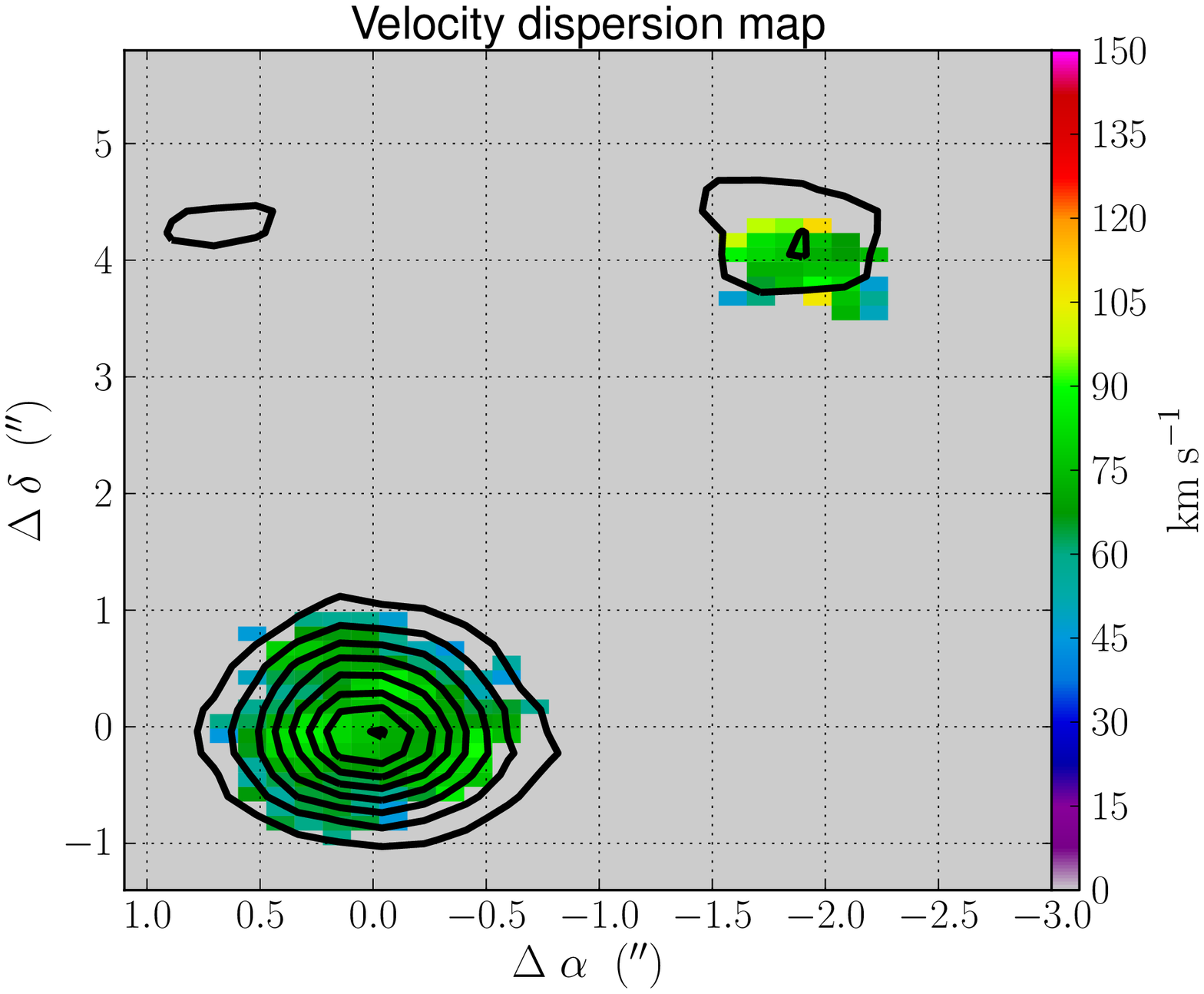}}
	\caption{The same as Fig.~\ref{src_220397579}, but for the MASSIV source 020116027 (major merger). 
		The principal galaxy is that in the south-east. The outer contour marks 
		the 2.02 ADU (3.5$\sigma_{\rm sky}$) isophote, while brighter isophotes 
		increase in 1.16 ADU (2$\sigma_{\rm sky}$) steps. 
		[{\it A colour version of this plot is available at the electronic edition}].}
	\label{src_020116027}
	\label{mbzfig}
	\end{figure*}


	\begin{figure*}[t!]
	\resizebox{0.32\hsize}{!}{\includegraphics{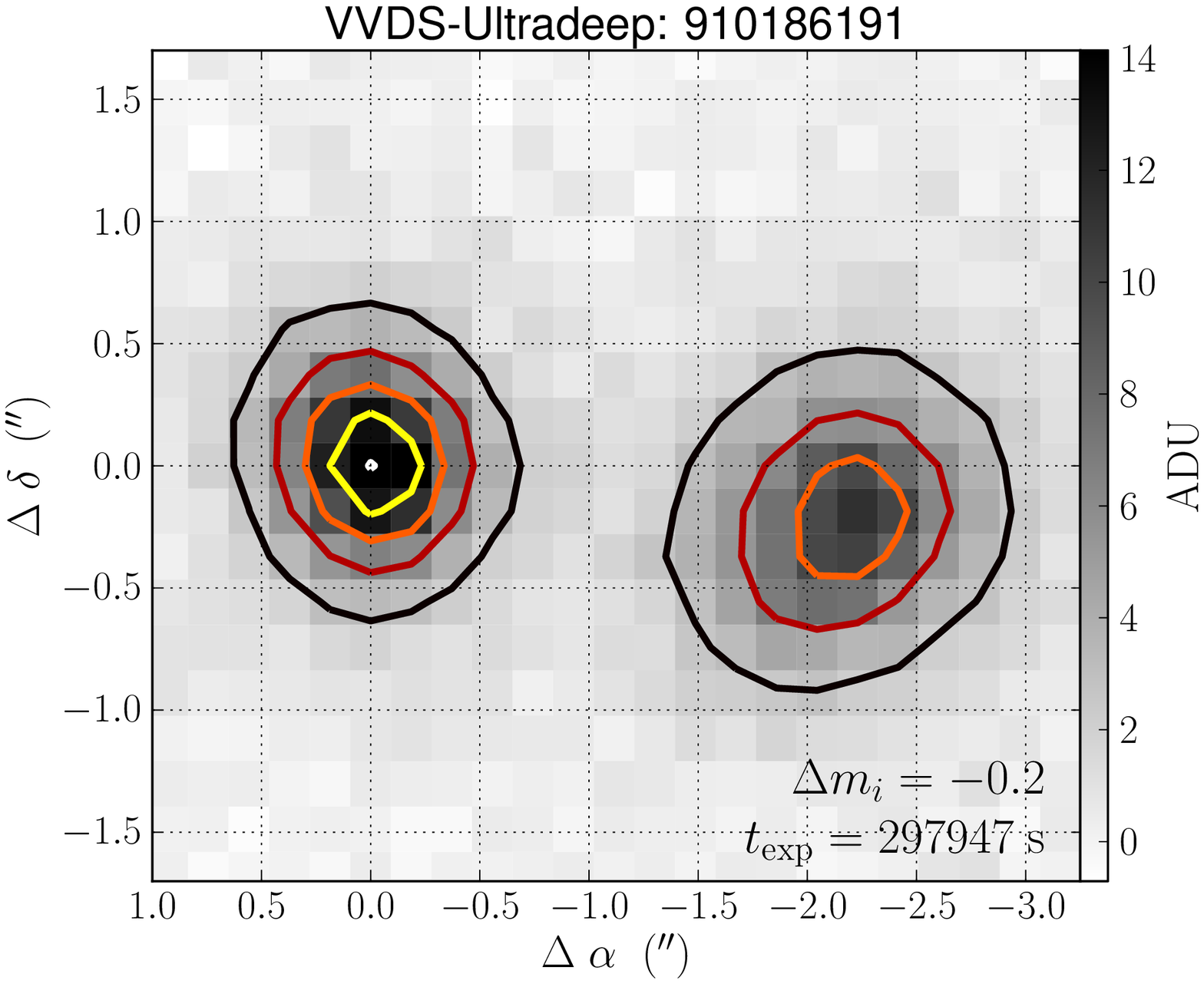}}
	\resizebox{0.32\hsize}{!}{\includegraphics{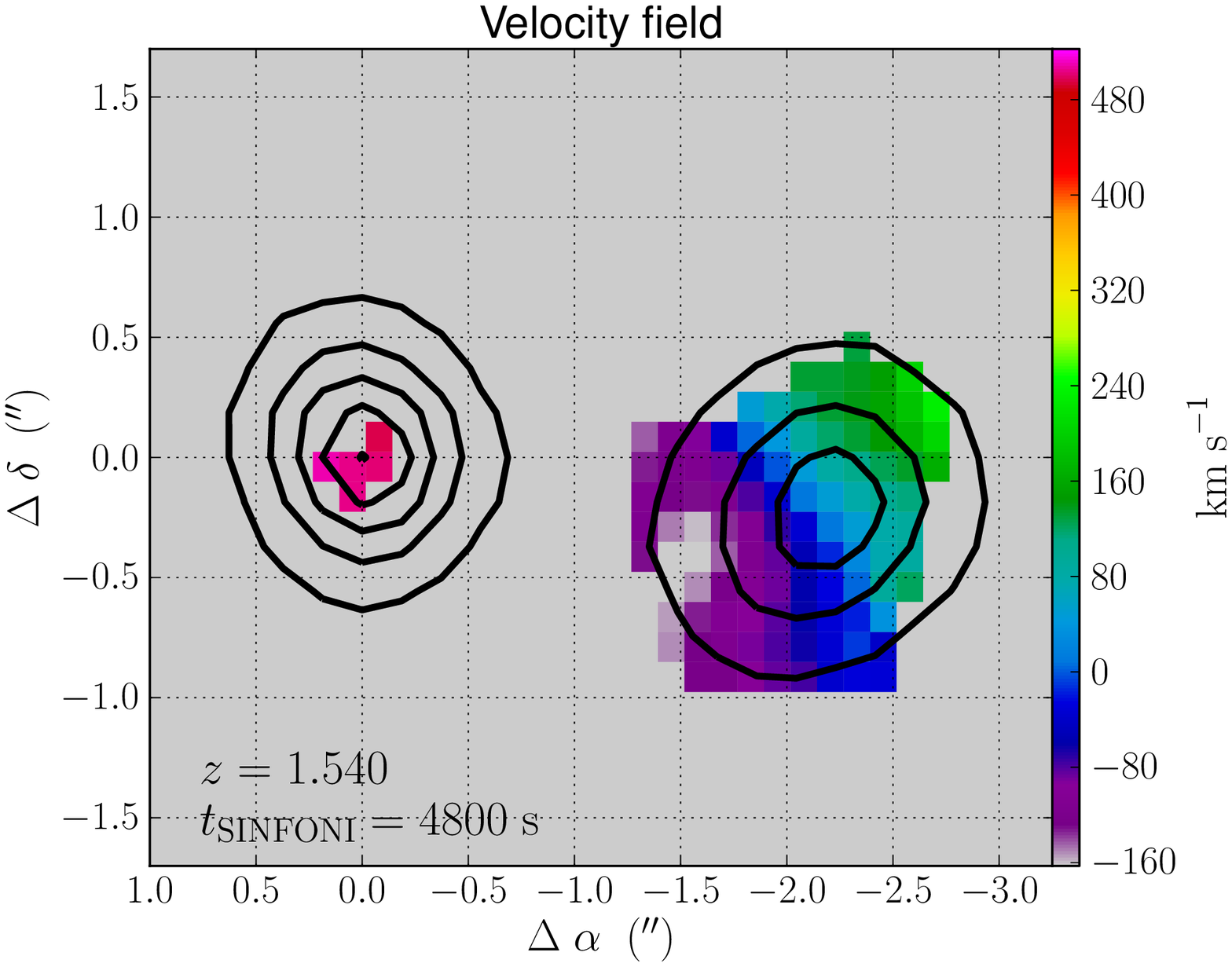}}
	\resizebox{0.32\hsize}{!}{\includegraphics{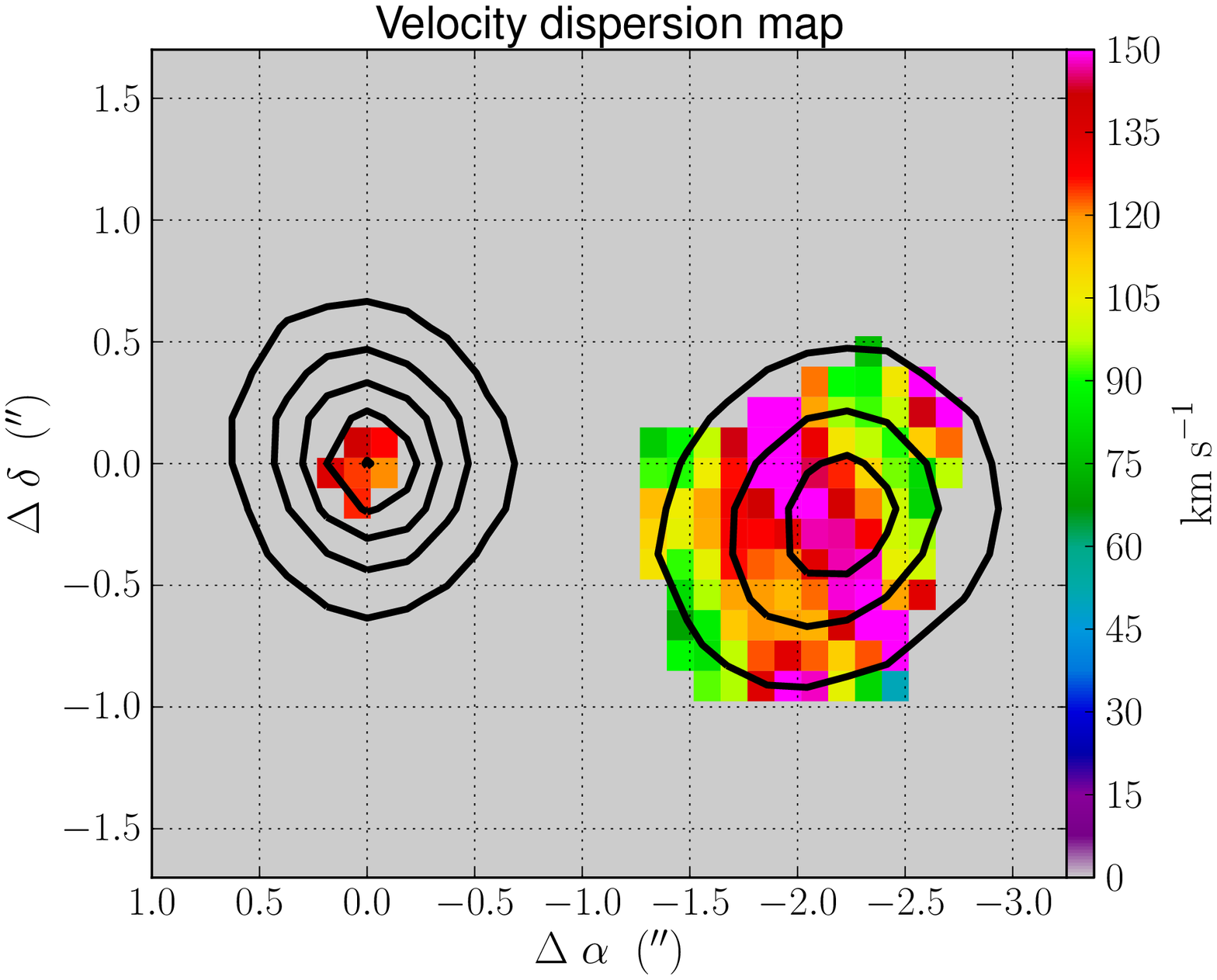}}
	\caption{The same as Fig.~\ref{src_020294045_i}, but for the MASSIV source 910186191 (major merger). 
		The outer contour marks the 2.83 ADU (8$\sigma_{\rm sky}$) isophote, 
		while brighter isophotes increase in 2.83 ADU (8$\sigma_{\rm sky}$) steps. 
		[{\it A colour version of this plot is available at the electronic edition}].}
	\label{src_910186191}
	\end{figure*}
	

	\begin{figure*}[t!]
	\resizebox{0.32\hsize}{!}{\includegraphics{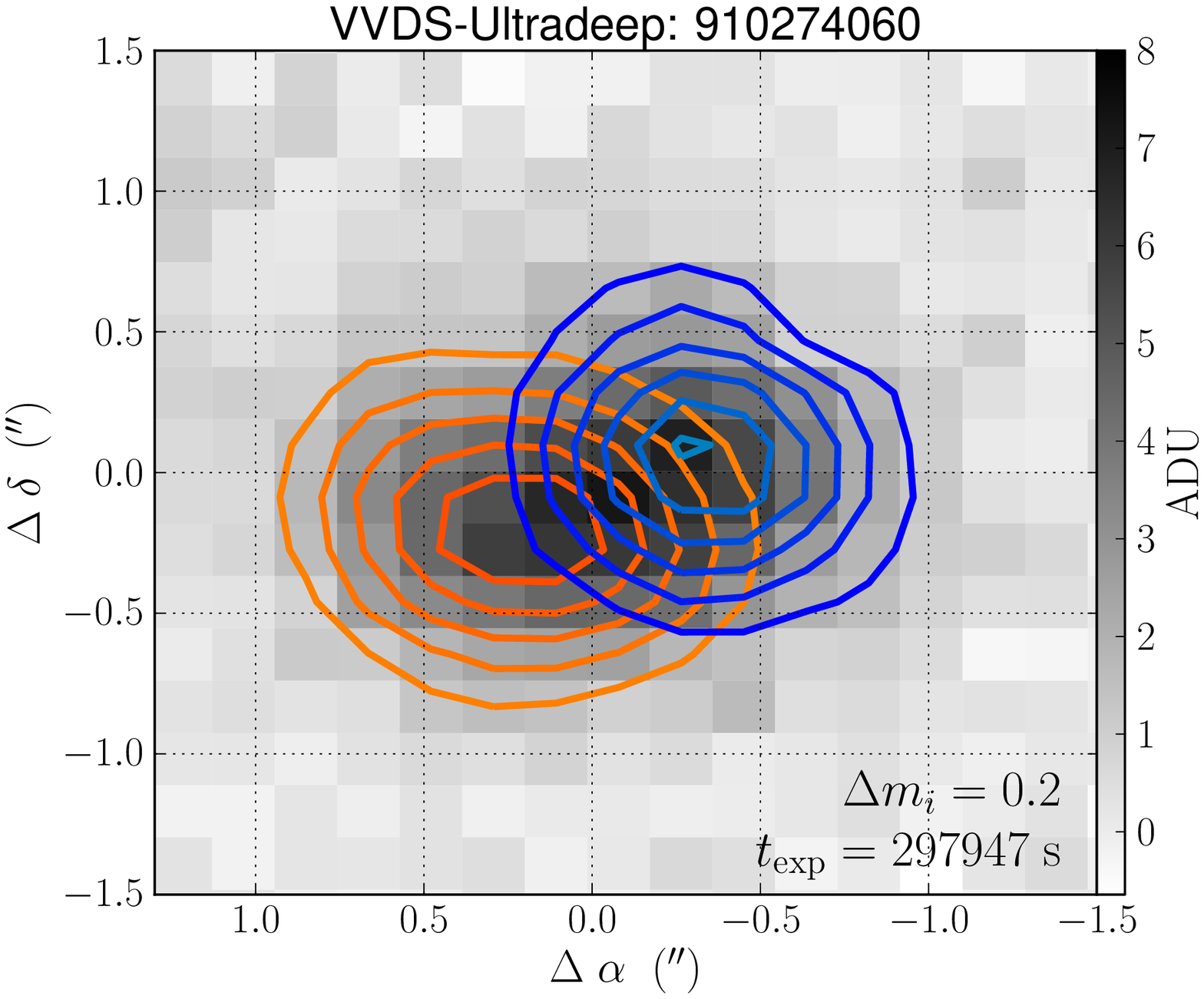}}
	\resizebox{0.32\hsize}{!}{\includegraphics{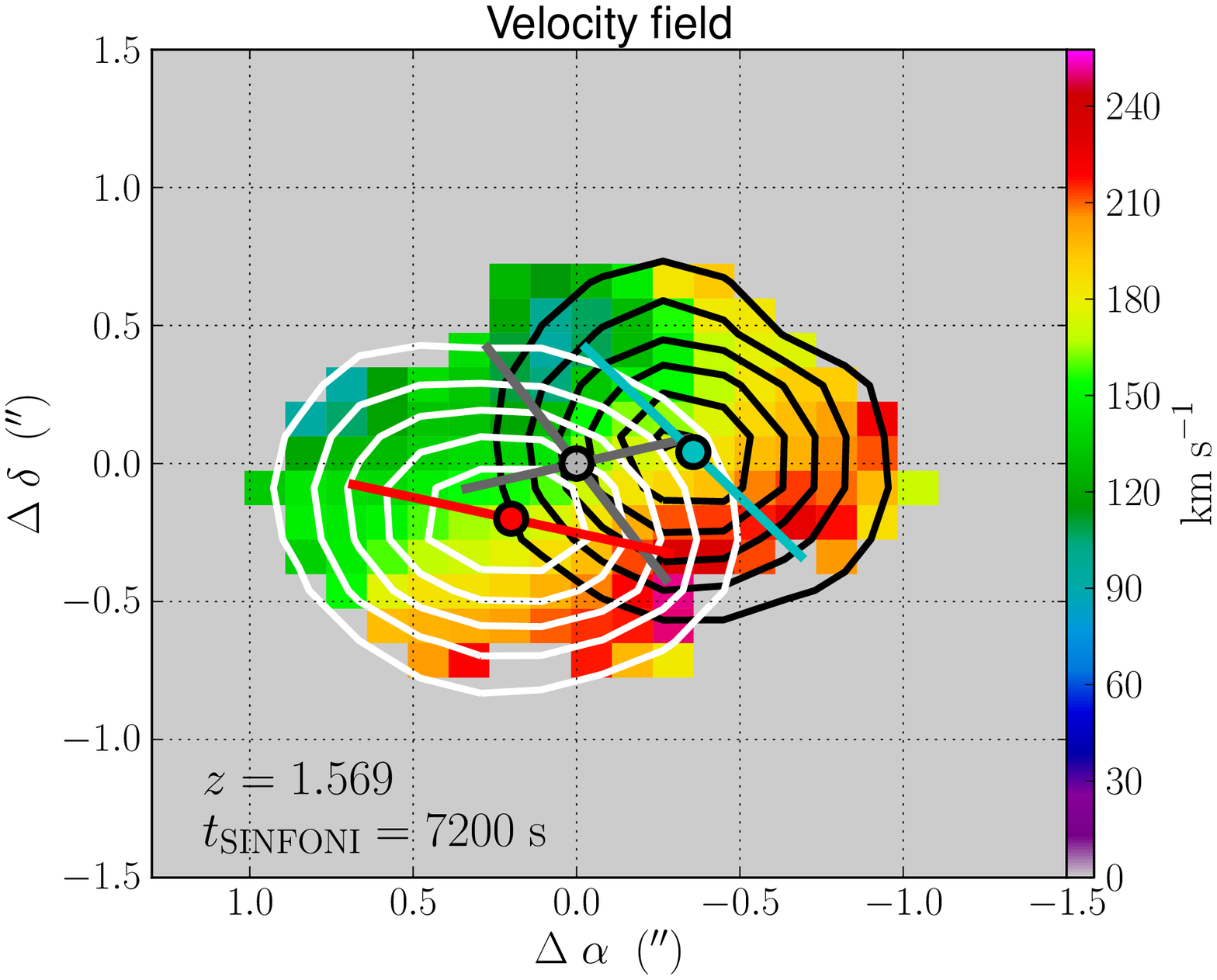}}
	\resizebox{0.32\hsize}{!}{\includegraphics{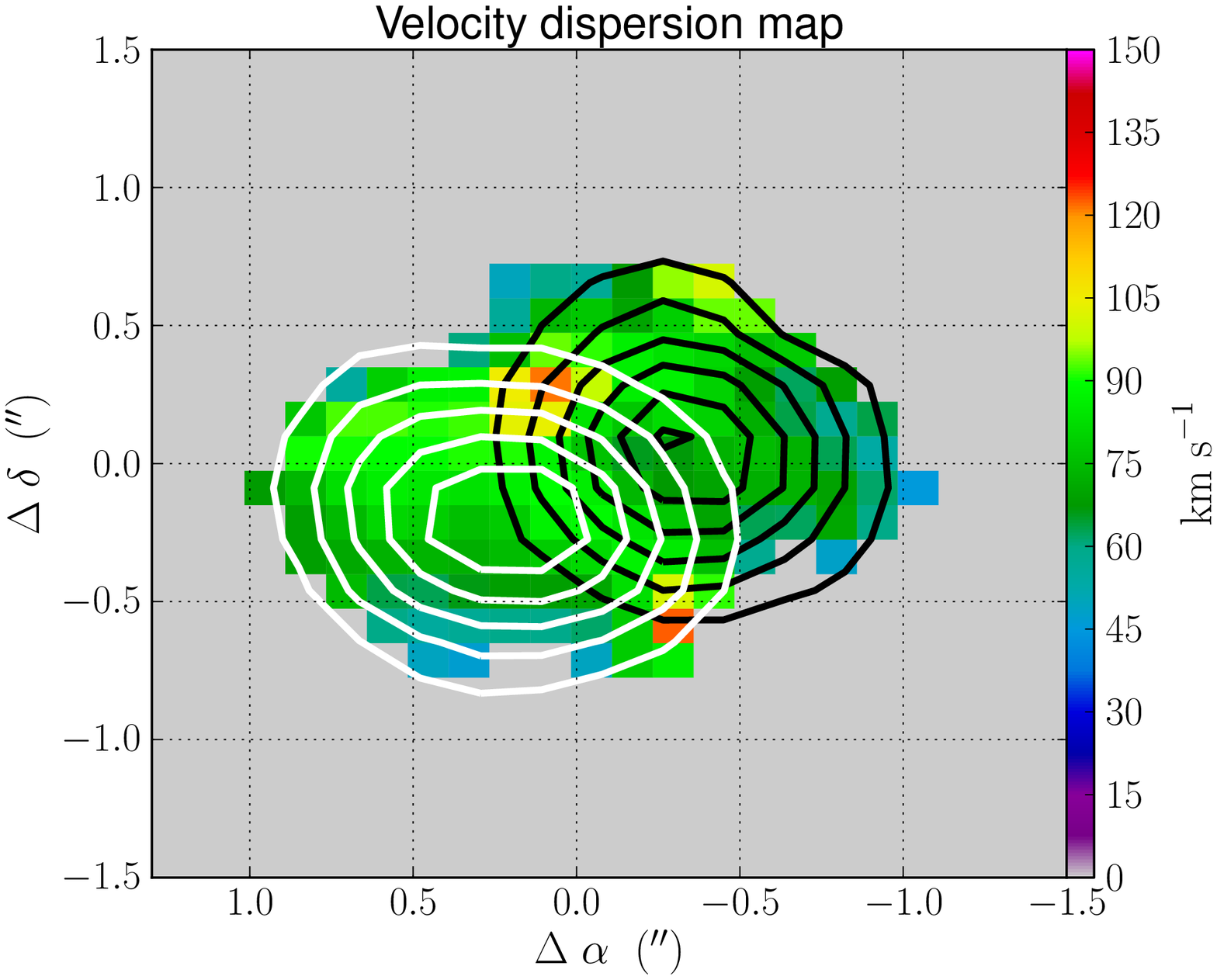}}
	\caption{The same as Fig.~\ref{src_020294045_i}, but for the MASSIV source 910274060 (major merger). 
		The outer contour marks the 1.20 ADU (3$\sigma_{\rm sky}$) isophote, while brighter 
		isophotes increase in 1.00 ADU (2.5$\sigma_{\rm sky}$) steps. 
		The grey dot in the central figure marks the centre of the system given 
		by the kinematical model, while the red/blue dot marks the photometric centre of 
		the principal/companion galaxy given by the two-component model of the source. 
		The long bars mark the position angle of the previous models. 
		The short bar marks the photometric position angle given by the one-component model of the source. 
		[{\it A colour version of this plot is available at the electronic edition}].}
	\label{src_910274060}
	\end{figure*}


\begin{itemize}
	
	
	\item 020167131 (Fig.~\ref{src_020167131}). {\bf Major merger}. The velocity map shows two
		different components with a separation of $r_{\rm p} = 15.2h^{-1}$ kpc
		and a relative velocity of $\Delta v \sim 130$ km s$^{-1}$. The line targeted
		in this case is $[\ion{O}{iii}]\lambda5007$, which explains the low signal.
		The companion galaxy is toward the south-east, while the luminosity differences 
		are $\Delta m_i = 0.2$ and $\Delta m_{K_{\rm s}} = 0.1$.
		We classify the system as a major merger.


	\item 020218856 (Fig.~\ref{src_020218856}). {\it No major merger}. A faint companion is detected
		in the velocity map about 1.6 arcsec from the principal
		galaxy. There is no continuum detection of this
		companion at the depth of the $i-$band image. 
		We followed
		the same steps than for the source 140096645 to estimate the
		detection probability of companion galaxies, finding that we
		are 100\% complete up to $m _{i,2} = 26$. We note that the curves
		for extended, normal and compact galaxies are similar, reflecting that 
		the size of the PSF is larger than the assumed
		size of the fake sources.
		Because a major companion should
		be brighter than $m^{\rm MM}_{i,2} = 25.32$ 
		(vertical line in the left panel of Fig.~\ref{src_020218856}), 
		and the non detection in the continuum means that
		the companion should have $\Delta m_i \gtrsim 2$, we do not classify the system
		as a major merger.


	\item 020240675 (Fig.~\ref{src_020240675}). {\it No major merger}. 
		A faint companion is detected
		in the velocity map about 2.4 arcsec from the principal
		galaxy. There is no continuum detection of this
		companion at the depth of the $i-$band image.
		We followed
		the same steps than for the source 020218856 to estimate
		the detection curves of companion galaxies. We find similar
		completeness curves.
 		Because a major companion should
		be brighter than $m^{\rm MM}_{i,2} = 25.15$ 
		(vertical line in the left panel of Fig.~\ref{src_020240675}), 
		and the non detection in the continuum means that
		the companion should have $\Delta m_i \gtrsim 2.5$, we do not classify the system
		as a major merger.


	\item 020283083 (Fig.~\ref{src_020283083}). {\bf Major merger}. The velocity map suggests two
		projected components separated by 3.8$h^{-1}$ kpc and $\sim 5$ km~s$^{-1}$, 
		with an extended region in the north-west. The GALFIT model with two components 
		reproduces the shape of the velocity map and suggests that the extended 
		region is due to the companion galaxy. 
		The luminosity difference from the models is $\Delta m_{i} = 0.7$. 
		Thus, we classify this system as a major merger.


	\item 020283830 (Fig.~\ref{src_020283830}). {\it No major merger}. The velocity map shows two
		different components with a separation of $r_{\rm p} = 8.5h^{-1}$ kpc and a relative velocity 
		of $\Delta v \sim 500$ km s$^{-1}$. The companion only presents six detected pixels in 
		the H$\alpha$ map, already suggesting a minor companion. 
		We used GALFIT to model
		the system, finding $\Delta m_{i} = 1.9$. However, the model of the
		companion provided by GALFIT is a point-like source, thus
		overestimating the luminosity of the extended companion.
		The measurement with SExtractor in the image with the principal galaxy subtracted suggests 
		$\Delta m_{i} = 2.1$. Since the companion is fainter than the major merger limit, 
		we do not classify the system as a major merger.


	\item 020465775 (Fig.~\ref{src_020465775}). {\bf Major merger}. The velocity map suggests two
		projected components. The GALFIT model with two components finds 
		that the companion galaxy is toward the north-west of the principal galaxy, 
		at $r_{\rm p} = 3.6h^{-1}$ kpc and $\Delta v \sim 40$ km s$^{-1}$. 
		The H$\alpha$ emission of this system is located in the central part of the galaxies, 
		since the $i-$band models match the velocity map at $9\sigma_{\rm sky}$ level, 
		i.e., we do not detect emission from the outer parts of the galaxies. 
		The position of the companion explains the
		abnormal velocity pattern and the high velocity dispersion
		peak in the maps. The luminosity difference of the system is
		$\Delta m_i = 0.7$, so we classify it as a major merger.


	\item 220376206 (Fig.~\ref{src_220376206}). {\it No major merger}. The velocity map shows two
		different components with a separation of $r_{\rm p} = 13.4h^{-1}$ kpc
		and a relative velocity of $\Delta v \sim 400$ km s$^{-1}$. The companion,
		located toward the north, presents ten detected pixels and its
		velocity is inconsistent with that expected 
		from the velocity field of the principal galaxy. 
		We find $\Delta m_i = 2.4$, so the system is not a major merger.


	\item 220544103 (Fig.~\ref{src_220544103}). {\bf Major merger}.  The velocity map suggests two
		projected components. The southern component presents a
		large velocity gradient and defines the kinematical centre
		of the system, while the northern component is more extended 
		and has a nearly flat velocity field. The GALFIT
		model with two components recovers the configuration in
		the velocity map and suggests that the southern component
		is edge-on, while the northern component is nearly face-on,
		thus explaining the observed high and null velocity gradients. 
		The separation between the components is
		$r_{\rm p} = 5h^{-1}$ kpc, while their relative velocity is $\Delta v \sim 75$ km s$^{-1}$. 
		From the GALFIT models we estimate $\Delta m_i = -1.1$, and
		we classify the system as a major merger.


	\item 910154631 (Fig.~\ref{src_910154631}). {\bf Major merger}. The velocity map suggests two
		projected components, with the companion toward the north-west. 
		The $i-$band image shows another two well separated
		sources close to the MASSIV target. These sources are not
		detected in H$\alpha$. To avoid contamination from these sources
		in the $i-$band photometry, we performed a four component fitting,
		with two components for the MASSIV target and one component for each
		nearby source.
		The GALFIT model finds the second component of the MASSIV target at
		the expected position, but we only detect the southern half
		of the $i-$band source in H$\alpha$. 
		We explored the reduced data cube of the source, and we find 
		(i) there are two clear velocity gradients in the cube, reinforcing the presence
		of two different components and (ii) there is an OH sky-line
		in the channels in which we expect the northern part of the
		companion, explaining the non detection in the maps. 
		In addition, the higher velocity dispersion of the MASSIV target occurs in the 
		expected overlapping region between both components. 
		We conclude that this is a close pair system
		with $r_{\rm p} = 4.2h^{-1}$ kpc, $\Delta v \sim 130$ km s$^{-1}$ and $\Delta m_i = 0.8$.
		Thus, we classify the system as a major merger.


	\item 910296626 (Fig.~\ref{src_910296626}). {\bf Major merger}. The velocity map shows two
		different components with a separation of $r_{\rm p} = 12.1h^{-1}$ kpc
		and a relative velocity of $\Delta v \sim 165$ km s$^{-1}$. The companion is
		located toward the north-east. We find $\Delta m_i = -0.1$
		and $\Delta m_{K_{\rm s}} = -0.2$, so the system is a major merger.


	\item 910337228 (Fig.~\ref{src_910337228}). {\bf Major merger}. The velocity map shows two
		different components with a separation of $r_{\rm p} = 9.5h^{-1}$ kpc
		and a relative velocity of $\Delta v \sim 220$ km s$^{-1}$. The companion
		is toward the west and has $\Delta m_i = 1.4$. Thus, we
		classify the system as a major merger.
\end{itemize}

In summary for this redshift range, we classify 7 of the 11
close pair candidates as major mergers. This translates to a gas-rich major merger fraction of
$f_{\rm MM} = 0.201^{+0.080}_{-0.051}$ at $\overline{z_{\rm r}}_{,2} = 1.32$.


\subsection{Close pair candidates at $1.5 \leq z < 1.8$}
The weighted mean redshift of the third redshift bin is $\overline{z_{\rm r}}_{,3}=1.54$. 
This is a redshift range where there is no measurement of the merger fraction
from spectroscopic close pairs yet. 
We identify 3 close pair candidates over 12 galaxies:

\begin{itemize}


	\item 020116027 (Fig.~\ref{src_020116027}). {\bf Major merger}. The velocity map shows two
		different components with a separation of $r_{\rm p} = 26.8h^{-1}$ kpc
		and a relative velocity of $\Delta v \sim 100$ km s$^{-1}$. 
		The companion is toward the north-west. The luminosity difference is
		$\Delta m_i = 0.7$, suggesting a major merger. 
		The difference in the $K_{\rm s}$ band is $\Delta m_{K_{\rm s}} = 0.5$, 
		confirming the previous major merger classification.


	\item 910186191 (Fig.\ref{src_910186191}). {\bf Major merger}. The velocity map shows two
		different components with a separation of $r_{\rm p} = 12.7h^{-1}$ kpc and a 
		relative velocity of $\Delta v \sim 450$ km s$^{-1}$. The MASSIV
		target is only detected in 6 pixels because there is an OH sky-line in 
		the position of H$\alpha$ at its redshift. 
		The companion, located toward the west, is well detected. 
		We find $\Delta m_i = -0.2$, this is, the companion is slightly brighter 
		than the principal galaxy. 
		However, the MASSIV target is barely detected in the $K_{\rm s}$ band, 
		with $\Delta m_{K_{\rm s}} = -2.4$, suggesting a low mass system. 
		This is the only system in which the classification in the two bands is different. 
		Fortunately, both sources are VVDS targets, and we have an estimation of their stellar 
		masses from SED fitting. The difference in stellar mass is $\mu \sim 1/3$, 
		so we classify the system as a major merger.


	\item 910274060 (Fig.~\ref{src_910274060}). {\bf Major merger}. The velocity map is consistent with 
		one single component. However, the position angle
		(PA) from the $i-$band photometry, PA = $105^{\circ}$ (North has PA = $0^{\circ}$ 
		and East has PA = $90^{\circ}$), is nearly perpendicular to that
		from the kinematical modelling, PA = $33^{\circ}$. 
		This suggests a complex system, so we
		performed the GALFIT modelling with two sources. 
		We recover well two sources, one in the north and the other in the
		south. The photometric PAs of these two sources, provided
		by the GALFIT fitting, are now in better agreement with the
		kinematical one (PA$_1 = 75^{\circ}$, PA$_2 = 41^{\circ}$), 
		supporting that this is a close pair system. The separation between
		the components is 3.4$h^{-1}$ kpc, with $\Delta v \sim 10$ km s$^{-1}$, and
		the luminosity difference is $\Delta m_{i} = 0.2$. 
		Thus, we classify the system as a major merger.

\end{itemize}

In summary, we identify the 3 candidates as major mergers.
This translates to a gas-rich major merger fraction of $f_{\rm MM} = 0.323^{+0.201}_{-0.107}$.
Note that in this range our merger candidates have $r^{\rm max}_{\rm p} = 30h^{-1}$
kpc to improve the statistics. Applying Eq.~(\ref{ffMM20}) we estimate
$f_{\rm MM} = 0.220^{+0.137}_{-0.073}$ for $r^{\rm max}_{\rm p} = 20h^{-1}$ kpc 
at $\overline{z_{\rm r}}_{,3} = 1.54$.

\begin{figure}[t!]
\resizebox{\hsize}{!}{\includegraphics{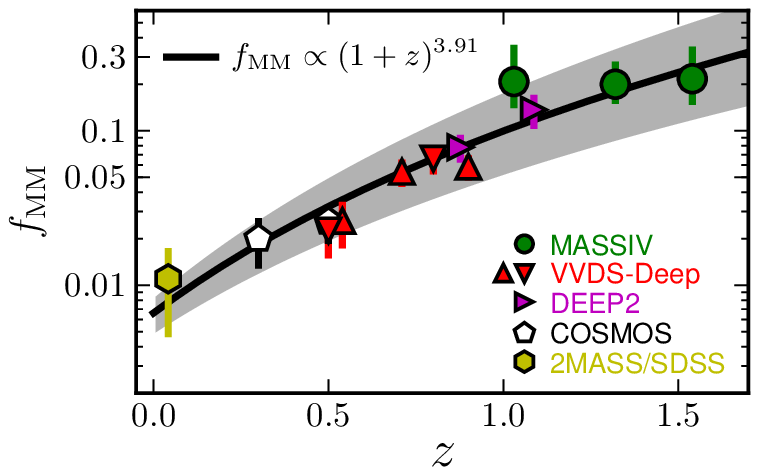}}
\caption{Gas-rich major merger fraction of $\overline{M}_{\star} \sim 10^{10-10.5}\ M_{\odot}$ galaxies 
as a function of redshift. Circles are from this MASSIV data set, 
triangles are from \citet{deravel09} and 
inverted triangles are from \citet{clsj11mmvvds}, both in VVDS-Deep, 
right-pointing triangles are from \citet{lin08} in DEEP2 redshift survey,
pentagons are from \citet{xu12} in the COSMOS field, 
and the hexagon is from \citet{xu12} in 2MASS/SDSS. 
The solid line is the least-squares fit of a power-law function, 
$f_{\rm MM} = 0.0066\times(1+z)^{3.91}$, to the data. 
The grey area marks the $3\sigma$ confidence interval in the fit.
[{\it A colour version of this plot is available at the electronic edition}].}
\label{fffig}
\end{figure}

Our results alone, summarised in Table~\ref{fftab}, suggest a constant
major merger fraction of $f_{\rm MM} \sim 0.21$ at $0.9 < z < 1.8$ for 
$r^{\rm max}_{\rm p} = 20h^{-1}$ kpc close pairs (Fig.~\ref{fffig}). 
{\it This merger fraction at $z > 1$ is higher by an order of magnitude 
than in the local universe, where $f_{\rm MM} \sim 0.01 - 0.03$}
\citep{patton00,depropris07,patton08,domingue09,darg10I,xu12}. 
This is the first main result of the present paper. 
We compare our major merger fractions with others in the literature in Sect.~\ref{ffevol}.

\begin{table*}
\caption{Gas-rich major merger fraction and rate of star-forming galaxies at $0.9 < z < 1.8$ in the MASSIV sample.}
\label{fftab}
\begin{center}
\begin{tabular}{lcccccccc}
\hline\hline\noalign{\smallskip}
$z_{\rm r}$ & $N$  & $N_{\rm p}$ & $\overline{z_{\rm r}}$ & $\log\,(\overline{M}_{\star}/M_{\odot})$ & $r_{\rm p}^{\rm max}$ & $T_{\rm MM}$ & $f_{\rm MM}$ & $R_{\rm MM}$\\
\noalign{\smallskip}
    &    &            &                     &                           &    ($h^{-1}$ kpc)             & (Gyr)         &               & (Gyr$^{-1}$)\\
\hline
\noalign{\smallskip}
$0.94 \leq z \leq 1.06$    & 18  & 3 & $1.03$ & 10.17 & 20 & 1.80 & $0.208^{+0.152}_{-0.068}$ & $0.116^{+0.084}_{-0.038}$\\\noalign{\smallskip}
$1.2  \leq z < 1.5$        & 30  & 7 & $1.32$ & 10.57 & 20 & 1.37 & $0.201^{+0.080}_{-0.051}$ & $0.147^{+0.058}_{-0.037}$\\\noalign{\smallskip}
$1.5  \leq z < 1.8$        & 12  & 3 & $1.54$ & 10.09 & 30 & 2.54 & $0.323^{+0.201}_{-0.107}$ & $0.127^{+0.079}_{-0.042}$\\\noalign{\smallskip}
\hline
\end{tabular}
\end{center}
\end{table*}

\section{The gas-rich major merger rate in MASSIV}\label{mrmassiv}
In this section we estimate the gas-rich major merger rate ($R_{\rm MM}$), defined
as the number of mergers per galaxy and Gyr, of star-forming
galaxies at $0.9 < z < 1.8$. We remind here the steps to transform 
a merger fraction to a merger rate. Following \citet{deravel09}, 
we define the major merger rate as
\begin{equation}
R_{\rm MM} = C_{\rm m}\,f_{\rm MM}\,T_{\rm MM}^{-1},\label{mreq}
\end{equation}
where the factor $C_{\rm m}$ is the fraction of the observed close pairs
that finally merge in a typical time scale $T_{\rm MM}$. The typical
merger time scale can be estimated by cosmological and $N-$body simulations. 
In our case, we compute the major merger
time scale from the cosmological simulations of \citet{kit08}, based on the Millennium simulation \citep{springel05}. 
This major merger time scale refers 
to major mergers ($\mu \geq 1/4$ in stellar mass), and depends mainly on 
$r^{\rm max}_{\rm p}$ and on the stellar mass of the principal galaxy, 
with a weak dependence on redshift in our range of interest 
\citep[see][for details]{deravel09}. We measured the median-weighted 
stellar mass from MASSIV sources in each of the three redshift
bins under study, and estimated the merger time scale for these
stellar masses. These time scales already include the factor $C_{\rm m}$
\citep[see][]{patton08,bundy09,lin10,clsj11mmvvds}, so we take $C_{\rm m} = 1$ 
in the following. In addition, \citet{clsj11mmvvds} show that the time
scales from \citet{kit08} are equivalent to that
from the $N-$body/hydrodynamical simulations by \citet{lotz08t}. 
However, we stress that these merger time scales have an
additional factor of two uncertainty in their normalization \citep[e.g.,][]{hopkins10mer,lotz11}. 
We summarise the stellar masses, the merger time scales and the gas-rich major merger rates in Table~\ref{fftab}. 
As for the merger faction, MASSIV data suggests a nearly constant major merger 
rate at $0.9 < z < 1.8$, $R_{\rm MM} \sim 0.12$ Gyr$^{-1}$ (Fig.~\ref{mrfig}). 
We study in detail the evolution of the major merger rate at $z \lesssim 1.5$ in Sect.~\ref{mrevol}.


\section{The redshift evolution of the gas-rich major merger fraction and rate up to $z \sim 1.5$}\label{discussion}
In this section we use the MASSIV results at $z > 1$ to expand the
study of the gas-rich major merger fraction (Sect.~\ref{ffevol}) and rate (Sect.~\ref{mrevol})
from spectroscopic close pairs to the redshift desert. Then, we
explore the importance of gas-rich major mergers in the assembly of the red 
sequence since $z \sim 1.5$ in Sects.~\ref{nmer} and \ref{mergerole}.

\subsection{The redshift evolution of the gas-rich major merger fraction}\label{ffevol}
In this section we compare the merger fraction from MASSIV
with those from previous works. Because the merger fraction evolution 
depends on mass \citep[e.g.,][]{deravel09,deravel11}, 
luminosity \citep[e.g.,][]{deravel09,clsj10pargoods}
and colour \citep[e.g.,][]{lin08,chou10}, we focus on samples with
$\overline{M}_{\star} \sim 10^{10-10.5}\ M_{\odot}$ (\citealt{salpeter55} IMF) to minimise systematics. 
In addition, this mass regime is greatly dominated by gas-rich (wet) mergers, 
as those that we observe in MASSIV, at least
at $z \gtrsim 0.2$ \citep{lin08,deravel09,chou10}.

We define the major ($\mu \geq 1/4$) merger fraction normalised to $r_{\rm p}^{\rm max} = 20h^{-1}$ kpc as
\begin{equation}
f_{\rm MM}(20,1/4) = C_{\rm p}\,F(\mu)\,\bigg( \frac{20h^{-1}\ {\rm kpc}}{r_{\rm p}^{\rm max}} \bigg)^{0.95} f_{\rm m}(r_{\rm p}^{\rm max},\mu),\label{ffMM20}
\end{equation}
where the factor $C_{\rm p} = r_{\rm p}^{\rm max}/(r_{\rm p}^{\rm max} - r_{\rm p}^{\rm min})$ accounts for 
the missing close companions at small radii in those studies with $r_{\rm p}^{\rm min} > 0$ \citep[e.g.,][]{bell06} and the factor $F(\mu)$ translates the merger fraction for a given $\mu$ to
the major merger fraction. The merger fraction depends on $\mu$ as $f_{\rm m}(\geq \mu) \propto \mu^{s}$ \citep[e.g.,][]{clsj11mmvvds}, that implies $F(\mu) = (4\mu)^{-s}$. We take $s = -0.9\pm0.4$, a value derived from the observational estimations of \citet{clsj11mmvvds,clsj12sizecos} and \citet{xu12}. The search radius dependence of the major merger fraction, $f_{\rm MM} \propto r_{\rm p}^{0.95}$, is the observational one found by \citet{clsj11mmvvds} in the VVDS.
With Eq.~(\ref{ffMM20}) we avoid systematic differences due to the close pair 
definition when comparing different works.

\citet{deravel09} study the major merger fraction of 
$M_{\star} \geq 10^{9.75}\ M_{\odot}$ 
($\overline{M}_{\star} \sim 10^{10.25}\ M_{\odot}$) galaxies in VVDS-Deep
by spectroscopic close pairs, while \citet{clsj11mmvvds}
provide the major merger fraction of blue (star-forming) galaxies
with $\overline{M}_{\star} \sim 10^{10.55}\ M_{\odot}$ in the same sample. 
In both studies $r_{\rm p}^{\rm max} = 100h^{-1}$ kpc. 
\citet{lin08} report the number of companions of $-21 \leq M_{B} + 1.3z \leq -19$
galaxies ($\overline{M}_{\star} \sim 10^{10.25}\ M_{\odot}$) with $10h^{-1}$ kpc
$\leq r_{\rm p} \leq 30h^{-1}$ kpc in three DEEP2 redshift survey \citep{deep2} fields.
Their principal and companion sample are the same, so they miss major companions near to
the selection boundary. Thus, we apply an extra factor 1.74 to Eq.~(\ref{ffMM20})
to account for these missing companions \citep[see][for details]{lin04}.
\citet{xu12} measure the fraction of galaxies
in close pairs with $\mu \geq 1/2.5$ in the COSMOS\footnote{http://cosmos.astro.caltech.edu}
(Cosmological Evolution Survey, \citealt{cosmos}) and SDSS\footnote{http://www.sdss.org/} 
(Sloan Digital Sky Survey, \citealt{sdssdr7}) surveys for 
$\overline{M}_{\star} \sim 10^{10.2}\ M_{\odot}$ galaxies. 
We applied a factor 0.5 to pass from 
their number of galaxies in close pairs to the number of close pair systems in the sample 
(C.~K. Xu, private communication), and a factor $F(1/2.5) = 1.5\pm0.3$ to obtain the
major merger fraction. All these published (gas-rich) major merger fractions 
are shown as a function of redshift in Fig.~(\ref{fffig}), 
together with the values derived from MASSIV.

We parametrise the redshift evolution of the (gas-rich) major merger fraction with a power-law, 
\begin{equation}
f_{\rm MM} = f_{\rm MM, 0}\,(1+z)^{m}.
\end{equation}
The least-squares fit to all the data in Fig.~\ref{fffig}
yields $f_{\rm MM,0} = (6.6 \pm 0.6) \times 10^{-3}$ and $m = 3.91 \pm 0.16$. We find
good agreement between all works, with the MASSIV point at
$z \sim 1$ being higher than expected from the fit, but consistent within errors
with the measurement of \citet{lin08} at that redshift. In the next section
we show that this difference disappears when the stellar mass of
the samples is taken into account, emphasizing the importance
of comparing results from similar parent samples.

\begin{figure}[t!]
\resizebox{\hsize}{!}{\includegraphics{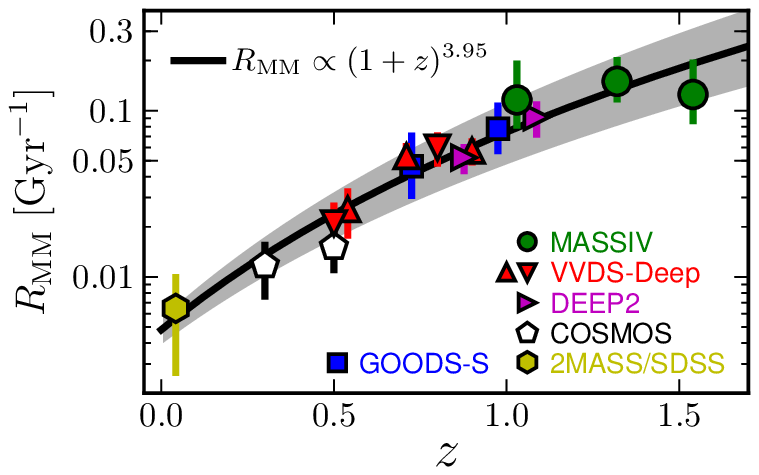}}
\caption{Gas-rich major merger rate of $\overline{M}_{\star} \sim 10^{10 -10.5}\ M_{\odot}$
galaxies as a function of redshift. Circles are from this MASSIV data set, 
triangles are from \citet{deravel09} 
and inverted triangles are from \citet{clsj11mmvvds}, both in VVDS-Deep,
right-pointing triangles are from \citet{lin08} in DEEP2 redshift survey,
squares are from \citet{clsj09ffgoods} in GOODS-S from morphological criteria, 
pentagons are from \citet{xu12} in the COSMOS field, 
and the hexagon is from \citet{xu12} in 2MASS/SDSS. 
The solid line is the least-squares fit of a power-law function, 
$R_{\rm MM} = 0.0048\times(1+z)^{3.95}$, to the data.
The grey area marks the $3\sigma$ confidence interval in the fit.
[{\it A colour version of this plot is available at the electronic edition}].}
\label{mrfig}
\end{figure}

\subsection{The redshift evolution of the gas-rich major merger rate}\label{mrevol}
We use Eq.~(\ref{mreq}) to translate the original (i.e., without any 
normalization in $r_{\rm p}^{\rm max}$) major merger fractions reported in previous
section into merger rates. We show them in Fig.~\ref{mrfig}. The good
agreement between different works is remarkable, reinforcing
the idea that the merger time scales used account properly for
the dependence of the merger fraction both on $r_{\rm p}^{\rm max}$ and on stellar mass.

We also show the major merger rate from morphological criteria derived by \citet{clsj09ffgoods}. 
They measure the gas-rich merger fraction of $\overline{M}_{\star} \sim 10^{10.5}\ M_{\odot}$ galaxies 
from asymmetries ($A$) in the GOODS\footnote{http://www.stsci.edu/science/goods/} \citep{goods} 
South field and take into account
the effect of observational errors in $z$ and $A$, that lead to overestimations 
in the major merger fraction by a factor of two-three \citep{clsj09ffgs,clsj09ffgoods}, 
using maximum likelihood techniques developed in \citet{clsj08ml}. Other studies find 
good agreement between the asymmetry-based major
merger rates from \citet{clsj09ffgoods} and those from
close pair statistics \citep{clsj10pargoods,lotz11,deravel11}, 
confirming the robustness of their methodology.

We also parametrise the redshift evolution of the (gas-rich) major merger rate with a power-law, 
\begin{equation}
R_{\rm MM} = R_{\rm MM, 0}\,(1+z)^{n}.\label{mrevoleq}
\end{equation}
The least-squares fit to all the data in Fig.~\ref{mrfig}
yields $R_{\rm MM, 0} = (4.8 \pm 0.3)\times10^{-3}$ Gyr$^{-1}$ and $n = 3.95 \pm 0.12$. 
The agreement between different works points out the importance of
comparing results from similar parent samples to avoid systematics 
\citep[see also][for an extensive discussion on this topic]{lotz11}. 
We also point out that the merger rate and the merger
fraction of star-forming galaxies show a similar evolution with redshift.

The second main result in this paper is that {\it the major merger
rate is well described by a power-law function up to $z \sim 1.5$}.
However, we note that our MASSIV data seem to indicate a flattening 
of the merger rate's evolution beyond $z \sim 1$. Previous studies 
from morphological criteria \citep[e.g.,][]{conselice03,conselice08} 
and from photometric pairs \citep{ryan08} suggest that the power-law parametrization 
is not longer valid at $z \gtrsim 1.5$, where a lower merger fraction than expected from the
low$-z$ evolution is measured, possibly indicating a maximum in
the major merger rate at $z \sim 2$ \citep[e.g.,][]{conselice06ff,ryan08,clsj09ffgoods}. 
Our new measurements agree with this picture and measurements from spectroscopic
close pairs beyond $z \sim 1.5$ are needed to test the early evolution of the merger fraction. 

The power-law index $n = 3.95 \pm 0.12$ is higher than several
previous measurements in the literature \citep[e.g.,][]{bridge10,lotz11}, 
as well as our major merger fraction evolution, $m = 3.91 \pm 0.16$. 
However, our results refer to $\overline{M}_{\star} \sim 10^{10-10.5}\ M_{\odot}$ star-forming galaxies, 
and it is known that the merger
fraction and rate evolve faster for blue, star-forming galaxies than
for the red and global populations \citep[e.g.,][]{lin08,deravel09,chou10,clsj11mmvvds}.

\subsection{Number of gas-rich mergers since $z = 1.5$}\label{nmer}
We can obtain the average number of gas-rich major mergers per star-forming galaxy between $z_2$ and $z_1 < z_2$ as
\begin{equation}
N_{\rm MM}(z_1,z_2) = \int_{z_1}^{z_2} \frac{R_{\rm MM}\,{\rm d}z}{(1+z)H_0E(z)},\label{NMM}
\end{equation}
where $E(z) = \sqrt{\Omega_{\Lambda} + \Omega_{\rm m}(1+z)^3}$ in a flat universe. 
Using the merger rate parametrisation in Eq.~(\ref{mrevoleq}), 
we obtain $N_{\rm MM}(0,1.5) = 0.35\pm0.04$. Interestingly, half of this merging 
activity happens at $z > 1$, with $N_{\rm MM}(1,1.5) = 0.18\pm0.02$ 
and $N_{\rm MM}(0,1) = 0.17\pm0.02$. Because the cosmic time lapse in these redshift 
intervals is 1.55 Gyr and 7.7 Gyr, respectively, the average merger activity 
was higher at $1 < z < 1.5$ than at $1 < z$ by a factor of five. 
In the next section we further explore the consequences of this very different major merger 
activity above and below $z \sim 1$ for the assembly of the red sequence.

\subsection{Testing the major merger origin of massive early-type galaxies}\label{mergerole}
The number density of massive ($M_{\star} \gtrsim 10^{11}\ M_{\odot}$) early-type
galaxies (E/S0, ETGs in the following) has increased with cosmic time since 
$z \sim 3$ \citep[e.g.,][]{pozzetti10,buitrago12}, with ETGs being 
the dominant population among massive galaxies only since
$z \sim 1$ \citep{vergani08,buitrago12,vanderwel11,vandokkum11}. 
Gas-rich major mergers have been proposed as an efficient mechanism to transform star-forming
late-type galaxies into red ETGs 
\citep[e.g.,][]{naab06ss,rothberg06a,rothberg06b,hopkins08ss,rothberg10,bournaud11},
so the comparison between the observed number density evolution of ETGs ($\rho_{\rm ETG}$) 
and the major merger history of star-forming galaxies imposes important 
constraints on the role of mergers in galaxy evolution.

In their work, \citet{robaina10} and \citet{man12}
integrate the observed major merger rate over cosmic time and
predict the evolution of the number density of massive galaxies
assuming that all the merger remnants are new massive galaxies. 
They suggest that major mergers are common enough to explain 
the number density evolution of massive galaxies (ETGs + spirals) 
at $z < 1$ and $z < 3$, respectively.
\citet{eliche10I} model the evolution of the luminosity function backwards
in time for bright galaxies, selected according to their colours
(red/blue/total) and their morphologies. They find that the observed 
luminosity functions' evolution can be naturally explained
by the observed gas-rich and dry major merger rates, and that
50-60\% of massive ETGs in the local universe were formed by
major mergers at $0.8 < z < 1$, with a small number evolution
since $z = 0.8$.

\begin{figure}[t!]
\resizebox{\hsize}{!}{\includegraphics{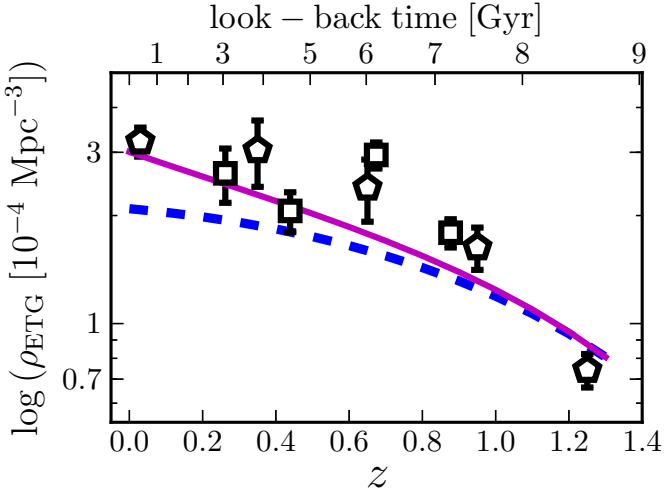}}
\caption{ Number density evolution of massive ($M_{\star} \geq 10^{11.25}\ M_{\odot}$) 
ETGs (E/S0) as a function of redshift from \citet[][pentagons]{buitrago12} and 
\citet[][squares]{pozzetti10}.
The dashed line is the expected number density evolution due to gas-rich 
(wet) major mergers from our model. The solid line is the expected number density
evolution due to wet major and dry mergers (both major and minor) from our model.
Mergers are common enough to drive the number density evolution of
massive ETGs since $z \sim 1.3$. 
[{\it A colour version of this plot is available at the electronic edition}].}
\label{rhofig}
\end{figure}

In this section we implement a model to explore the role
of mergers in the number density evolution of massive ETGs
since $z \sim 1.3$. As reference values, we use the number densities
of massive ($M_{\star} \geq 10^{11.25}\ M_{\odot}$) ETGs provided by
\citet{buitrago12}. They perform a consistent morphological
study by visual inspection between $z = 0$ and $z = 3$, combining
the SDSS, POWIR (Palomar Observatory Wide-Field Infrared, \citealt{powir}) 
and GNS\footnote{http://www.nottingham.ac.uk/astronomy/gns/} 
(GOODS NICMOS Survey, \citealt{gns}) surveys. We also use the number 
densities of massive spheroidal galaxies with $M_{\star} \geq 10^{11.25}\ M_{\odot}$, 
selected by automatic indices in the zCOSMOS\footnote{http://www.astro.phys.ethz.ch/zCOSMOS/} 
\citep{lilly07} survey, provided by \citet{pozzetti10}. 
We show these number densities in Fig.~\ref{rhofig}.

In our model we assume that, after the final coalescence of
the merging galaxies, a cosmic time $\Delta t = 0.5$ Gyr is necessary
for the merger remnant to be classified as an ETG
\citep[see][for a detailed summary of this
topic]{eliche10I}. This implies that the new ETGs which appeared between
$z_{\rm max}$ and $z_{\rm min}$ came from the merger activity in the range 
$z \in [z_1, z_2)$, where $z_1 = z_{\rm min} + \Delta z$, 
$z_2 = z_{\rm max} + \Delta z$, and $\Delta z$ is the
redshift interval that spans $\Delta t$ in each case. Therefore, we take
$z_{\rm max} = 1.3$ as the upper redshift in our model because that implies
$z_2 \sim 1.5$, the redshift limit of the present merger rate study 
(see Sect.~\ref{mrevol}).

The number density of new ETGs with stellar mass 
$M_{\star} \geq M_{\star, {\rm lim}}$ from gas-rich major merger events appeared in 
the range $z_{\rm min} \leq z < z_{\rm max}$ is
\begin{equation}
\rho_{\rm wet}(z_{\rm min},z_{\rm max},M_{\star,\rm lim}) = \!\!\int_{z_1}^{z_2}\!\!\!\int_{0}^{\infty}\!\!\!\!\Phi\,R_{\rm MM} E_{\rm MM} f_{\rm LTG}\,{\rm d}M_{\star} {\rm d}z,\label{rhowet}
\end{equation}
where the different elements in the integral are (Fig.~\ref{seedfig}):

\begin{itemize}
\item the global mass function, parametrised with a Schechter function
	\begin{equation}
		\Phi\,(z,M_{\star}) = \frac{\phi^{*}(z)}{M_{\star}^{*}(z)}\,\bigg( \frac{M_{\star}}{M_{\star}^{*}(z)} \bigg)^{\alpha(z)} \exp \bigg(-\frac{M_{\star}}{M_{\star}^{*}(z)}\bigg).\label{lfunc}
	\end{equation}
We assumed the Schechter function parameters from \citet{pgon08} and used 
their parametrisation with redshift provided by \citet{clsj10pargoods}, 
\begin{eqnarray}
\log (\phi^{*}(z)/{\rm Mpc}^{-3}) = -2.72 - 0.56(z - 0.5),\\
\log (M^{*}_{\star}(z)/M_{\odot}) = 11.23 + 0.13(z - 0.5),\\ 
\alpha(z) = -1.22 - 0.04(z-0.5).
\end{eqnarray}

\item The gas-rich major merger rate, $R_{\rm MM}(z)$, 
as measured in Sect.~\ref{mrevol}. We assumed that it is independent of the stellar mass.

\item The merger efficiency function, $E(z,M_{\star}, M_{\star,{\rm lim}}, \mu_{\rm max}, \mu_{\rm min})$. This function provides the probability that a gas-rich merger with a mass ratio $\mu_{\rm min} < \mu \leq \mu_{\rm max}$ produces an early-type remnant more massive than $M_{\star,{\rm lim}}$. The merger efficiency function takes several effects into account:

\begin{itemize}
\item The stellar mass difference $\mu$ needed to reach $M_{\star,{\rm lim}}$ for a given $M_{\star}$. For example, and regarding major mergers with $\mu \geq 1/4$, the $\mu_{\rm max} = 1$ mergers of galaxies with $M_{\star} \geq 10^{10.7}\ M_{\odot}$ will provide a massive remnant with $M_{\star} \geq M_{\star,{\rm lim}} = 10^{11}\ M_{\odot}$, while for $\mu_{\rm min} = 1/4$ only galaxies with $M_{\star} \geq 10^{10.9}\ M_{\odot}$ can be progenitors of massive ETGs. Thus, mergers below $M_{\star,{\rm lim}}/(1 + \mu_{\rm min})$ do not contribute to the assembly of massive ETG, while all mergers above $M_{\star,{\rm lim}}/(1 + \mu_{\rm max})$ contribute. Between $\mu_{\rm min}$ and $\mu_{\rm max}$, we used the dependence of the merger fraction on the mass ratio $\mu$, $f_{\rm MM} \propto \mu^{s}$ \citep[e.g.,][]{clsj11mmvvds}, to estimate the effective merger rate at $M_{\star} = M_{\star,{\rm lim}}/(1 + \mu)$.

\item The extra stellar mass in the final remnant due to the star formation during the merger, $f_{\rm sf}$. This extra stellar mass simply decreases the limiting mass for a given $\mu$ described in the previous item by a factor $(1 + f_{\rm sf})^{-1}$.

\begin{figure}[t!]
\resizebox{\hsize}{!}{\includegraphics{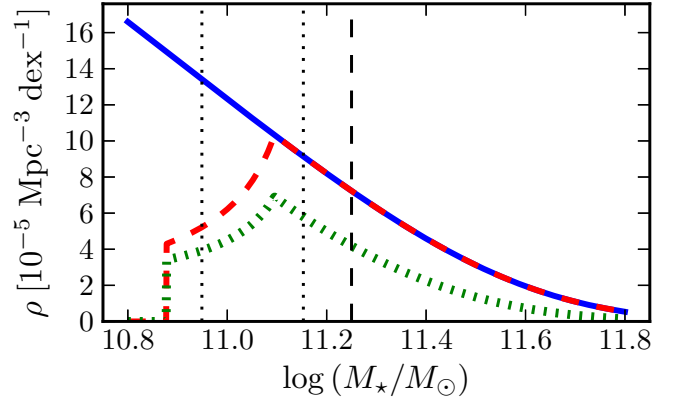}}
\caption{Number density distribution from the different functions in Eq.~(\ref{rhowet}) at
$z = 1$. The solid line is the stellar mass function ($\Phi$) multiplied by the
major merger rate ($R_{\rm MM}$). The dashed line includes the major merger efficiency
function ($E_{\rm MM}$). The dotted line includes the fraction of late-type galaxies 
($f_{\rm LTG}$) and provides the final function that we integrate over stellar
mass and cosmic time to obtain $\rho_{\rm wet}$. The vertical dashed line marks the
limiting stellar mass that define massive galaxies, 
$M_{\star,\rm lim} = 10^{11.25}\ M_{\odot}$,
while the dotted lines mark the stellar masses $M_{\star,1} = 10^{10.15}\ M_{\odot}$ 
and $M_{\star,2} = 10^{11.05}\ M_{\odot}$ when $f_{\rm sf} = 0$ 
(see text for details).
[{\it A colour version of this plot is available at the electronic edition}].}
\label{seedfig}
\end{figure}

\item The fraction of remnants of a gas-rich major merger that are ETGs. We express this fraction as $1 - f_{\rm disc}$, where $f_{\rm disc}$ is the fraction of remnants that rebuild a disc component and are classified as late-type galaxies after the merger (see Sect.~\ref{modelun}, for further details).
\end{itemize}

Formally, the merger efficiency function is defined as
\begin{equation}
E = \left\{\begin{array}{ll}
1 - f_{\rm disc}, & \quad {\rm if}\ M_{\star} \geq M_{\star,1}, \\
(1 - f_{\rm disc}) \times (\mu/\mu_{\rm min})^{s}, & \quad {\rm if}\ M_{\star,2} \leq M_{\star} < M_{\star,1},\\
0, & \quad {\rm if}\ M_{\star} < M_{\star,2},
\end{array}\right.
\end{equation}
where 
\begin{eqnarray}
M_{\star,1} = \frac{M_{\star, \rm lim}}{(1+\mu_{\rm min})(1+f_{\rm sf})},\\ 
M_{\star,2} = \frac{M_{\star, \rm lim}}{(1+\mu_{\rm max})(1+f_{\rm sf})},\\
\mu = \frac{M_{\star, \rm lim}}{M_{\star}(1+f_{\rm sf})} - 1.
\end{eqnarray}
For convenience, we define the major merger efficiency in Eq.~(\ref{rhowet}) as $E_{\rm MM} \equiv E(z,M_{\star},M_{\star,{\rm lim}}, \mu_{\rm max} = 1, \mu_{\rm min} = 1/4)$. We assumed that $f_{\rm disc} = 0$, i.e., all the merger remnants are ETGs (see Sect.~\ref{modelun}), and that $s = -0.9$ as in Sect~\ref{ffevol}. To estimate $f_{\rm sf}$ we used the parametrisation of the gas fraction as a function of stellar mass and cosmic time provided by \citet{rodrigues12},
\begin{equation}
f_{\rm gas}(z,M_{\star}) = \frac{M_{\rm gas}}{M_{\star} + M_{\rm gas}} = \frac{1}{1 + [M_{\star}/10^{A(t)}]^{\,B(t)}},\label{fgas}
\end{equation}
where $t$ is the cosmic time between redshift $z$ and the present,
\begin{eqnarray}
A(t) = 9.15 + 0.13t,\\
B(t) = 0.5 + 13.36\times \exp(-38.02/t).
\end{eqnarray}
$A(t)$ represents the stellar mass at a given time for which the gas
fraction is equal to 50\%. The parameter increases linearly with
lookback time. $B(t)$ corresponds to the slope of the function.
Then, we assumed the prescriptions in \citet{hopkins09disksurvive} to 
estimate the amount of the initial gas mass that is transformed into stars during 
the merger, $f_{\rm burst} = f_{\rm gas}(1-f_{\rm gas})2\mu/(1+\mu)$. Finally,
\begin{equation}
f_{\rm sf} = M_{\rm burst}/M_{\star} = \frac{2}{3}f_{\rm gas},
\end{equation}
where the factor 2/3 is the efficiency for $\mu = 1/2$ mergers, which is the 
typical major merger mass ratio \citep{clsj11mmvvds,clsj12sizecos}.

\item The fraction of late-type galaxies (spirals and irregulars), $f_{\rm LTG}(z,M_{\star})$, is the number of potential gas-rich merger progenitors over the total population. We parametrise this fraction as
	\begin{equation}
		f_{\rm LTG}(z,M_{\star}) = 0.56 + 0.16z - 0.58[\log (M_{\star}/M_{\odot}) - 11].\label{fltgzmass}
	\end{equation}
We estimated the redshift dependence from the $f_{\rm LTG}$ reported by \citet{buitrago12}
at $z < 3$ (Fig.~\ref{fltgzmass_fig}, top panel), and the mass dependence from the late-type fractions in the SDSS
presented in \citet[][Fig.~\ref{fltgzmass_fig}, bottom panel]{bernardi10}. 
We assumed $f_{\rm LTG} = 1$ when Eq.~(\ref{fltgzmass}) is higher than one, and $f_{\rm LTG} = 0$ when it is negative.
\end{itemize}

\begin{figure}[t!]
\resizebox{\hsize}{!}{\includegraphics{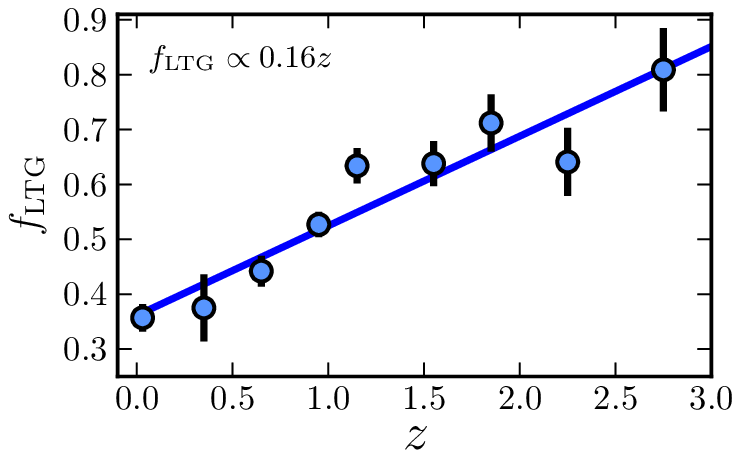}}
\resizebox{\hsize}{!}{\includegraphics{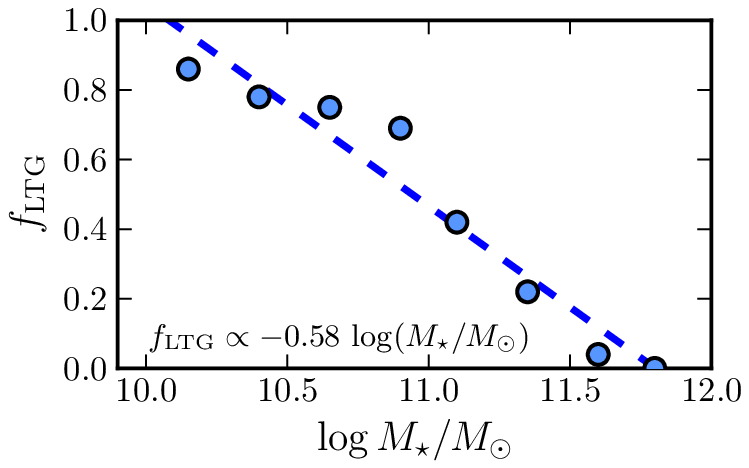}}
\caption{Fraction of late-type galaxies as a function of redshift (top panel) and stellar mass at $z \sim 0$ (bottom panel).
The redshift data points are from \citet{buitrago12}, and the stellar mass ones from \citet{bernardi10} in the SDSS. The line in both panels is the best least-squares linear fit to the data.
[{\it A colour version of this plot is available at the electronic edition}].}
\label{fltgzmass_fig}
\end{figure}

Finally, we define the fraction of new massive ETGs due to gas-rich (wet) mergers as
\begin{equation}
f_{\rm wet}(z_{\rm min},z_{\rm max}) = \frac{\rho_{\rm wet}(z_{\rm min},z_{\rm max},10^{11.25}\ M_{\odot})}{\rho_{\rm ETG}(z_{\rm min}) - \rho_{\rm ETG}(z_{\rm max})}.
\end{equation}

Our merger model finds $f_{\rm wet}(0,1.3) \sim 50$\%, while this fraction 
increases up to $f_{\rm wet}(z,1.3) \gtrsim 90$\% when we focus on
the high$-z$ regime, $z \gtrsim 0.8$ (dashed line in Fig.~\ref{rhofig}). This indicates that
gas-rich major mergers are common enough at $z \gtrsim 0.8$ to explain 
the observed increase in the number density of massive
ETGs. However, at $z \lesssim 0.8$ only one third of the evolution can be
accounted by these mergers, $f_{\rm wet}(0,0.8) \sim 40$\%. That supports
the idea that gas-rich major merging is the main process involved
in the assembly of the red sequence at $z \gtrsim 1$
\citep[e.g.,][]{ilbert10,clsj10megoods,eliche10I,prieto13}.

In addition to wet mergers, dry mergers can also increase
the number density of massive ETGs. In this case they promote
ETGs with $M_{\star} < 10^{11.25}\ M_{\odot}$ to the massive regime. 
We estimate the contribution of both major and minor dry mergers as

\begin{eqnarray}
\rho^{+}_{\rm dry}(z_{\rm min},z_{\rm max},M_{\star,\rm lim}) =\!\!\int_{z_1}^{z_2}\!\!\!\int_{0}^{M_{\star,\rm lim}}\!\!\!\!\!\!\!\!\Phi\,R_{\rm MM}^{\rm ETG} E_{\rm MM} f_{\rm ETG}\,{\rm d}M_{\star} {\rm d}z \nonumber\\ + \int_{z_1}^{z_2}\!\!\!\int_{0}^{M_{\star,\rm lim}}\!\!\!\!\!\!\!\!\!\Phi\,R_{\rm mm}^{\rm ETG} E_{\rm mm} f_{\rm ETG}\,{\rm d}M_{\star} {\rm d}z, 
\end{eqnarray}
where $R_{\rm MM}^{\rm ETG}$ ($R_{\rm mm}^{\rm ETG}$) is the major (minor) merger rate 
of massive ETGs galaxies from \citet{clsj12sizecos}, $f_{\rm ETG} = 1 - f_{\rm LTG}$, the minor merger efficiency
function is defined as $E_{\rm mm} \equiv E(z,M_{\star},M_{\star,{\rm lim}}, \mu_{\rm max} = 1/4, \mu_{\rm min} = 1/10)$,
and we assumed $f_{\rm sf} = 0$ in both major and minor merger efficiency functions. 
Note that the integration in mass space only reach $M_{\star,\rm lim}$ and
that the merger rates from \citet{clsj12sizecos} include
also mixed mergers (ETGs - LTGs). However, dry mergers between two already 
massive ETGs {\it decrease} the number density $\rho_{\rm ETG}$. We take this into account with 
the following function,
\begin{eqnarray}
\rho^{-}_{\rm dry}(z_{\rm min},z_{\rm max},M_{\star,\rm lim}) =\!\!\int_{z_1}^{z_2}\!\!\!\int^{\infty}_{M_{\star,\rm lim}}\!\!\!\!\!\!\!\!\Phi\,R_{\rm MM}^{\rm ETG} \epsilon_{\rm MM} f_{\rm ETG}\,{\rm d}M_{\star} {\rm d}z \nonumber\\ + \int_{z_1}^{z_2}\!\!\!\int^{\infty}_{M_{\star,\rm lim}}\!\!\!\!\!\!\!\!\!\Phi\,R_{\rm mm}^{\rm ETG} \epsilon_{\rm mm} f_{\rm ETG}\,{\rm d}M_{\star} {\rm d}z, 
\end{eqnarray}
where in this case the merger efficiency function has the form
\begin{equation}
\epsilon = \left\{\begin{array}{ll}
0.65, & \quad {\rm if}\ M_{\star} \geq M_{\star,3}, \\
0.65\times(\mu/\mu_{\rm min})^{s} & \quad {\rm if}\ M_{\star,4} \leq M_{\star} < M_{\star,3},\\
0, & \quad {\rm if}\ M_{\star} < M_{\star,4},
\end{array}\right.
\end{equation}
where 
\begin{eqnarray}
M_{\star,3} = \mu_{\rm min}^{-1}\,M_{\star, \rm lim},\\ 
M_{\star,4} = \mu_{\rm max}^{-1}\,M_{\star, \rm lim},\\
\mu = \frac{M_{\star, \rm lim}}{M_{\star}}.
\end{eqnarray}
and the factor 0.65 is the fraction of companions of massive galaxies that are 
already early-type/red galaxies \citep{clsj12sizecos,newman12}. As previously, 
we define the major merger efficiency as 
$\epsilon_{\rm MM} \equiv \epsilon(z,M_{\star},M_{\star,{\rm lim}}, \mu_{\rm max}~=~1, \mu_{\rm min}~=~1/4)$ and the minor merger efficiency as 
$\epsilon_{\rm mm} \equiv \epsilon(z,M_{\star},M_{\star,{\rm lim}}, \mu_{\rm max}~=~1/4, \mu_{\rm min}~=~1/10)$.

Analogous to the wet merger case, we define
\begin{eqnarray}
f_{\rm dry}(z_{\rm min},z_{\rm max}) = \frac{\rho_{\rm dry}(z_{\rm min},z_{\rm max},10^{11.25}\ M_{\odot})}{\rho_{\rm ETG}(z_{\rm min}) - \rho_{\rm ETG}(z_{\rm max})} \nonumber\\
= \frac{\rho^{+}_{\rm dry}(z_{\rm min},z_{\rm max},10^{11.25}\ M_{\odot}) - \rho^{-}_{\rm dry}(z_{\rm min},z_{\rm max},10^{11.25}\ M_{\odot})}{\rho_{\rm ETG}(z_{\rm min}) - \rho_{\rm ETG}(z_{\rm max})}
\end{eqnarray}
to explore the role of dry mergers in massive ETGs assembly since $z = 1.3$. 
We find $f_{\rm dry}(0,1.3) \sim 40$\%. As shown in Fig.~\ref{rhofig}, 
dry mergers are more important at recent cosmic times due to the increase in 
the number density of ETGs, in contrast with the diminishing importance of gas-rich mergers. 
For example, we have $f_{\rm dry}(0.8,1.3) \sim 15$\%, while $f_{\rm dry}(0,0.8) \sim 45$\%.

The combined effect of gas-rich and dry mergers, $f_{\rm tot} = f_{\rm wet} + f_{\rm dry}$, 
is able to explain most of the evolution in $\rho_{\rm ETG}$ since $z = 1.3$, with
$f_{\rm tot}(0,1.3) \sim 90$\%. 
Thus, our model suggests merging as the main process in the assembly of massive ETGs 
since $z = 1.3$. Two thirds of the number density evolution is due to gas-rich major 
mergers, while one third is coming from major and minor dry mergers. 

The measurement of the merger rate at $z \gtrsim 1.5$ is needed to fully constraint 
the role of gas-rich major mergers in the early assembly of the red sequence, 
as well as the possible contribution of cold gas accretion in this mass assembly.

\subsubsection{Fast and slow rotators in the local universe}
The results from the SAURON project \citep{sauron}
propose a kinematical classification of morphological ETGs into
fast and slow rotators \citep{emsellem07}. Recently, the
ATLAS$^{\rm 3D,}$\footnote{http://www-astro.physics.ox.ac.uk/atlas3d/} 
\citep{atlas3d} survey has observed a representative sample 
of 260 nearby ETGs, finding that the fraction
of slow rotators increases with the dynamical mass \citep{emsellem11}. 
The cosmological simulation analysed by \citet{khochfar11} 
suggests that the main difference between fast and
slow rotators is their average number of major mergers, with fast
rotators having undergone one major merger, while slow rotators have undergone two. 
We use this fact and the merging model developed in the previous
section, to predict the fraction of slow rotators
in the local universe, $f_{\rm slow}$. We simply assumed that wet major
mergers produce fast rotators (first major merger event), dry major mergers produce slow
rotators (second major merger event), and dry minor mergers do not change the kinematical
state of ETGs. In addition, we took all the ETGs at $z = 1.3$ as
fast rotators.

With the previous assumptions, we expect $f_{\rm slow} \sim 60$\%, in
good agreement with the ATLAS$^{\rm 3D}$ result of $f_{\rm slow} \sim 47 - 75$\%
for ETGs with a dynamical mass $M_{\rm dyn} \geq 10^{11.25}\ M_{\odot}$ \citep{emsellem11}. 
Thus, our model also reproduces the local fraction of slow/fast rotators of
massive ETGs galaxies, reinforcing the conclusions of the previous
section.

\subsubsection{Uncertainties in the model}\label{modelun}
The model presented in previous sections has set most of its
parameters from observational results. However, there are two unconstrained 
parameters that could affect our conclusions. The
first parameter is $f_{\rm disc}$, the fraction of gas-rich major mergers
whose remnant rebuild a disc component and do not contribute
to the increase in $\rho_{\rm ETG}$. We assumed $f_{\rm disc} = 0$, 
and in the following we justify this selection. \citet{hopkins09disksurvive}
find that a high gas fraction prevents the destruction of the disc component after
a major merger in their $N-$body/hydrodynamical simulations (but see \citealt{bournaud11}
for a different point of view). These simulations suggest
that disc rebuild could be an efficient process when the gas fraction is
$f_{\rm gas} \gtrsim 50$\%.
However, we have focused in the massive end of the
galaxy population, where the gas fractions are lower. Thanks to
Eq.~(\ref{fgas}), we can estimate the gas fraction of the gas-rich mergers in our model. 
We find $f_{\rm gas} \lesssim 30$\%, justifying the assumed $f_{\rm disc} = 0$. 
At lower masses than explored in the present section, the gas fraction is
higher, and disc rebuild could be an important process in the formation 
of bulge-dominated spirals \citep[e.g.,][]{huertas10,puech12}. 
However, we cannot discard positive values of $f_{\rm disc}$ for massive galaxies, 
as we see below.

The second parameter is the assumed merger time scale,
which typically has a factor of two uncertainty in their normalization 
\citep[e.g.,][]{hopkins10mer}. The $T_{\rm MM}$ from \citet{kit08} are 
typically longer than others in the literature
\citep[e.g.,][]{patton08,lin10} or similar to those
from $N-$body/hydrodynamical simulations \citep{lotz10t,lotz10gas}.
Thus, we expect, if anything, a shorter $T_{\rm MM}$, which implies a
larger number density of ETGs due to mergers. Nevertheless,
the good description of the $\rho_{\rm ETG}$ evolution with our merger
model strongly suggests that mergers are indeed the main process 
in massive ETGs assembly. Thus, a lower $T_{\rm MM}$ (i.e., a
higher merger rate that translates to an excess of ETGs) could
be compensated by a positive value of $f_{\rm disc}$, that reduces the
number density of ETGs due to mergers.

Future observational studies will be important to better constraint the
parameters of our model, and further theoretical efforts are needed to
understand the uncertainties in the assumed parameters.

\section{Summary and conclusions}\label{conclusion}
Using SINFONI/VLT 3D spectroscopy, we have been able
to measure, for the first time with spectroscopically-confirmed close pairs,
the gas-rich major merger fraction and merger rate at around the
peak in star formation activity at $0.9 < z < 1.8$, from the MASSIV
sample of star-forming galaxies with a stellar mass range $M_{\star} = 10^{9} - 10^{11}\ M_{\odot}$. 
In this redshift range we identify 20 close
pairs, and classify 13 as major mergers based on a separation
$r_{\rm p} \leq 20h^{-1} - 30h^{-1}$ kpc and a relative velocity 
$\Delta v \leq 500$ km~s$^{-1}$.

We find that the gas-rich major merger fraction is high, $20.8^{+15.2}_{-6.8}$\%, 
$20.1^{+8.0}_{-5.1}$\%, and $22.0^{+13.7}_{-7.3}$\% for $r_{\rm p} \leq 20h^{-1}$ kpc 
close pairs in redshift ranges $z = [0.94, 1.06], [1.2, 1.5)$, and $[1.5, 1.8)$, respectively.
When compared to measurements at redshifts $z < 1$, the evolution 
of the (gas-rich) merger fraction can be parametrised as 
$f_{\rm MM} = 0.0066 \times (1 + z)^{m}$ with $m = 3.91 \pm 0.16$. 
We note that the evolution between $z = 1$ and $z \sim 1.5$ seems 
to flatten out compared to lower redshifts.

The merger rate has been derived using merger time scales
from the literature: we find that the gas-rich merger rate is $0.116^{+0.084}_{-0.038}$~Gyr$^{-1}$, 
$0.147^{+0.058}_{-0.037}$~Gyr$^{-1}$, and $0.127^{+0.079}_{-0.042}$~Gyr$^{-1}$ 
at $z = 1.03, 1.32$, and $1.54$, respectively, for merger time scales of $T_{\rm MM} \sim 1.5$ Gyr. 
We then find that the (gas-rich) merger rate's evolution for galaxies with stellar mass
$\overline{M}_{\star} = 10^{10-10.5}\ M_{\odot}$ over $z = [0, 1.5]$ scales as 
$(1 + z)^n$ with $n = 3.95 \pm 0.12$.

Using these measurements, we developed a simple model to estimate the contribution 
of gas-rich major mergers to the growth of galaxies on the red sequence. 
We infer that $\sim35$\% of the star-forming galaxies 
with stellar mass $\overline{M}_{\star} = 10^{10}-10^{10.5}\ M_{\odot}$ 
have undergone a major merger since $z \sim 1.5$.
The number of major merger events was about 5 times higher at $1 < z < 1.5$ 
compared to $z < 1$. Assuming that each gas-rich major merger produces a new
early-type galaxy, we infer that the number of gas-rich major mergers is
large enough at $z > 1$ to explain the increase in the number density 
of massive ETGs, supporting a picture where gas-rich (wet)
merging is the main process building-up the red sequence. 
While gas-rich mergers become rarer towards lower redshifts, the number 
of dry mergers is steadily increasing, and the combination of
these two processes accounts for most, if not all, of the increase
in the number density of massive ETGs since $z \sim 1.3$. Two-thirds of this
number density evolution is due to wet major mergers, while
one-third is coming from major and minor dry mergers. These
results add further evidence to a picture where merging is a major 
process driving mass assembly into the massive red sequence galaxies.

We note that minor merging is definitely present in the
MASSIV sample (see Sect.~\ref{ffmassiv}). However, due to incompleteness
in detecting these faint companions, we are not able to assess a
minor merger rate at these epochs from our data. In the global
picture of red sequence assembly, we emphasise that a simple
extrapolation of the minor merger rate measured up to $z \sim 1$ by
\citet{clsj11mmvvds}, would indicate that from $z \sim 1.5$ to
the present, minor mergers with $1/10 \leq \mu < 1/4$ are not 
common enough to significantly move spiral galaxies into the red
sequence.

To get a complete picture of the life-time effect of major merging 
on massive galaxies, accurate measurements of the
merger fraction and merger rate are needed beyond $z \sim 2$.
Spectroscopic surveys will remain an important element to provide 
secure identification of merging systems at these early
epochs.

\begin{acknowledgements}
We dedicate this paper to the memory of our six IAC colleagues and friends who
met with a fatal accident in Piedra de los Cochinos, Tenerife, in February 2007,
with a special thanks to Maurizio Panniello, whose teachings of \texttt{python}
were so important for this paper.

We thank the comments and suggestions of the anonymous referee, and L. Pozzetti and
F. Buitrago for kindly provide their number densities. This work has been supported 
by funding from ANR-07-BLAN-0228 and ERC-2010-AdG-268107-EARLY
and partially supported by the CNRS-INSU and its Programme National 
Cosmologie-Galaxies (France) and by the French ANR grant ANR-07-JCJC-0009.
\end{acknowledgements}

\bibliography{biblio}
\bibliographystyle{aa}

\begin{appendix}

\section{Spectroscopic Success Rate in the VVDS-Wide fields}\label{ssrvvds}
We computed the Spectroscopic Success Rate ($SSR$) in the VVDS-Wide 14h and 22h 
fields for a given redshift range by following the prescriptions in \citet{ilbert06}: 
we compared the number of galaxies with flag = 4, 3 and 2 over the total number 
of galaxies (those with flag = 4, 3, 2, 1 and 0). Because flag = 1 galaxies have 
a 50\% reliability and no redshift information is available for flag = 0 galaxies, 
we took advantage of the latest photometric redshifts from CFHTLS survey to estimate 
the number of galaxies with flag = 1 and 0 that belong to the redhsift range of interest. 
With the previous steps we estimate the $SSR$ in the 22h field as
\begin{equation}
SSR_{\rm 22h}(z_{\rm r}) = \frac{N_{\rm 22h, spec}(z_{\rm r})}{N_{\rm 22h, spec}(z_{\rm r}) + N_{\rm 22h, phot}(z_{\rm r})},
\end{equation}
where $N_{\rm 22h, spec}(z_{\rm r})$ is the number of galaxies with 
flag = 4, 3 and 2 in the redshift range $z_{\rm r}$, and $N_{\rm 22h, phot}(z_{\rm r})$ 
is the number of galaxies with flag = 1 and 0, and with a photometric redshift 
in the same redshift range.  As the available imaging data in the 14h
ﬁeld is not as deep as the CFHTLS, we show in the following that
we can assume that the $SSR$ in the 14h field follows the same
distribution as in the 22h field, for which CFHTLS photometry
is available.

First we checked the properties of the stars in these two
VVDS-Wide fields. The fraction of stars in the 22h (14h) field is
66\% (51\%) for flag = 4 sources, 38\% (26\%) for flag = 3 sources,
24\% (23\%) for flag = 2 sources, and 11\% (12\%) for flag = 1
sources. So the fraction of stars is similar in both fields for flag
= 2 and 1 sources, while higher in the 22h field for flag = 4 and
3. We checked that the normalised distributions of the VVDS-Wide 
stars as a function of their observed $I_{\rm AB}-$band magnitude
in both fields are similar for each flag. We note that stars with
high confidence flag are brighter than the low confidence ones,
as expected. The fraction of bright stars ($I_{\rm AB} \leq 21$) 
in the 22h (14h) field is 75\% (67\%) for flag = 4 stars, 49\% (43\%) 
for flag = 3 stars, 30\% (29\%) for flag = 2 stars, and 19\% (16\%) 
for flag = 1 stars. These fractions suggest that there is a higher density
of bright stars in the 22h area than in the 14h one, that translates
in a higher fraction of stars for flag = 4 and 3 sources, while
faint stars have similar densities in both fields, leading to a similar 
fraction of stars with flag = 2 and 1. Because of this, we
assumed that the fraction of stars among flag = 0 sources, that
we estimate in the 22h field from the CFHTLS photometry, is
similar in 22h and 14h fields.

The distribution of galaxies at $z \geq 0.9$, the redshift range in which we 
are interested on, are also similar in both fields. Because of this, we assumed 
that the photometric distribution of galaxies with flag = 1 and 0 is similar in both fields, 
and we use those in 22h field to estimate $N_{\rm 14h,phot}$ and the $SSR$ in the 14h field:
\begin{equation}
SSR_{\rm 14h}(z_{\rm r}) = \frac{N_{\rm 14h,spec}(z_{\rm r})}{N_{\rm 14h,spec}(z_{\rm r}) + f_{\rm 22h}(z_{\rm r})\times n_{\rm 14h}(1,0)},
\end{equation}
where $n_{\rm 14h}(1,0)$ is the total number of sources (galaxies and stars) 
with flag = 1 and 0 in the 14h field, 
and $f_{\rm 22h}(z_{\rm r}) = N_{\rm 22h,phot}(z_{\rm r})/n_{\rm 22h}(1,0)$ is the fraction 
of sources with a photometric redshift in the redshift range $z_{\rm r}$ over 
the total population of sources with flag = 1 and 0 in the 22h field.

\section{GALFIT residual images of blended MASSIV close pairs}\label{galres}

\begin{figure*}[!ht]
	\resizebox{\hsize}{!}{\includegraphics{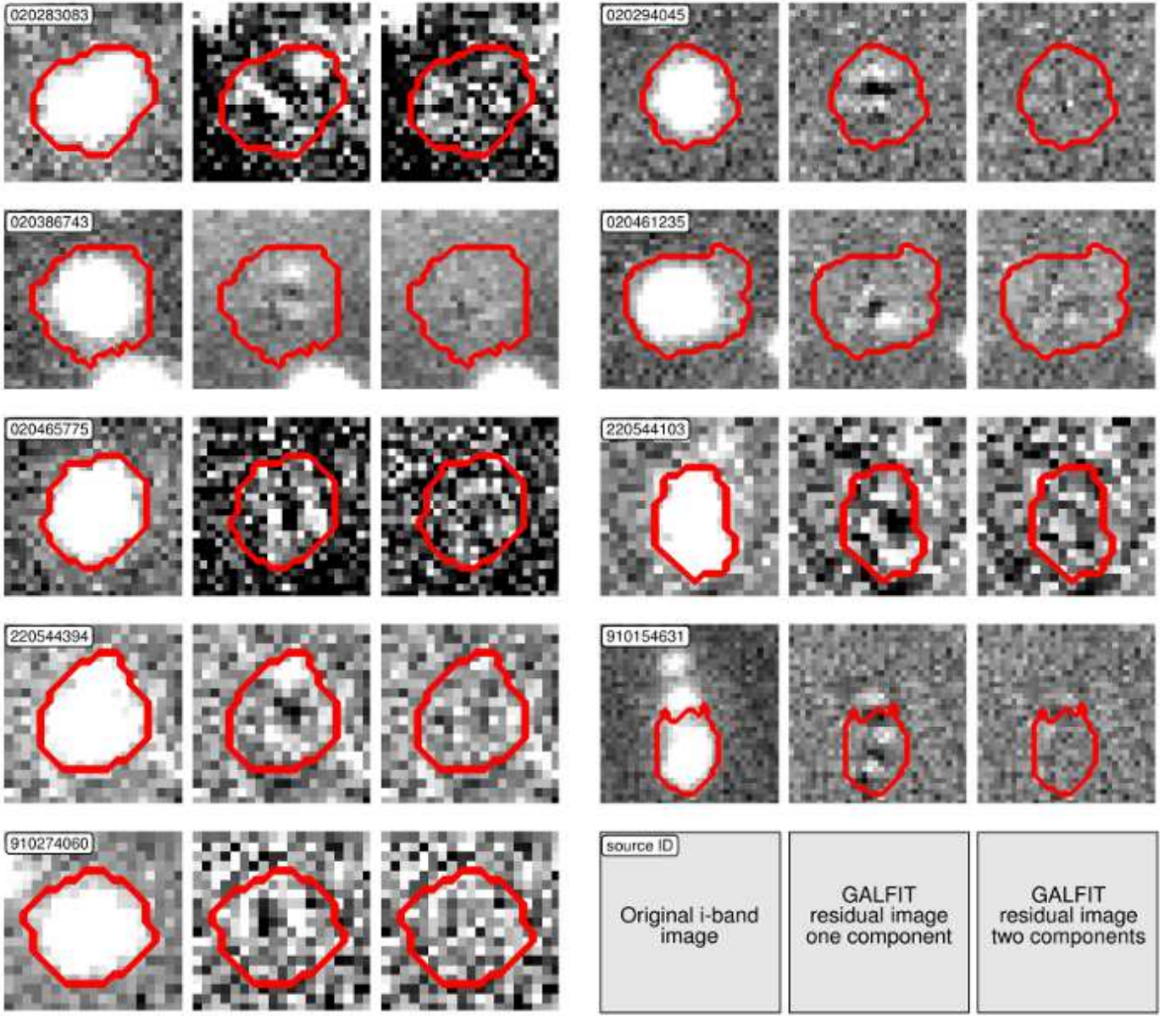}}
	\caption{Original $i-$band image ({\it left panels}) and residual images from the GALFIT fit with one ({\it central panels}) and two components ({\it right panels}) of those MASSIV close pair candidates with overlapping components in the kinematical maps. The grey scale has been chosen to enhance the light residuals in the images. The solid contour marks the targeted MASSIV source to guide the eye. The ID of the showed MASSIV source is labelled in each left panel. [{\it A colour version of this plot is available at the electronic edition}].}
	\label{galfitres}
\end{figure*}

In this Appendix we detail the modelling in the $i$ band of those MASSIV close pair candidates with overlapping components in the kinematical maps. We used GALFIT v3.0 \citep{galfit3} to model the light distribution with two independent S\'ersic components.
We set the initial positions of the sources using the information from the kinematical maps, and we did not impose any constraint on the other initial parameters of the fit (i.e., luminosity, effective radius, S\'ersic index, position angle and inclination). Because of the minimisation process preformed by GALFIT in the fitting, the best model with two components should not be unique and the convergence to a given solution should depend on the initial values of the parameters defined by the user. We checked, by exploring randomly the space of initial values, that the number of good solutions is at most two. Where two good solutions exist, the one that better reproduces both the velocity field and the velocity dispersion map is preferred. To illustrate the need of a second component to describe the blended close pair systems in MASSIV, we show in Fig.~B.1 the original image in the $i$ band and the residual image from GALFIT with one and two components. In all the cases the residual map from the one component fit suggests that a second component is present. This Figure also demonstrates that both the seeing ($\sim0.7\arcsec$) and the depth of the $i-$band images used in the present study are good enough to characterise sources with two overlapping components.

\section{Completeness of the close pair sample}\label{maglim}

\begin{figure}[!ht]
\resizebox{\hsize}{!}{\includegraphics{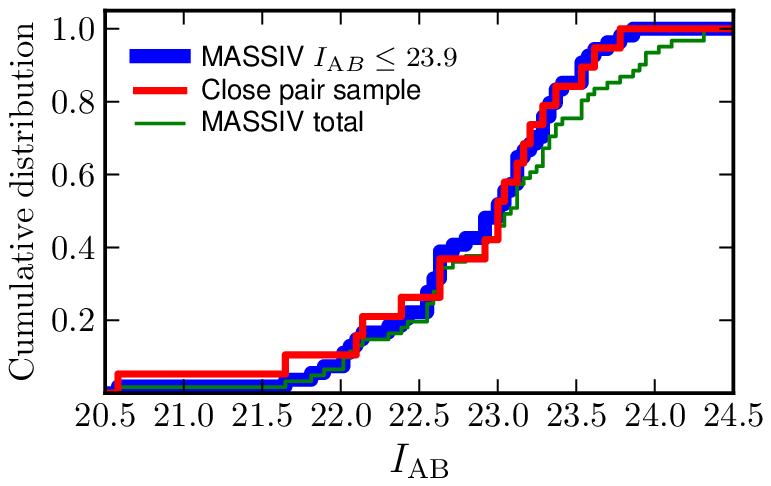}}
\resizebox{\hsize}{!}{\includegraphics{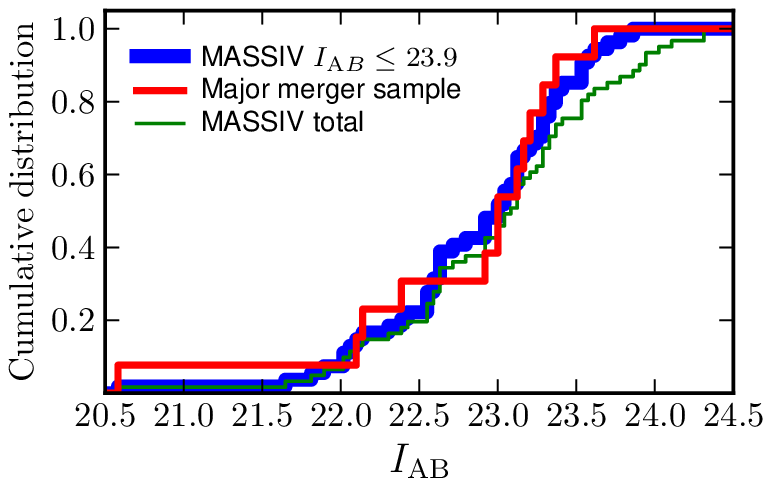}}
\resizebox{\hsize}{!}{\includegraphics{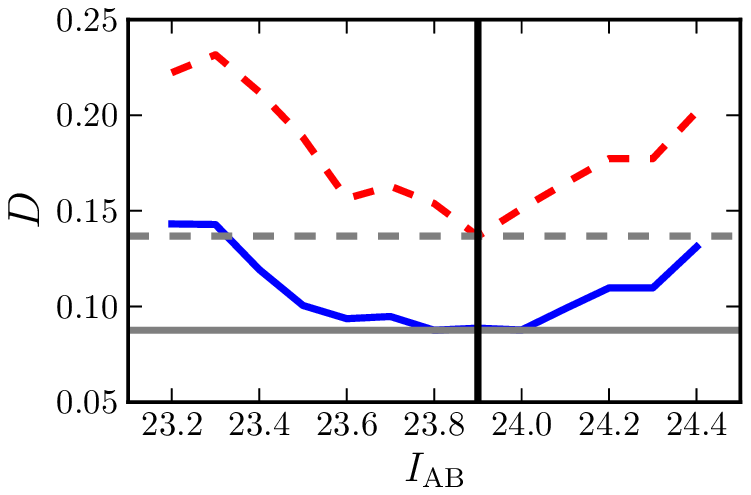}}
\caption{{\it Top}: cumulative distribution in the apparent $I_{\rm AB}$ magnitude 
of the total MASSIV sample (green line), the close pair sample (red line) and
the total MASSIV sample with $I_{\rm AB} \leq 23.9$ (blue line). 
{\it Middle}: the same than before, but the red line shows the 
distribution of the major merger sample. 
{\it Bottom}: Kolmogorov-Smirnov estimator $D$
of the total MASSIV sample distribution against the close pair (blue
solid line) and the major merger (red dashed line) distributions. We
computed $D$ for those galaxies brighter than a given $I_{\rm AB}$ magnitude.
The solid (dashed) horizontal line marks the minimum in the $D$ distribution 
for close pairs (major mergers). The vertical solid line marks the
$I_{\rm AB}$ magnitude at the minimum value of $D$, $I_{\rm AB,comp} = 23.9$. 
[{\it A colour version of this plot is available at the electronic edition}].}
\label{completefig}
\end{figure}

Throughout the present paper we have assumed that the close pair
systems detected in the MASSIV sample are representative for
galaxies with $I_{\rm AB} \leq 23.9$. The MASSIV sample comprises
galaxies fainter than this luminosity limit, and in this appendix
we justify the boundary applied in our analysis.

The VVDS parent samples of the MASSIV sample are randomly selected 
in the $I_{\rm AB}$ band \citep{lefevre05}. Because of
this, we studied the distribution of MASSIV sources as a function
of the $I_{\rm AB}-$band magnitude to obtain clues about the completeness 
of the MASSIV sample in our close pair study. We show
the cumulative distribution of all MASSIV sources and of those
with a close companion (both major and no major) in the top panel
of Fig.~\ref{completefig}. Obviously, we are able to detect single MASSIV
galaxies at fainter magnitudes than the close pairs ($I_{\rm AB} = 24.4$ vs 23.8)
because in the latter case we have to detect both the principal
source and the companion galaxy, which is usually fainter.

Lets assume for a moment that the detection curve of the
MASSIV sources is a step function that is 1 for $I_{\rm AB} \leq I_{\rm AB,lim}$
and 0 for fainter magnitudes. In this ideal case, $I_{\rm AB,lim}$ is defined
by the fainter galaxy (close pair) detected. Thus, we should only
be able to detect close pairs at $I_{\rm AB} \leq 23.8$, and the measured
merger fraction in the total MASSIV sample ($I_{\rm AB} \leq 24.4$) will
be lower than the real merger fraction because of the missing
faint close pairs with $I_{\rm AB} > 23.8$. Of course, the detection curve
in the MASSIV sample is more complicated than a step function,
and could depend on redshift, geometry, luminosity, etc. Instead
of trying to estimate the detection curve for MASSIV sources and
for close companions to recover statistically the missing close
pairs \citep[see][for examples of this kind of correction]{patton00,deravel09}, 
we define the $I_{\rm AB}-$band magnitude up to which the detected close pairs are 
representative of the total MASSIV sample, named $I_{\rm AB,comp}$, and in
our study we only keep those sources brighter than $I_{\rm AB,comp}$ to
ensure reliable merger fractions.

The MASSIV sample is a representative subsample of the
global star-forming population \citep{massiv1}, and we expect 
the MASSIV galaxy pairs to be also a random sample of this 
global population. Therefore, the distributions in
the $I_{\rm AB}$ band of the total and the close pair MASSIV sources
should be similar when the close pairs detection was slightly affected
by incompleteness issues. Following this idea, we performed
a Kolmogorov-Smirnov (KS) test over the total and close pair
MASSIV galaxies with $I_{\rm AB} \leq I_{\rm AB,lim}$, and explored different 
values of $I_{\rm AB,lim}$, from 23.0 to 24.5. Then, the completeness magnitude 
$I_{\rm AB,comp}$ was defined by the minimum in the KS estimator $D$, i.e.,
where both distributions have the lower probability to be different. 
This exercise provides the curve in the bottom panel of Fig.~\ref{completefig}, 
that states $I_{\rm AB,comp} = 23.9$. We repeated this procedure
with the major merger sample, and we obtain a similar distribution of $D$ values, 
reinforcing our choice.

In summary, in the estimation of the merger fraction we only
use those MASSIV galaxies with $I_{\rm AB} \leq I_{\rm AB,comp} = 23.9$. This
ensures that the close pair sample is a random subsample of the
total MASSIV one, and we avoid any bias related with the shallower 
detection curve of close pairs compared with that of the
total MASSIV sample.

\end{appendix}

\end{document}